\documentclass{aa}  

\usepackage{graphicx}
\usepackage{subcaption}
\usepackage{mwe}
\usepackage{booktabs}
\usepackage{tabularx}
\usepackage{natbib}
\usepackage{xcolor}
\bibpunct{(}{)}{;}{a}{}{,}
\usepackage{gensymb}
\usepackage{graphicx}
\usepackage{txfonts}
\usepackage{siunitx}

\newcommand{\mockalph}[1]{}

\begin{document}

   \title{First constraints on the AGN X-ray luminosity function at $z\sim6$ from an eROSITA-detected quasar}

   \author{J. Wolf\,\thanks{jwolf@mpe.mpg.de}
          \inst{1,2},
          K. Nandra
          \inst{1},
          M. Salvato
          \inst{1}, 
          T. Liu
          \inst{1},
          J. Buchner
          \inst{1},
          M. Brusa
          \inst{3,4},
          D. N. Hoang
          \inst{5},
           V. Moss
          \inst{6,7},
           R. Arcodia
          \inst{1},
          \newline
          M. Br\"uggen
          \inst{5},
          J. Comparat
          \inst{1},
          F. de Gasperin
          \inst{5},
          A. Georgakakis
          \inst{8},
          A. Hotan
          \inst{9},
          G. Lamer
          \inst{10},
          A. Merloni
          \inst{1},
          A. Rau
          \inst{1},
          \newline
          H. J. A. Rottgering
          \inst{11},
          T. W. Shimwell
          \inst{12,11},
          T. Urrutia
          \inst{10},
          M. Whiting
          \inst{6},
          \and
          W. L. Williams
          \inst{12}
          }
          
    \titlerunning{eROSITA high-z quasar}
    \authorrunning{Wolf et al.}

   \institute{Max-Planck-Institut f\"{u}r extraterrestrische Physik, Gie\ss enbachstra\ss e 1, 85748 Garching, Germany
         \and
         Exzellenzcluster ORIGINS, Boltzmannstr. 2, D-85748 Garching, Germany 
         \and
         Dipartimento di Fisica e Astronomia dell’Università degli Studi di Bologna, via P. Gobetti 93/2, 40129 Bologna, Italy
         \and
         INAF/OAS, Osservatorio di Astrofisica e Scienza dello Spazio di Bologna, via P. Gobetti 93/3, 40129 Bologna, Italy
         \and
          Hamburger Sternwarte, University of Hamburg, Gojenbergsweg 112, 21029 Hamburg, Germany 
         \and 
         ATNF, CSIRO Astronomy and Space Science, PO Box 76, Epping, New South Wales 1710, Australia
         \and
         Sydney Institute for Astronomy, School of Physics A28, University of Sydney, Sydney, NSW 2006, Australia
        \and
         Institute for Astronomy and Astrophysics, National Observatory of Athens, V. Paulou and I. Metaxa, 11532, Greece
         \and
         ATNF, CSIRO Astronomy and Space Science, PO Box 1130, Bentley, WA 6102, Australia
         \and
          Leibniz-Institut für Astrophysik Potsdam, An der Sternwarte 16, 14482 Potsdam, Germany
         \and
         Leiden Observatory, Leiden University, PO Box 9513, 2300 RA Leiden, The Netherlands
         \and
         ASTRON, the Netherlands Institute for Radio Astronomy, Postbus 2, 7990 AA, Dwingeloo, The Netherlands}

   \date{Received October 20, 2020; accepted XX}

  \abstract
   {High-redshift quasars signpost the early accretion history of the Universe.
  The penetrating nature of X-rays enables a less absorption-biased census of the population of these luminous and persistent sources compared to optical/near-infrared(NIR) colour selection. The ongoing SRG/eROSITA X-ray all-sky survey offers a unique opportunity to uncover the bright end of the high-z quasar population and probe new regions of colour parameter space.}
  { We searched for high-z quasars within the X-ray source population detected in the contiguous  $\rm \sim 140 \, deg^2$ field observed by eROSITA  during the performance verification phase. With the purpose of demonstrating the unique survey science capabilities of eROSITA, this field was observed at the depth of the final all-sky survey.
   The blind X-ray selection of high-redshift sources in a large contiguous, near-uniform survey with a well-understood selection function can
be directly translated into constraints on the X-ray luminosity function (XLF), which encodes the luminosity-dependent evolution of accretion through cosmic time.}
   {We collected the available spectroscopic information in the eFEDS field, including the sample of all currently known optically selected z > 5.5 quasars and cross-matched secure Legacy DR8 counterparts of eROSITA-detected X-ray point-like sources with this spectroscopic sample. 
   }
   {We report the X-ray detection of eFEDSU J083644.0+005459, 
   an eROSITA source securely matched to the well-known quasar SDSS J083643.85+005453.3 (z=5.81).  
   The soft X-ray flux of the source derived from eROSITA is consistent with previous Chandra observations. The detection of SDSS J083643.85+005453.3 allows us to place the first constraints on the XLF at $z>5.5$ based on a secure spectroscopic redshift. Compared to extrapolations from lower-redshift observations, this favours a relatively flat slope for the XLF at $z\sim 6$
   beyond $L_{*}$, the knee in the luminosity function. In addition, we report the detection of the quasar with LOFAR at 145 MHz and ASKAP at 888 MHz. The reported flux densities confirm a spectral flattening at lower frequencies in the emission of the radio core, indicating that SDSS J083643.85+005453.3 could be a (sub-) gigahertz peaked spectrum source. The inferred spectral shape and the parsec-scale radio morphology of SDSS J083643.85+005453.3 indicate that it is in an early stage of its evolution into a large-scale radio source or confined in a dense environment. We find no indications for a strong jet contribution to the X-ray emission of the quasar, which is therefore likely to be linked to accretion processes. }
   {Our results indicate that the population of X-ray luminous AGNs at high redshift may be larger than previously thought. From our XLF constraints, we make the conservative prediction that eROSITA will detect $\sim 90$ X-ray luminous AGNs  at redshifts $5.7<z<6.4$ in the full-sky survey (De+RU). While subject to different jet physics, both high-redshift quasars detected by  eROSITA so far are radio-loud; a hint at the great potential of combined X-ray and radio surveys for the search of luminous high-redshift quasars.}

   \keywords{quasars: individual --
                Galaxies: high-redshift --
                X-rays: galaxies
               }

   \maketitle
%
\section{Introduction}

Active  galactic  nuclei  (AGNs) are the brightest persistent beacons in the universe and sign-post the population of accreting super-massive black holes (SMBHs) and their evolution throughout cosmic time. The  detection  of quasars at z $\rm > 5.5$ in the past two decades is an intriguing development, because their associated black hole masses challenge our understanding of the formation and initial growth of SMBHs. Testing black hole seed models requires a  complete  census  of  high-redshift  AGNs  encoded  in  well-constrained  luminosity  functions. In this redshift regime, X-rays in the soft band ($\rm 0.2-2.3 \, keV$ for eROSITA) probe the restframe hard X-ray emission of the distant sources ($\rm \sim 1.3-15 \, keV$ at $z>5.5$). The soft X-ray selection of AGNs therefore suffers less from absorption biases. However, so  far, optical and infrared dropout-selected  AGNs at $z > 5.5$  \citep[e.g][]{fan01,willott09,venemans13,reed15,banados16,matsuoka16,wang17, banados18} significantly  outnumber  X-ray  selected ones because of the lack of sufficiently wide and deep X-ray surveys backed up by homogeneous ancillary multi-wavelength data. Currently, about $ 345$ sources have been discovered in dedicated optical/near-infrared(NIR) surveys. Chandra  and  XMM-Newton  pointed observations of  known quasars in the range $z=5.7-7.54$  have  led  to  the  detection  of  an  X-ray  signal  for  only approximately $ 30$ of  these  objects 
\citep[e.g.][]{brandt02,nanni17,vito19,pons20}. However, such X-ray follow-up samples suffer from the selection biases of the optical selection because of tight colour--magnitude constraints and absorption.

   For  the  study  of  accretion  history, absorption biases can be avoided by constructing  the  X-ray  luminosity  function  \citep[XLF, ][]{hasinger05,ueda14,vito14,miyaji15,aird15,georgakakis15,buchner15,Khorunzhev18,ananna19} from  a  purely  X-ray-selected sample. However,  only  three  X-ray-selected  AGNs  have  been  identified  at z $>5$ so far \citep[in  the  Chandra  Deep  Fields,  COSMOS  and XMM-XXL][]{barger03,marchesi16,menzel16} with the most distant being at z=5.3. These  surveys  suffer  from  the  small  cosmological  volume  they  probe. The full-sky survey currently being carried out with the extended ROentgen Survey with an Imaging Telescope Array (eROSITA) on board the
Spectrum-Roentgen-Gamma (SRG) mission \citep{predehl20}  will allow us to overcome these limitations and probe the bright end of the XLF at high redshifts with a limiting flux of $\sim 10^{-15} \, \mathrm{erg \, s^{-1} \, cm^{-2}}$. For comparison, the second ROSAT all-sky survey catalogue \citep[2RXS][]{boller16} reached a depth of $\sim 10^{-13} \, \mathrm{erg \, s^{-1} \, cm^{-2}}$. Indeed, as early as the early months of the first eROSITA all-sky survey, \citet{medvedev20a,medvedev20b} reported the detection of the X-ray ultra-luminous source SRGE J142952.1+544716, which is matched to the z=6.18 quasar CFHQS J142952+544717 (henceforth CFHQJ14).

In the present work, we exploit the contiguous area of $\rm \sim 140 \, deg^2$ observed by eROSITA during the calibration and performance verification phase with the purpose of demonstrating the science capabilities of the all-sky survey after 4 years (eRASS:8, $\rm \sim 2.3\, ks$): the eROSITA Final Equatorial Depth Survey (eFEDS).
We report here the blind detection of a high-redshift X-ray source (eFEDSU J083644.0+005459), that we could identify as the well-known quasar SDSS J083643.85+005453.3 (z=5.81, \citealt{fan01}, henceforth SDSSJ08). This quasar was initially discovered by \textit{i}-band dropout selection in the main Sloan Digital Sky Survey (SDSS). 
CFHQJ14 and SDSSJ08 are the highest-redshift X-ray-selected AGNs known to date.
Taking advantage of the synergy between the eROSITA detection and new radio data from the LOw-Frequency Array \citep[LOFAR;][]{vanhaarlem13} and the Australian Square Kilometre Array Pathfinder \citep[ASKAP,][Hotan et al. submitted]{johnston08} Survey With ASKAP of GAMA-09 + X-Ray (SWAG-X, Moss et al. in prep) programme, we investigate the origin of the X-ray emission from SDSSJ08 and find further evidence for a confined jetted radio structure. We consequently discuss how the detection of the quasar in a contiguous survey constrains models of the space density of X-ray-emitting AGNs.

The optical counterpart determination procedure for eFEDS sources is outlined in Section 2. The identified quasar and the extraction and reduction of the eROSITA spectral data are presented in Section 3. In Section 4, we investigate the multi-wavelength properties of eFEDSU J083644.0+005459/SDSSJ08 using ancillary data. The new radio data from LOFAR and ASKAP are presented in Section 5. In Section 6, we derive constraints on the X-ray AGN space density at high redshift from the detection of SDSSJ08. 
We discuss the origin of the X-ray emission in Section 7. After comparing the detection to expected source counts from optical surveys, we conclude by making a prediction for the expected number of z>5.7 quasars which will be found in eRASS:8.    
We assume a flat $\rm \Lambda CDM$ cosmology \citep{planck18}: $\Omega_m=0.31$, $\Omega_\Lambda=0.69$ and $H_0=68 \, \mathrm{km\, s^{-1} \, Mpc^{-1}}$.  Unless stated otherwise, uncertainties are given at the $68 \%$ confidence level.
 \section{Optical counterparts to eFEDS sources}

\begin{figure}
\includegraphics[width=9 cm]{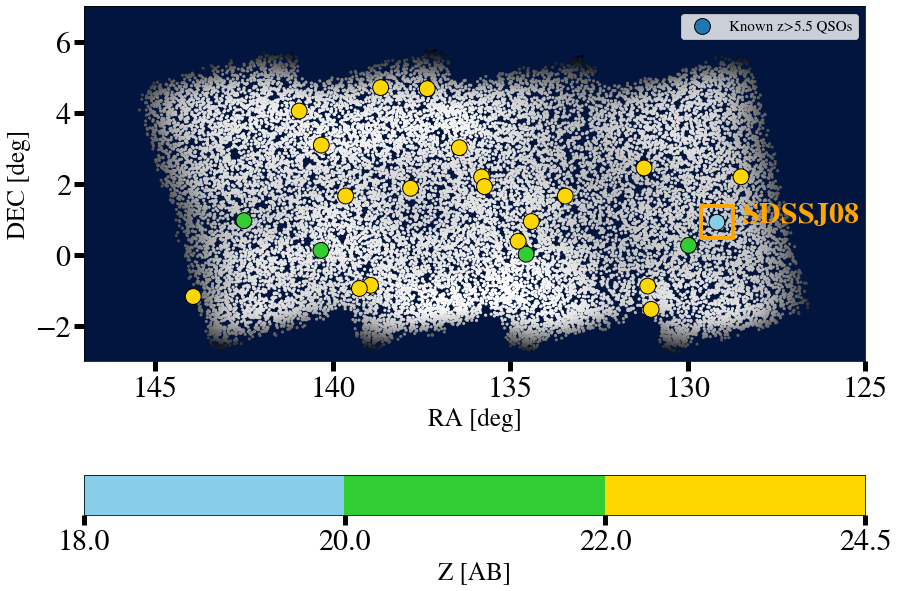}
\caption{Detected point-like sources in eFEDS colour-coded according to their spatial density (brighter is denser). The visible difference in source density is due to the non-uniform exposure of the eFEDS field (Brunner et al. in prep). Known z>5.5 QSOs in the footprint are shown as circles and are colour-coded according to their z-band magnitude. The detected quasar is marked by an orange square.  }
\label{fig:mon}
\centering
\end{figure}

The eFEDS survey was carried out by eROSITA between 3$^{\rm }$ and 7 November 2019 to a nominal depth of approximately $\rm 2.3 \, ks$, corresponding to a flux limit of $F_{0.5-2 \, \mathrm{keV}} \sim 10^{-14} \, \mathrm{erg\, s^{-1}\, cm^{-2}}$. The source detection is performed in the 0.2-2.3 keV band using a sliding box algorithm from the eROSITA Science Analysis Software System (eSASS). The catalogue is presented in  Brunner et al. (in preparation). In Fig.~\ref{fig:mon}, the footprint of the survey is presented. The resulting eFEDS source catalogue contains 27910 sources from which 27369 are classified as point-like.

\label{section:source_det} 

The field is embedded in the footprint of the Legacy Survey DR8 (LS8) survey, which provides photometry in the \textit{g}, \textit{r,} and \textit{z} bands and in mid-infrared wavebands via forced photometry at the optical positions on Wide-field Infrared Survey Explorer (WISE) images \citep[unWISE data release, ][]{wright10,schlafly19}. With close-to uniform 5$\sigma$ depths $g\sim 24.0$, $r\sim23.4$, and $ z \sim22.5$ (AB magnitudes), LS8 ensures the determination of secure counterparts for eFEDS sources to a high level of completeness. 

The eFEDS point-like sources were cross-matched to LS8 optical counterparts in a two-method approach, which will be detailed further in Salvato et al. (in prep.). All LS8 sources within $30''$ of an X-ray source are considered as potential counterparts. 
There are an average of approximately $ 20$ LS8 sources within this radius for each X-ray source at the depth of eFEDS.

A careful treatment of astrometry and additional photometric information is needed to associate each source with its correct optical counterpart. The counterpart identification was performed using the Bayesian cross-matching algorithm NWAY\footnote{https://github.com/JohannesBuchner/nway} \citep{salvato18}. In addition to the positional offset and positional uncertainty, it uses a multi-dimensional photometric prior which was modelled using a Random Forest classifier. The prior was defined using 23058 X-ray sources from the XMM-Newton serendipitous survey \citep[3XMM DR8,][]{rosen16} and the Chandra Source Catalogue \citep[CSC 2.0,][]{evans20} with comparable fluxes to the eFEDS sources and secure counterparts.
In parallel a similar multi-dimensional photometric prior was applied to the classical Likelihood Ratio technique\footnote{https://github.com/ruizca/astromatch}  \citep{sutherland92}.
The reconciliation of both approaches delivers a highly reliable set of optical counterparts to the eFEDS sources. Tests on a  validation set of simulated eFEDS-like sources indicate that the chosen approaches reach $\sim 96 \%$ purity and $\sim 96 \%$ completeness (Salvato et al, in prep.).
\section{X-ray properties of SDSSJ08}

\begin{figure*}[ht]
\centering
\includegraphics[width=\textwidth]{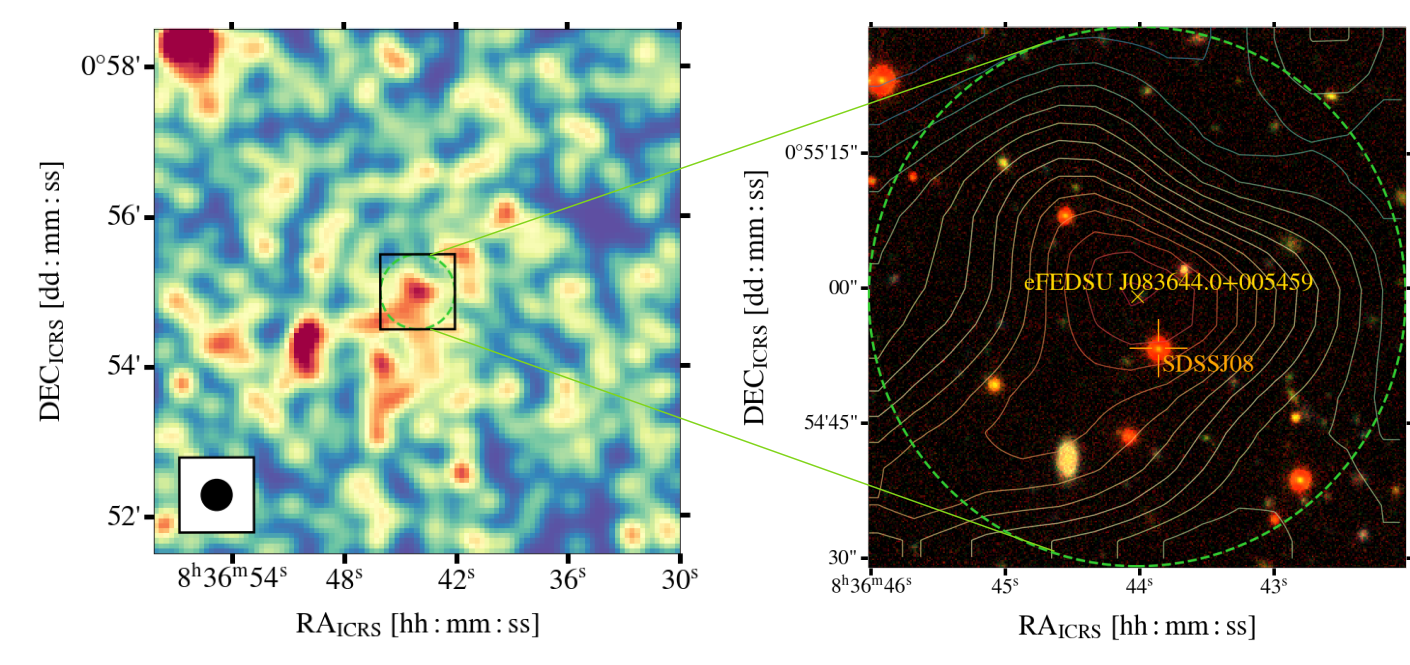}

\caption{(Left)  $7'$ image of the eROSITA events centred at the optical position of SDSSJ08 (full band 0.2-10 keV). The circle has a radius of 30''. The square shows the size of the field shown in Fig. \ref{fig:radio_image}. (Right)  G, I, and Y 60'' x 60'' HSC (PDR2) image centred at the coordinates of the SDSSJ08 associated
eFEDS source. The orange cross shows the optical  position of the quasar. 
The eROSITA contours are derived from the eFEDS image smoothed with a Gaussian kernel. The black dot in the lower left corner shows the eROSITA FWHM of the eROSITA PSF in survey mode (12'').}
\label{fig:matches}
\centering
\end{figure*}

By matching eFEDS LS8 optical counterparts to our compilation of all spectroscopic entries in the field we were able to determine that the X-ray emission from the eROSITA source eFEDSU J083644.0+005459 is associated with the z=5.81 SDSS quasar SDSS J083643.85+005453.3 (hereafter SDSSJ08). At the current level of spectroscopic completeness, it is the highest redshift eFEDS source identified so far. The quasar lies at a distance of 6.3" from the eROSITA position and is the brightest source within a radius of 30".
The LS8 counterpart to the eROSITA source matches SDSSJ08 within $0.1''$. The eROSITA-eFEDS and Hyper-Suprime Cam \citep[HSC,][]{aihara18} images of the matching region are shown in Fig. \ref{fig:matches}. A summary of the match is presented in Table \ref{crossmatch}.
The source belongs to an up-to-date list of 24 spectroscopically confirmed $z>5.5$ quasars in the eFEDS footprint. which were all discovered in dedicated optical searches \citep{fan01,venemans15,matsuoka18b,matsuoka18a, matsuoka19}. In Fig. 1, we have colour-coded these sources according to their z-band magnitude, which clearly reveals a decrease in space-density with increasing optical brightness.
SDSSJ08 is by far the brightest of all z>5.5 quasars in the field. It has previously been observed in X-rays in a follow-up program of high-redshift Sloan quasars \citep{brandt02}. The measured flux is higher than the average soft flux limit of eROSITA in the field ($\rm \sim 8\times 10^{-15} \, erg \, cm^{-2}s^{-1}$). This high-redshift quasar is also the only radio-loud one in the list. 
 
 In the eFEDS catalogue, eFEDSU J083644.0+005459 has $14.4 \pm 5.0$ source model counts. These model counts are obtained by fitting the point-spread function (PSF) to a count rate image (the ratio of the spatial count distribution and on-axis exposure time corrected for vignetting). Background and exposure maps are used in the fitting procedure.
 The corrected exposure time for the source is $ML\_EXP=1179 \, s$.
 
\begin{table}[]
\begin{tabular}{@{}lcl@{}}
\toprule
eFEDS ID          & -            & eFEDSU J083644.0+005459  \\ 
$\rm RA_{eFEDS}$           & {[}deg{]}    & $129.1834$                 \\
$\rm DEC_{eFEDS}$          & {[}deg{]}    & $0.9164$                   \\
$\rm \sigma_{RADEC,eFEDS}$ & {[}arcsec{]} & $4.1$                      \\
$\rm DET_{LIKE}$           & -            & $11.00$                    \\
Counts (0.2-2.3 keV)         & - & $14.4 \pm 5.0$                      \\
LS8 objID/brickID          & -            & 926/336644               \\
$\rm Sep. X/LS8$           & {[}arcsec{]} & $6.3$                       \\
QSO ID                     & -            & SDSS J083643.85+005453.3 \\
QSO Redshift               & -            & 5.81                     \\
$\rm Sep. QSO/LS8$         & {[}arcsec{]} & \textless{}0.1           \\ \bottomrule
\end{tabular}
    \caption{Basic source and counterpart information. The coordinates of the eFEDS source are equatorial, with $\rm \sigma_{RADEC, \,eFEDS}$ being the $1\sigma$ X-ray positional uncertainty. The net counts and errors are obtained via photon-mode PSF fitting (Brunner et al. in prep.)}. The  Sep. X/LS8 measures the separation between the centroid of the detected X-ray source and the position of the LS8 counterpart. Sep. QSO/LS8 corresponds to the separation between the quasar optical position and the LS8 counterpart. 
    \label{crossmatch}
\end{table}

\subsection{Manual eROSITA spectrum extraction}
 The eSASS task \texttt{srctool} was used to extract source and background spectra, along with instrumental responses. The source coordinates as well as calibrated event files are passed to the extraction algorithm, with the background and source-extraction regions determined manually. The background extraction region was defined as an annulus of inner and outer radii $(60'',198'')$. The source region is delimited by a circle $30''$ in  radius centred at the X-ray position. Detected sources in the background region were excluded. The photons are collected over the full eROSITA band (0.2-10 keV). 
We obtain a total of 20 counts in the spectrum (source and background).

\subsection{X-ray spectral analysis}
\label{section:fit}

A spectral analysis was performed to infer the primary X-ray properties of the quasar (see Liu et al. in prep., for more details about the spectral analysis of eFEDS sources). We used the analysis software \texttt{BXA} \citep{buchner14}, which connects the X-ray spectral fitting tool \texttt{XSPEC} \citep[v12.11,][]{arnaud96} to the nested sampling algorithm MultiNest \citep{feroz09}.
The fit was performed in the 0.3 - 8.0 keV energy range. A simple redshifted power-law model only accounting for Galactic absorption was chosen to fit the extracted spectrum: \textit{tbabs*zpowerlw}. In addition we used a background model, which was trained on eFEDS AGN spectra using a  principal component analysis and scaled to the source and background extraction sizes following \citet[][see their appendix A]{simmonds18}.  The corresponding Galactic absorbing column density is taken from \citet{hi4pi16}: $N_\mathrm{H}= 4.8\times 10^{20} \, \mathrm{cm^{-2}}$. The power law is shifted to the spectroscopic redshift of the quasar. 
The low photon counting statistics limit our ability to accurately retrieve X-ray spectral parameters. Nevertheless, we allow the photon index $\Gamma$, the normalisation of the power law, and the normalisation of the background model to vary freely in the fit in order to retrieve realistic error bars on the measured X-ray fluxes. 
The best fit was determined with the C-statistic  \citep{cash79}.
 We assumed a flat uniform prior for the photon index, restricting the range to $\Gamma=1-3$. The resulting posterior parameter distributions are shown in Fig. \ref{fig:marginal}. We obtain a photon index of $2.20^{+0.49}_{-0.60}$. While $\Gamma$ is not well constrained, it tends to typical values of X-ray-detected, radio-quiet quasars in this redshift regime. From their joint spectral analysis of X-ray-detected  $z>6$ quasars, \cite{vito19} derived an average photon index of $\rm \Gamma = 2.20^{+0.22}_{-0.20}$, a value consistent with earlier results by \citet{nanni17} who performed the same exercise for a $z>5.7$ quasar sample. 
 
 \begin{figure}
\includegraphics[width=8 cm]{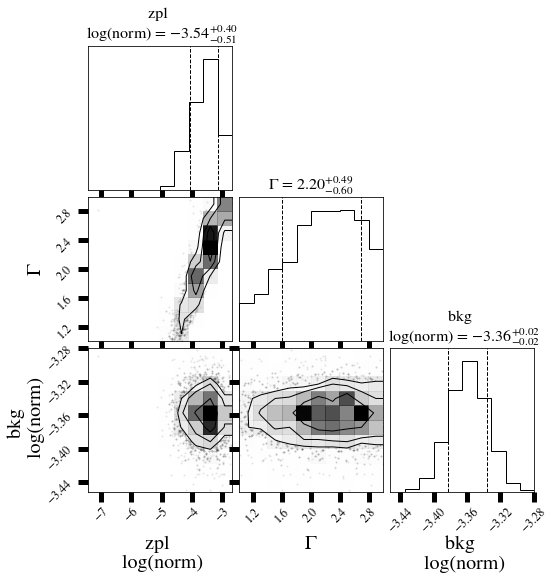}
\caption{Posterior marginal distributions of $\Gamma$,  normalisation of the power law (zpl), and normalisation of the PCA background model (bkg). The photon index remains poorly constrained but is consistent with typical X-ray spectral slopes of the radio quiet quasar population.} 
\label{fig:marginal}
\centering
\end{figure}

 \begin{figure}
\includegraphics[width=8 cm]{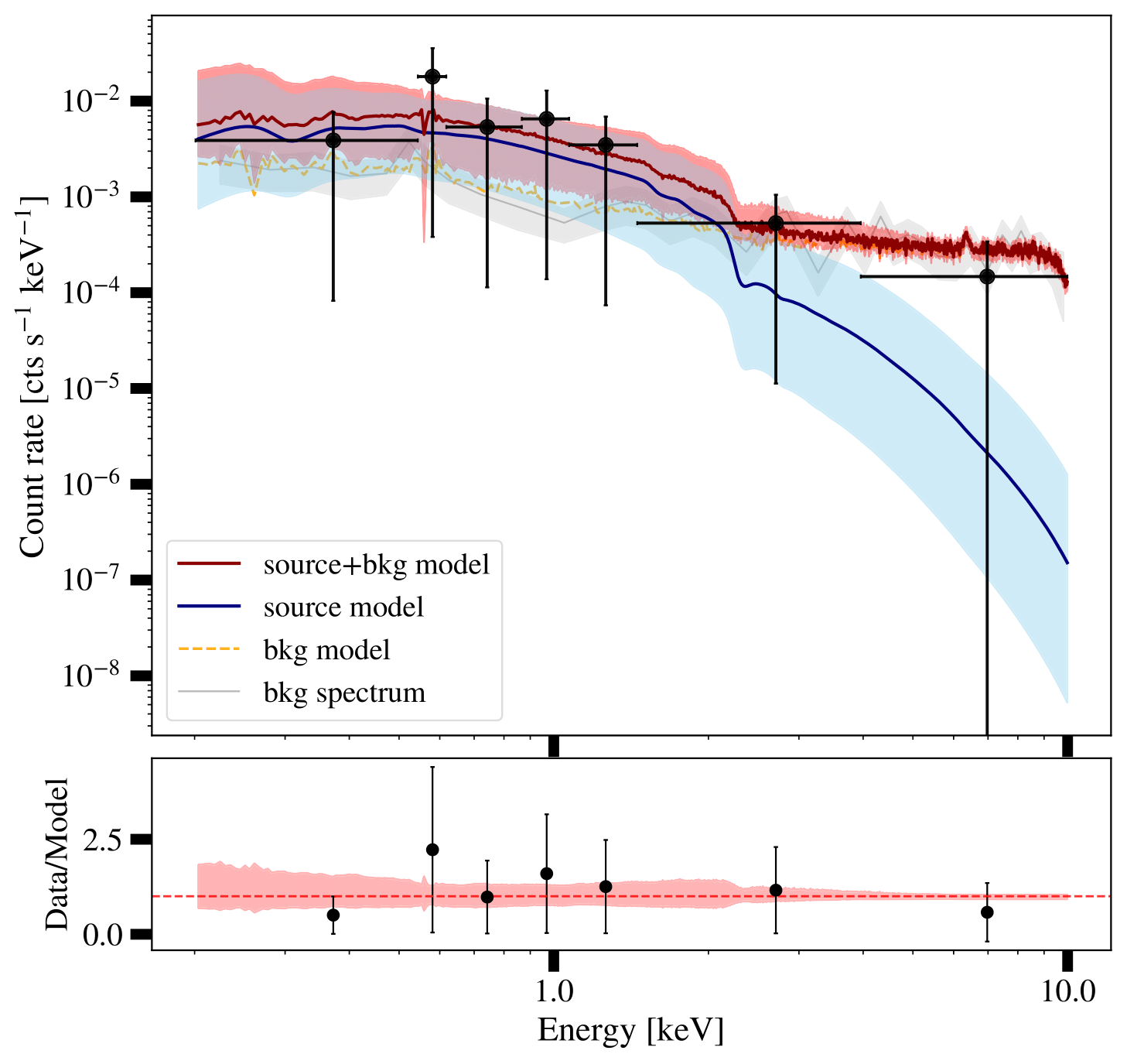}
\caption{X-ray spectrum for eFEDSU J083644.0+005459. The observed count rates are shown in black. The fit was performed in the range 0.3 - 8.0 keV. The fitted source model (blue) and combined source and background model (red) are also presented. The residuals are shown in the lower panel.}
\label{fig:marginal}
\centering
\end{figure}

We compute the soft band flux and intrinsic luminosity from the fitted model. The errors are propagated with \texttt{XSPEC} using the posterior samples. 
 The resulting soft-band flux-corrected for Galactic absorption, intrinsic $2-10$ keV luminosity, monochromatic luminosity at 2 keV, as well as the photon index and the two-point spectral X-ray to optical spectral index (see Section \ref{section:aox}) are displayed in Table \ref{tab:prop}. 

\begin{figure}
\includegraphics[width=8 cm]{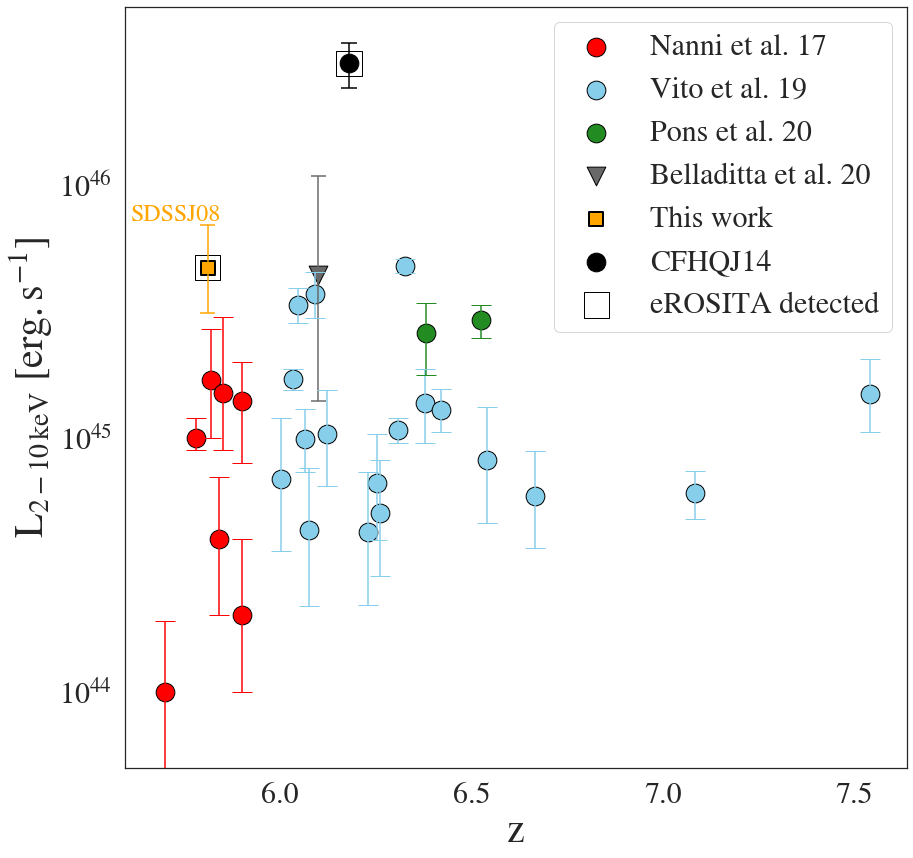}
\caption{Intrinsic hard X-ray luminosity as a function of redshift for X-ray detected $z>5.7$ quasars. SDSSJ08 lies at the X-ray luminous end of the sample. The luminosity of SDSSJ08 is computed from the eROSITA data. The quasar lies at the X-ray luminous end of the sample. eROSITA detected sources are marked by an empty square. In the case of overlapping sources in the sample of \citet{nanni17} and \citet{vito19}, only the data points from the latter are shown.}
\label{fig:Lxz}
\centering
\end{figure}

SDSSJ08 has an X-ray detection from a Chandra follow-up observation in 2002 \citep{brandt02}. 
Using a frozen power-law model with $\rm \Gamma=2$ and Galactic absorption $N_\mathrm{H}=4.4\times 10^{20} \, \mathrm{cm^{-2}}$ \citep{stark92} these authors obtained a soft-band flux of $F_{0.5-2.0 \, \mathrm{keV}}=1.05 \times 10^{-14} \, \mathrm{erg \, cm^{-2}\, s^{-1}}$.
The broad-band Chandra image taken at the optical position of the quasar reveals a single point-like source, which is strong evidence against any contamination in eROSITA from X-ray emission from any other source within a radius of 30''. Re-analysing the Chandra data with a slightly lower photon index ($\Gamma=1.9$), \citet{nanni17} 
derived the rest-frame intrinsic luminosity $L_{2.0-10.0 \mathrm{kev}}= 4.2^{+1.0}_{-1.4} \times 10^{45}\, \mathrm{erg \, s^{-1}}$. With respect to the complete sample of X-ray detected z>5.7 quasars \citep{nanni17,vito19,pons20}, this makes SDSSJ08 one of the four most X-ray luminous high-redshift quasars known to date, together with CFHQJ14 \citep{willott10,medvedev20a}, SDSS J010013.02+280225.8 \citep{wu15,ai16}, and PSO J030947.49+271757.31, the blazar discovered by \citet{belladitta20}. This can be seen in Fig. \ref{fig:Lxz}, where we present the redshift--luminosity plane for X-ray-detected $z>5.7$ quasars.
The luminosity and flux derived for SDSSJ08 are consistent with those reported by \citet{brandt02} and \citet{nanni17}. For a direct comparison with the results of \citet{brandt02}, we have also computed the unabsorbed flux in the 0.5-2 keV band, fixing $\Gamma=2$ and using the Galactic absorption quoted in \citet{stark92}. We obtain $F_{\rm{0.5-2\, keV}}=(1.01^{+0.42}_{-0.34}) \times 10^{-14} \,\mathrm{erg \, cm^{-2}\, s^{-1}}$, a value which is consistent with the previous Chandra results. We therefore find no evidence for X-ray variability in SDSSJ08 over a timescale of $\sim 20$ years.

\begin{table}[]
\centering
\bgroup
\def\arraystretch{1.5}
\begin{tabular}{@{}lcc@{}}
\toprule
X-ray property          & Units                                     & Value                   \\ \midrule
$^1F_{0.5-2\, \mathrm{keV}}$ & $\rm [10^{-14} \, erg \, cm^{-2}s^{-1}]$ & $9.9^{+3.7}_{-3.2}$ \\
$^2 L_{2-10\, \mathrm{keV}}$ & $\rm [10^{45} \, erg \,s^{-1}]$           &  $4.7^{+2.2}_{-1.6}$     \\
$ ^3 L_\nu$          & $\rm [10^{27} \, erg \,s^{-1}\, Hz^{-1}]$ & $7.0^{+6.0}_{-3.9}$     \\
$\rm ^4\alpha_{OX}$     & -                                         & $-1.57^{+0.10}_{-0.13}$ \\
$\rm ^5\Gamma$          & -                                         & $2.20^{+0.49}_{-0.60}$  \\ \bottomrule
\end{tabular}
\egroup
\caption{ Derived X-ray properties. $^1$Soft X-ray flux corrected for Galactic absorption. $^2$X-ray 2-10 keV rest-frame luminosity. $^3$Monochromatic rest-frame luminosity at 2 keV. $^4$: The X-ray to optical spectral slope. $^5$: Photon index.}
\label{tab:prop}
\end{table}

\subsection{X-ray loudness}
\label{section:aox}
The non-linear relation between X-ray and optical emission of quasars has been studied via the $\rm \alpha_{OX}$ parameter \citep{tananbaum79}, the optical to X-ray spectral index. This quantity measures the relative strength of UV continuum and coronal X-ray emission in the active core:

\begin{equation}
\mathrm{\alpha_{OX}}= 0.384\times \mathrm{log_{10}} \, \left( \frac{L_{\rm 2\, keV}}{ L_{\rm 2500 \, \si{\angstrom}}} \right)
\label{eq:aox}
,\end{equation}
where $L_\mathrm{\rm 2\, keV}$ and $L_\mathrm{ 2500 \, \si{\angstrom}}$ are the rest-frame monochromatic luminosities at $\rm 2 \, keV$ and $2500 \, \si{\angstrom}$.  
We computed the 2 keV monochromatic luminosity and its uncertainties from the posterior distribution of the $L_\mathrm{2-10\, keV}$ rest-frame luminosity and the associated photon index $\Gamma$ for each solution:

\begin{equation}
 L_{\mathrm{ 2\, keV}}= \frac{L_\mathrm{2-10\, keV}}{\int_{\nu_\mathrm{2 \, keV}}^{\nu_\mathrm{10 \, keV}} \nu^{1-\Gamma } \, d\nu } \nu_{2 \, \mathrm{keV}}^{1-\Gamma }
.\end{equation}

$L_\mathrm{2500 \, \si{\angstrom}}$ was extrapolated from the UV absolute magnitude $M_\mathrm{1450\, \si{\angstrom}}$ listed by \citet{jiang16}. An optical spectral slope of $\alpha=-0.3$ was assumed \citep[e.g. ][]{vito19}, corresponding to a correction $M_\mathrm{2500\,\si{\angstrom}} \approx M_\mathrm{1450\,\si{\angstrom}}-0.18$. The value of the X-ray to optical slope for  SDSSJ08 is given in Table \ref{tab:prop}. For the same quasar, \citet{nanni17} measured $\mathrm{\alpha_{OX}}=-1.61^{+0.03}_{-0.06}$, consistent with our estimated $-1.57^{+0.10}_{-0.13}$. There is a well-established anti-correlation between $\mathrm{\alpha_{OX}}$ and $L_\mathrm{2500 \, \si{\angstrom}}$ for $z<5$ AGNs, also investigated directly in the $L_{2\,\mathrm{keV}}-L_\mathrm{2500 \, \si{\angstrom}}$ plane \citep[e.g.][]{avni86,strateva05,just07,lusso10}.  Performing a linear regression in the $\mathrm{\alpha_{OX}}-L_\mathrm{ 2500 \, \si{\angstrom}}$ plane, \citet{nanni17} argue that their sample of 29 X-ray-detected $z>5.7$ quasars followed the anti-correlation well. From their sample of $z>6$ radio-quiet quasars, \citet{vito19} find no significant evolution of this trend with redshift. 

\begin{figure}
\includegraphics[width=8 cm]{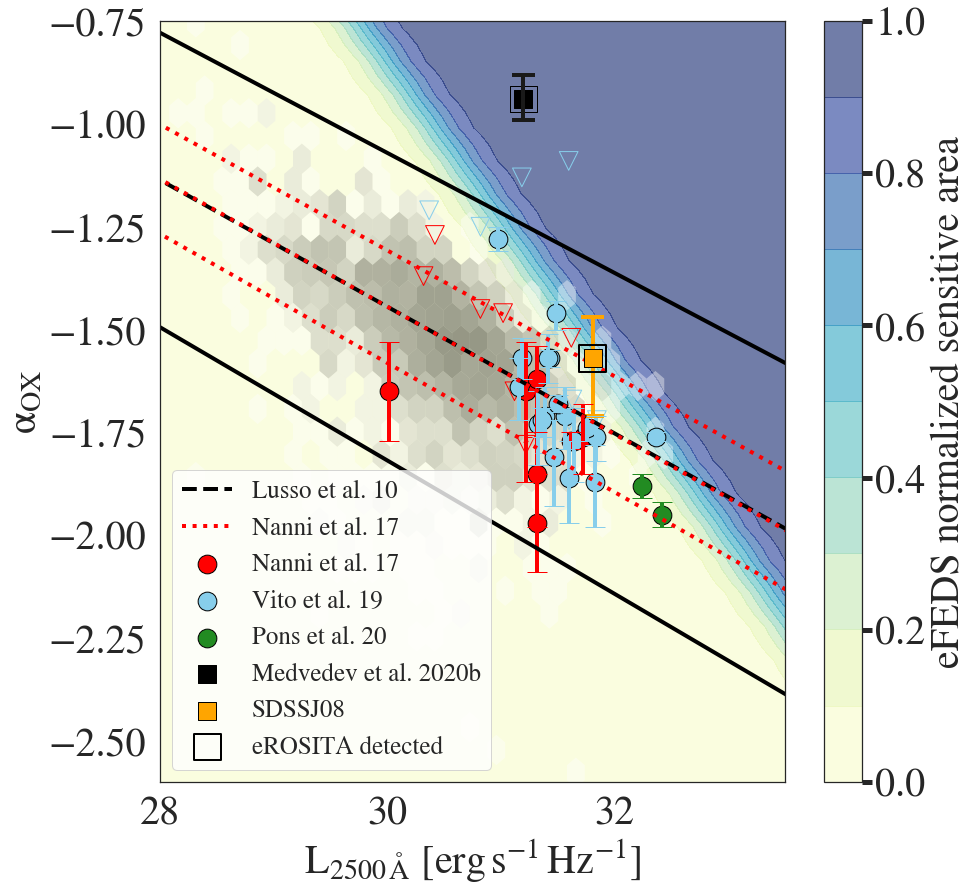}
\caption{X-ray-to-optical slope anti-correlates with the UV monochromatic luminosity at $2500 \, \si{\angstrom}$. SDSSJ08 (orange circle) is consistent with the best-fitting relation of \citet[][dashed line]{lusso10} and \citet[][red dotted line]{nanni17}. The $z > 5.7$ X-ray-detected sources from \citet[][restricted to $z<6$ and ignoring SDSSJ08]{nanni17}, \citet{vito19} and \citet{pons20} are shown. The empty triangles denote upper limits from undetected sources (same colour code as detected sources). The grey density scale shows a sample of 2685 XMM-Newton-detected $z<5$ SDSS quasars \citep{lusso16}. The contours show the eFEDS normalised, sensitive area derived from synthetic power-law spectra at $z=6$ with $\Gamma=2$ and Galactic absorption. We note that CFHQJ14 (black point) tends towards flatter $\rm \alpha_{OX}$ values than X-ray-detected quasars at similar UV luminosities. Its $\rm \alpha_{OX}$ was computed from the luminosity reported in \citet{medvedev20b}.}

\label{fig:aox}
\centering
\end{figure}

The eFEDS quasar shows a typical `X-ray loudness' with respect to the $\rm \alpha_{OX}$-luminosity relation. For CFHQJ14, \citet{medvedev20a} reported a significant deviation from the relation, arguing that this flattening of the spectral slope could be caused by an excess X-ray emission, possibly related to the inverse Compton scattering of cosmic microwave background photons off the jet \citep[iC-CMB,][]{tavecchio00,cellotti00}. Despite being radio-loud \citep{banados15}, SDSSJ08 does not display such an X-ray excess. We show how the two eROSITA-detected $z>5.7$ quasars are distributed with respect to this relation in Fig. \ref{fig:aox}. The $\alpha_{OX}-L_{2500 \si{\angstrom}}$ relation derived by \citet{lusso10} on a sample of \textit{XMM}-COSMOS AGNs is shown, as is the one obtained by \citet{nanni17} on their sample of $z>5.7$ quasars.

\section{Archival multi-wavelength properties}

\begin{figure*}[h]
\centering
\includegraphics[width=\textwidth]{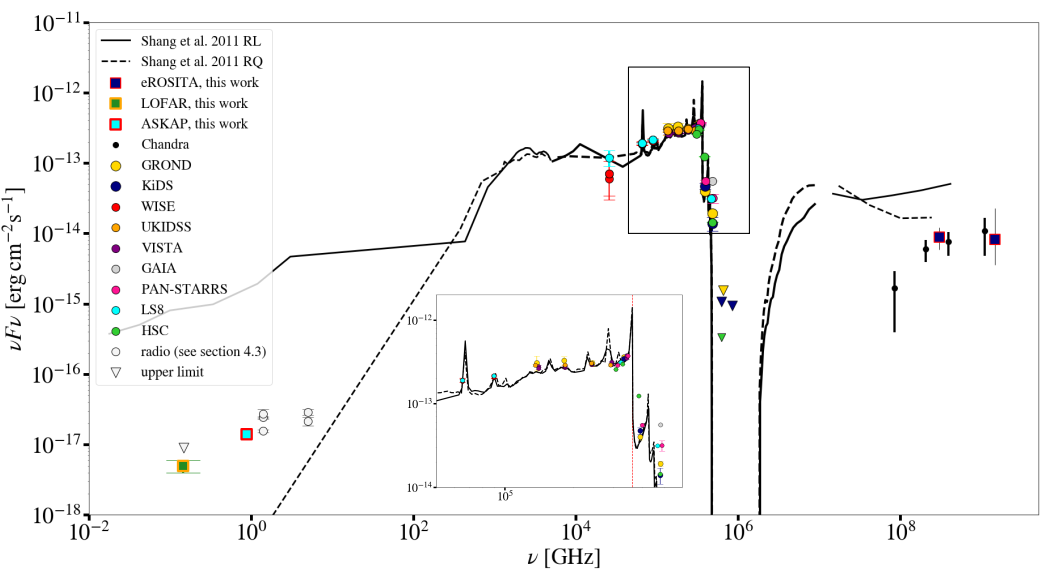}

\caption{SED of SDSSJ08 constructed using archival multi-wavelength data, together with the new eROSITA (observed broad band fluxes in 0.5-2 keV and 2-10 keV), LOFAR 145 MHz,  and ASKAP 888 MHz measurements. The markers show photometric points from various optical, X-ray, and radio surveys. Triangles denote upper limits. Composite SEDs from \citet{shang2011} for radio-loud and radio-quiet quasars are fitted to the SED at $z=5.81$ and corrected for absorption by the intergalactic medium  \citep{madau00}. 
SDSSJ08 does not present an X-ray excess typically observed in radio-loud quasars. }
\label{fig:SED}
\centering
\end{figure*}

\subsection{Optical selection and spectroscopy:}

SDSSJ08 was initially discovered through \textit{i}-dropout selection and consecutive spectroscopic confirmation in $\rm \sim 1550 \, deg^2$ of the SDSS main survey \citep{fan01}. It is part of a sample of 52 $z>5.4$ quasars that were found by exploiting imaging data in the SDSS main survey \citep{jiang16}. The $i-z>2.2$ dropout criterion selects $z>5.8$ quasars, because the neutral hydrogen absorption bluewards of the $\rm Ly\alpha$ line is shifted in the \textit{i}-band. A redshift of z = 5.81 was measured by \citet{kurk09} using VLT/ISAAC NIR observations. \citet{fan01} report a strong and broad $\rm Ly \alpha$ and $\rm N_V$ complex with an equivalent width of $\sim 70 \, \si{\angstrom}$. The quasar is extremely luminous with an absolute AB magnitude $ M_\mathrm{1450\si{\angstrom}}=-27.86$ \citep{jiang16}. Its black hole mass of $\rm (2.7\pm 0.6)\times 10^9 \, M_{\odot}$, was derived from the width of the broad $\rm MgII \, \lambda2800\si{\angstrom}$ line  \citep{kurk09}.  \citet{stern03} observed SDSSJ08 with the FLAMINGOS multi-object, NIR spectrograph at the 8 m Gemini-South Observatory and reported an optical power-law index of $\alpha= 1.55$ (measured over the rest-frame wavelength range $\lambda=1480-2510 \, \si{\angstrom}$). This red spectral slope is indicative of the presence of substantial amounts of dust in the environment of SDSSJ08. 

\subsection{ Spectral energy distribution}

 SDSSJ08 has been covered by  a number of imaging surveys 
 over the entire spectral energy distribution (SED)\footnote{Gaia \citep{gaia18}, Pan-STARRS1 \citep{chambers16}, HSC SSP, LS8. It has also been detected in NIR and mid-infrared (MIR) surveys: United Kingdom Infrared Telescope (UKIRT) Infrared Deep Sky Survey \citep[UKIDSS, ][]{lawrence07},  Visible and Infrared Survey Telescope (VISTA) Kilo-degree Infrared Galaxy Survey \citep[VIKING][]{kuijken19}, and WISE \cite[AllWISE data release,][]{wright10,cutri13}}.
In addition, we carried out simultaneous g$^{\prime}$, r$^{\prime}$, i$^{\prime}$, z$^{\prime}$, J, H, and K$_{\rm s}$-band photometry of SDSSJ08 with the Gamma-Ray Burst Optical/Near-Infrared Detector \citep[GROND, ][]{grainer08} at the MPG $2.2\,\mathrm{m}$ telescope at the ESO La Silla observatory. 
The resulting SED is presented in Fig. \ref{fig:SED}. We fitted the composites of radio-loud and radio-quiet quasars by \citet{shang2011} with the photometric code LePhare \citep{arnouts99,ilbert06}.
This figure can be directly compared to the SED of CFHQJ14 presented by \citet{medvedev20a}. For the fit, the redshift was fixed to its spectroscopic value.
We note that SDSSJ08 does not display the X-ray excess typically observed in radio-loud quasars \citep{wilkes87,shastri93,reeves97} and is more consistent with the radio-quiet template in the radio and X-ray part of the SED. The overall radio output is well below that of the radio-loud template. We summarise all available archival radio data for SDSSJ08 in the following section.

\subsection{Archival radio properties}

\begin{table}[]
\begin{tabular}{@{}llll@{}}
\toprule
Telescope & \begin{tabular}[c]{@{}l@{}}Frequency\\  {[}GHz{]}\end{tabular} & \begin{tabular}[c]{@{}l@{}}Flux density \\ {[}mJy{]}\end{tabular} & Survey/Ref.   \\ \midrule
GMRT       & 0.150                                                          & $< 6.1$ $(3\sigma)$                                               & $^1$TGSS-ADR1  \\
VLA        & 1.4                                                            & $1.11 \pm 0.15$                                                   & $^2$FIRST     \\
VLA        & 1.4                                                            & $2.1$                                                             & $^3$NVSS,     \\ 
           &                                                                &                                                                   &\citet{petric03}\\
VLA        & 1.4                                                            & $1.75 \pm 0.04$                                                   & \citet{petric03} \\
VLA        & 1.4                                                            & $1.96 \pm 0.31$                                                   & \citet{frey05}   \\
EVN        & 1.6                                                            & $1.1 $                                                            & \citet{frey03}   \\
VLA        & 5                                                              & $0.580\pm0.057$                                                   & \citet{petric03}\\           
VLA        & 5                                                              & $0.43 \pm 0.06$                                                   & \citet{frey05}   \\
EVN        & 5                                                              & $0.34$                                                            & \citet{frey05}  \\
MAMBO      & 250                                                            & $< 2.9$ $(3\sigma)$                                               & \citet{petric03}  \\
\bottomrule

\end{tabular}
\caption{Archival radio observations of SDSSJ08. $^1$:\citet{Intema2017}, \citet{coppejans17}; $^2$: \citet{becker95},\citet{white97}; $^3$: Due to the low resolution of NVSS ($45''$), the reported flux density of $2.5\pm0.5$ in \citet{condon98} is contaminated by a source located $10''$ to the south of SDSSJ08. We subtracted the contribution of the contaminated source (0.44~mJy; \citealt{petric03}) from the reported value.}
\label{tab:radio} 

\end{table}

All the radio fluxes measurements and upper limits associated to SDSSJ08 are reported in Table \ref{tab:radio}.
\citet{coppejans17} ascribe the tension between some of the 1.4 GHz flux-density measurements to resolution effects (e.g. NVSS has a resolution of $45''$) but could not entirely exclude variability. Assembling all available radio data on this quasar, these latter authors estimated the radio spectral slope of $\alpha_r = -0.89\pm 0.29$. In this work, sources with $\alpha < -0.5$ were classified as steep-spectrum sources. \citet{coppejans17} further note that their computed radio spectrum slope for SDSSJ08 would predict a flux density of $\rm \sim \, 12.0 \, mJy$ at 148 MHz, but \citet{coppejans16} argue that most steep-spectrum high-redshift quasars must have turnovers in their synchrotron spectra effectively making them MHz-peaked or GHz-peaked sources. The Giant Metrewave Radio Telescope (GMRT) 150 MHz All-sky Radio Survey \cite[TGSS-ADR1; ][]{Intema2017} does not detect SDSSJ08. Using the TGSS-ADR1 data, \citet{coppejans17} derived an upper limit of 6.1 mJy on the flux density of the source. If variability and resolution effects can be excluded, these latter authors conclude that this lower flux density could indicate a spectral turnover below 150 MHz in the observed frame ($\rm \sim 1 \, GHz$ rest-frame).
Using high-resolution imaging of SDSSJ08 from the European Very Long Baseline Interferometry (VLBI) Network at 5 GHz and simultaneous VLA observation, \citet{frey05} were able to demonstrate the compactness of the source down to an accuracy of 2 mas, thereby showing that the radio emission is concentrated within the central $\rm 40 \, pc$ of the AGN.  
\section{Confirmation of a sub-GHz spectral flattening with LOFAR and ASKAP}
 
   Within 6 months of the eROSITA observations, the entire eFEDS field was observed with LOFAR 145~MHz and ASKAP 888~MHz with dedicated programs. In the sections below, we describe the observations and report the detection of SDSSJ08 in the sub-GHz radio bands.
   
   \subsection{LOFAR 145 MHz observations}
   The eFEDS field was observed with LOFAR in the 120--168 frequency band (Project: LC13\_029) between February and May of 2020. The LOFAR data were processed with the data reduction pipelines (\texttt{PREFACTOR}\footnote{\url{https://github.com/lofar-astron/prefactor}}; \citealt{VanWeeren2016a,Williams2016a,deGasperin2019} and \texttt{ddf-pipeline}\footnote{\url{https://github.com/mhardcastle/ddf-pipeline}}; \citep{Tasse2014a,Tasse2014b,Smirnov2015,Tasse2018} that were developed by the LOFAR Surveys Key Science Projects \citep{Shimwell2017,Shimwell2019}. 
   Direction-independent (bandpass, instrumental delays) and direction-dependent (ionospheric, beam) effects were corrected for in the calibration.
   The flux scale of the LOFAR final image in the region of interest (i.e. 1.5 degree square region centred on SDSSJ08) was found to be consistent within $\sim15$ percent with the flux scale of the TGSS-ADR1~150~MHz data \citep{Intema2017}. In this paper, we use a conservative uncertainty of 20 percent for the flux scale of the LOFAR data. For details of the data reduction, we refer to Ghirardini et al. (submitted) and Hoang et al. (in prep.). 
   
\begin{figure}[h]
        \centering
        \includegraphics[width=0.49\textwidth]{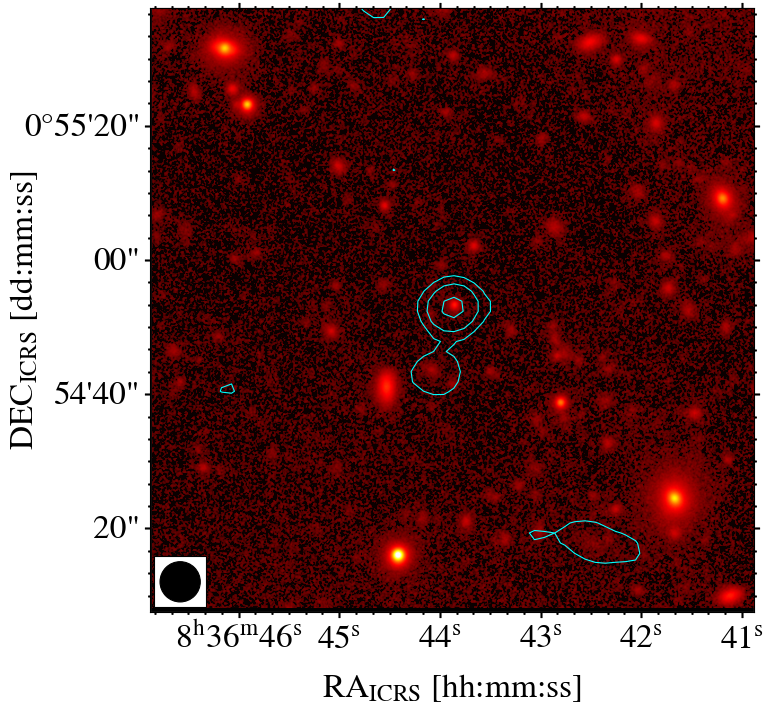}
        \caption
        {\small  LOFAR 145 MHz contours on top of 
        the HSC z-band image (right). The contour levels are $[1, 2, 4]\times 3\sigma$, where $\sigma=200 \, \mu{\rm Jy/beam}$. The beam size of  $6''\times6''$ is shown in the bottom left corner. The field corresponds to the square in Fig. \ref{fig:matches}.} 
        \label{fig:radio_image}
    \end{figure}

   In Fig. \ref{fig:radio_image} we report the detection of SDSSJ08 with LOFAR at 145 MHz. The emission peak is detected up to $14\sigma$, where $\sigma=200\,\mu{\rm Jy/beam}$. The flux density of SDSSJ08 at 145 MHz is $3.35 \pm 0.7\,{\rm mJy}$. In addition, our LOFAR image confirms the presence of a second fainter radio source with a flux density of $1.62\pm0.32\,{\rm mJy}$ to the south of SDSSJ08 (i.e. with an angular separation of $\sim 10''$). This source was found with high-frequency (1.4 and 5 GHz) observations and is associated to a lower redshift galaxy \citep{petric03,frey05}. Combining our measurement with the 1.4~GHz flux density (0.44~mJy) reported in \cite{petric03}, we find that the spectrum of the source has a spectral slope of $-0.57$. 

    \subsection{ASKAP SWAG-X 888 MHz observations}
    
   eFEDS was also observed by the ASKAP telescope as part of the SWAG-X Observatory Project (Moss et al. in prep) almost simultaneously with the eROSITA observations. In this paper, we characterise SDSSJ08 based on the continuum-only SWAG-X data observed at 888\,MHz in October 2019. This dataset comprises six ASKAP tiles for complete coverage of the eFEDS region, with 8\,hr integration per tile. Each tile was processed using ASKAPsoft with standard continuum settings (Whiting et al., in prep), including bandpass calibration, flagging, and self-calibration. Imaging was carried out using multi-scale, multi-frequency synthesis, resulting in average sensitivities across the full field of $\sim$50\,$\mu$Jy\,beam$^{-1}$. Sources and fluxes were extracted using Selavy \citep{whiting12}. The resolution of the image containing SDSSJ08 is 13.3 $\times$ 12.1\,arcsec, and at this resolution the two components seen with LOFAR are confused. The source extraction done as part of the pipeline processing fitted a single extended Gaussian component. A subsequent fit was performed that forced the size of the Gaussian components to be that of the PSF, and two components were fitted. These are spatially coincident with the components seen in LOFAR, and have fluxes of $1.575\pm0.008$~mJy (SDSSJ08) and $0.926\pm0.027$~mJy.

   \subsection{Low-frequency spectral flattening}
    \label{section:flattening}
   
\begin{figure}[h]
        \centering
        \includegraphics[width=0.43\textwidth]{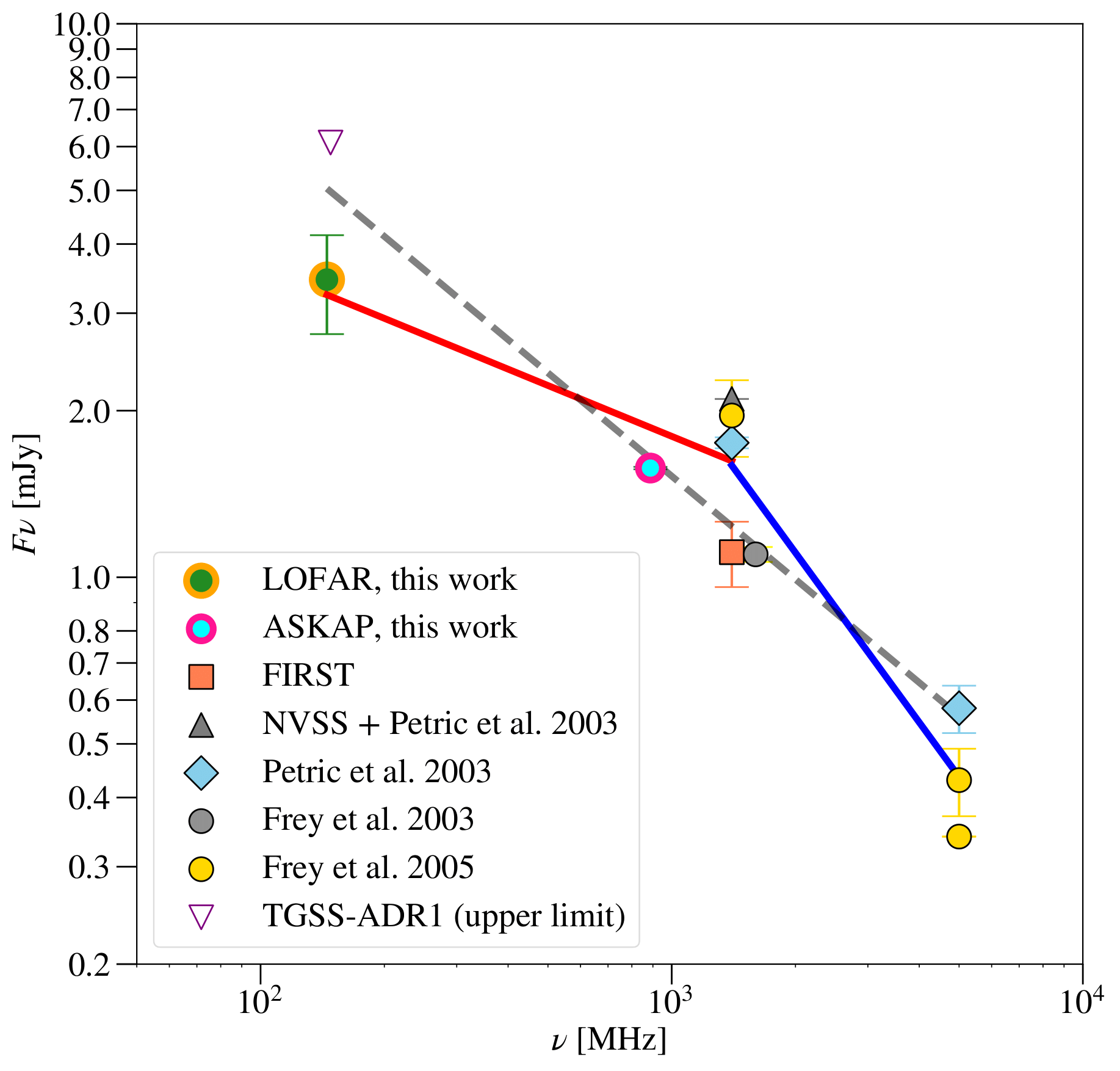}
        \caption
        {\small Radio spectrum of SDSSJ08. The coloured markers present radio measurements from the literature. The empty triangle shows the TGSS-ADR1 upper limit which is not used for spectral fitting. At frequencies above $\sim1$ GHz, the spectrum appears steep. The LOFAR 145 MHz (green and orange point) and the ASKAP 888 MHz (cyan and red point) flux densities reported in this work confirm a flattening of the spectrum at frequencies below $\sim1$ GHz. The red and blue lines show the best-fitting spectra with the indices of $-0.30\pm0.13$ and $-1.02\pm0.16$ in the frequency ranges below and above 1.4~GHz, respectively. The best-fitting line for all data points with an index of $-0.62\pm0.12$ is shown with the grey dashed line.}
        \label{fig:radio_spec}
        
    \end{figure}

   In Fig. \ref{fig:radio_spec} we present the radio spectrum of SDSSJ08 combining the measurements from new and ancillary data. The spectral slope becomes slightly flatter at lower frequencies, being $-0.30\pm0.13$ from 145~MHz to 1.4~GHz, compared to $-1.02\pm0.16$ from 1.4~GHz to 5~GHz. The SED confirms the spectral flattening at sub-GHz frequencies and hints at the peaked nature of the spectrum. However, no conclusion as to whether or not there is a spectral turnover at sub-GHz frequencies can be drawn from these data. 
   
   An alternative picture arises when considering  the discrepancy between the high signal-to-noise ratio 1.4 GHz measurements of \citet{petric03} and \citet{frey05} with the reported 888 MHz ASKAP flux density, which may be indicative of flux variability at $\rm \sim 1$\,GHz. Indeed, spectral variability has been observed for most steep spectrum sources \citep[e.g.][]{orienti07, mingaliev12}.
   The expansion of a young radio source can produce spectral variations that are observable  over the course of a decade \citep[for a review see Section 2.2. in][]{orienti16}. Such variability behaviour was observed by \citet{orienti08} in the GHz-peaked source $\rm RXJ1459+3337$. By compiling VLA observations over a period of 17 years, these latter authors report a shift of the turnover frequency towards lower frequencies, which they show to be well-explained by the adiabatic expansion of a homogeneous component. The expansion of this component results in a decrease of the source opacity, which effectively shifts the turnover frequency. 
   If we consider the contemporaneous ASKAP 888 MHz and LOFAR 145 MHz flux measurements (2019) separately from the \citet{frey05} VLA 1.4 GHz and 5 GHz observations (2003) for example, we obtain spectral slopes: $\alpha_{\rm 145-888\,MHz}=-0.42$ and $\alpha_{\rm 1.4-5\,GHz}=-1.02$.  This significant flattening observed between two observations separated by roughly 16 years could also be explained in the adiabatic expansion scenario. However, to answer questions relative to the turnover frequency, future observations at low frequencies, for example with LOFAR Low Band Antennas operating at 10--80~MHz, will be necessary. It is worth mentioning that the hypothesis of variability due to adiabatic expansion could also already be tested with contemporaneous flux measurements at 1.4 GHz.

\section{AGN space density at $z\sim 6$}
\label{section:AGNspacedens}
The  X-ray-selected AGNs with the highest and second-highest redshift to date are  CFHQJ14 and SDSSJ08  detected by eROSITA. By `X-ray-selected', we mean the blind, serendipitous detection of a source in a contiguous survey (in contrast to count extraction at known quasar coordinates). The extremely low expected space density of luminous X-ray sources at high redshifts requires wide surveys to reach within the epoch of re-ionisation. In  the  Chandra  Deep  Fields,  COSMOS  and XMM-XXL, only  three  X-ray-selected  quasars  have  been  identified  at z $>5$ \citep{barger03,marchesi16,menzel16}, the highest redshift being $z=5.3$ \citep{capak11}. The absence of $z>5.5$ X-ray-selected quasars can arise from technical difficulties such as the definition of appropriate source-extraction parameters (i.e. adapted to the detection of faint sources). However, a dominant factor is the probed cosmological volume which must be large enough to sample sources beyond the knee of the XLF at higher redshifts. 

 The evolution of AGNs selected in X-ray surveys has been extensively studied up to $z=5$ \citep[e.g.][]{hasinger05,ueda14,miyaji15,buchner15,georgakakis15,aird15,ananna19}. \citet{vito18}  investigated the AGN space density in the CDF-N and CDF-S up to $z=6$, focusing on the $\mathrm{log} \, (L_{2-10 \, \mathrm{keV}}/\mathrm{(erg/s)})<44$ regime, and particularly addressing the question of the evolution of the slope at the low-luminosity end. The highest spectroscopically confirmed redshift in their sample is $z=5.186$ \citep{vignali02}. The detection of eFEDSU J083644.0+005459 in a contiguous and flux-limited survey of near-homogeneous exposure allows us to impose  constraints on the space density of X-ray-selected AGNs at $z\sim 6$ based on secure spectroscopic data. 

In the following section, we derive a binned estimate of the XLF in the range $5.7<z<6.4$. In a complementary step, we compute number estimates from extrapolated fits to the XLF from the literature and verify whether they are consistent with the detection of eFEDSU J083644.0+005459. 

\subsection{Binned estimate of the XLF}

In population studies, the XLF traces the space density evolution of AGNs as a function of redshift and luminosity, while accounting for the effect of obscuration; it must be constructed from a purely X-ray-selected sample.

The XLF $\phi$ expresses the number $N$ of objects per unit comoving volume $V$ and X-ray luminosity $L_\mathrm{X}$:

\begin{equation}
    \mathrm{\phi} = \frac{\mathrm{d}^2 N}{\mathrm{d} V \, \mathrm{d} \, \mathrm{log} \, L_\mathrm{X}} = \frac{\,\mathrm{d\Phi}}{\mathrm{d}\, \mathrm{log} L_\mathrm{X}} (z,\, \mathrm{log} \, L_\mathrm{X})
.\end{equation}

\begin{figure}
\includegraphics[width=8 cm]{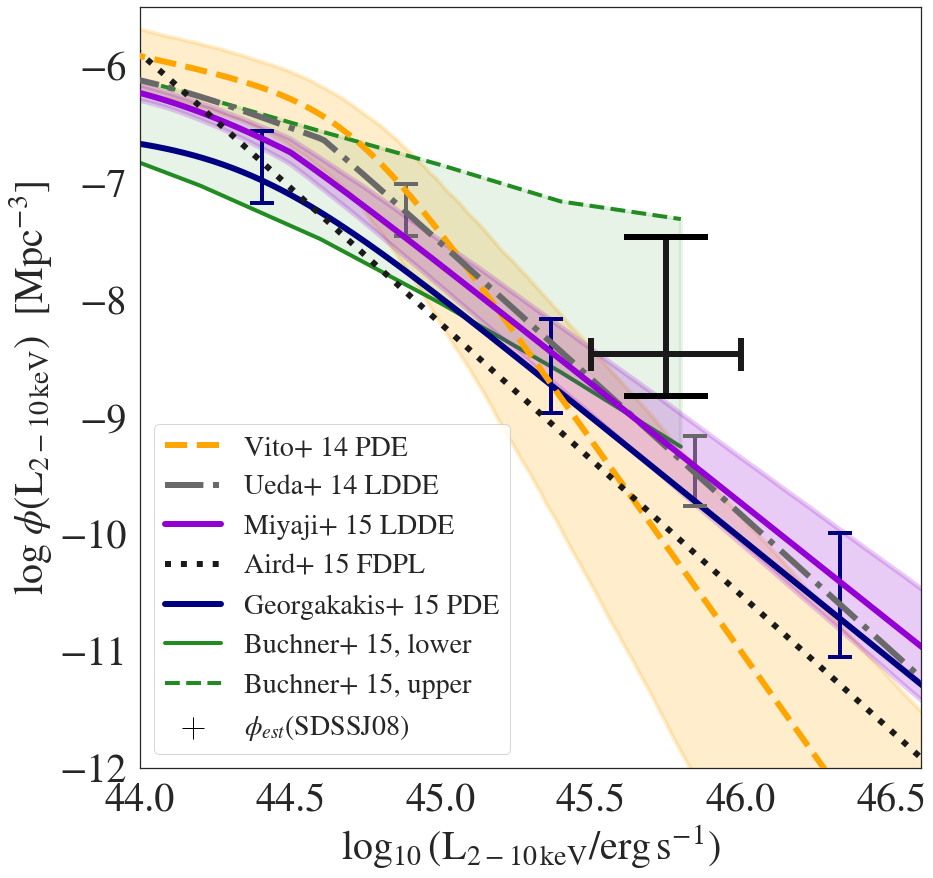}
\caption{Parametric models of the XLF extrapolated to $z=6.05$. The coloured error bars and shaded areas correspond to $1\sigma$ uncertainties on the nominal models of the same colour. The nonparametric XLF from B15 re-scaled to the comoving volume in the range $z=5.7-6.4$ is also shown in green. The binned estimate of the AGN space density as derived from the detection of SDSSJ08 is shown by the black error-bars.}
\label{fig:xlf}
\centering
\end{figure}

 Following \citet{page00}, under the assumption that the XLF changes little in a given bin of redshift and luminosity $(\Delta z, \Delta L_\mathrm{X})$ one can estimate the XLF from the number $N$ of detected sources in this bin as:

\begin{equation}
    \phi_{est} = \frac{N}{\int \int A(\mathrm{log} \, L_\mathrm{X}, z)\frac{\mathrm{d}V}{\mathrm{d}z}\mathrm{d}\,z \, \mathrm{d}\, \mathrm{log} L_\mathrm{X}}\, ,
    \label{eq:estimate}
\end{equation}

where $ A(\mathrm{log} \, L_\mathrm{X}, z)$ is the effective area of the survey sensitive to $L_\mathrm{X}$ at redshift $z,$ and $\mathrm{d}V/\mathrm{d}z$ is the differential comoving volume. The exact treatment of the eFEDS sensitivity is presented in Appendix \ref{section:sensitivity}. The statistical uncertainty on $\phi_{est}$ is given by the Poisson error on $N$ \citep[e.g.][]{gehrels86} normalised by the comoving sensitive volume. We compute $\phi_{est}$ in the bin $\Delta z =5.7-6.4$ and $\Delta \mathrm{log} \, (L_\mathrm{X}/(\mathrm{erg/s}))=45.5-46$, which contains SDSSJ08. The chosen redshift range corresponds to the selection function of bright SDSS quasars \citep{jiang16} and arises from colour-selection criteria.  The resulting binned XLF point is shown in Fig. \ref{fig:xlf}.

The XLF can be parametrised as a double power law:

\begin{equation}
    \phi_m = \frac{K}{\left( L_\mathrm{X}/L_* \right)^{\gamma_1} +\left( L_\mathrm{X}/L_* \right)^{\gamma_2} }
\label{eq:model}
,\end{equation}
where $K$ is the normalisation, $\gamma_1$ and $\gamma_2$ the slopes of the power-law components, and $L_*$ the break luminosity. This double power law is modified by a redshift evolution-term which can either be applied to the normalisation or the break luminosity (or both). 

We extrapolated fitted parametric models of the XLF from various authors to redshift $z=6.05$ (i.e. the central redshift of $\Delta z$), in order to compare $\phi_{est}$ with the expected number density from studies at lower redshift. For the comparison with our binned estimate, we have chosen to extrapolate XLFs which were measured up to $z<5$ by \citet[U14]{ueda14}, \citet[V14]{vito14}, \citet[B15]{buchner15}, \citet[M15]{miyaji15}, \citet[A15]{aird15}, and \citet[G15]{georgakakis15}. In these works, the XLF was measured from samples of soft-X-ray-selected AGNs found in Chandra\footnote{CDF-S, CDF-N, AEGIS, ECDF-S,C-COSMOS}, XMM-Newton\footnote{XMM-XXL, XMM-COSMOS, SXDS}, \textit{SWIFT}/BAT, MAXI, ASCA, and ROSAT data. The best-fitting model from each of these works is retained here.

V14 and G15 reported that a pure density evolution (PDE) best fits their data. This model assumes an evolutionary term, parametrised as a multiplicative factor to the normalisation $K$. 
 U14 assumed a luminosity-dependent density evolution (LDDE). Based on previous observations of a decline in the comoving number of $\mathrm{log} \, (L_{2-10 \, \mathrm{keV}}/\mathrm{(erg/s)})> 44$ AGNs at higher redshifts \citep{brusa09,civano11,hiroi12}, two cut-off redshifts were introduced in the parametrisation of the evolutionary term. M15 also used LDDE, accounting for the probability distribution of photometric redshifts in addition to absorption effects.
  A15 introduced a flexible double power law (FDPL) as parametrisation of the XLF. This allows any parameter in Eq. \ref{eq:model} to evolve with redshift. This redshift dependence is modelled by polynomials of $\mathrm{log} \, (1+z)$.
The extrapolations are presented in Fig. \ref{fig:xlf}. For U14, V14, M15, and G15, uncertainties are derived by sampling from the $1\sigma$ confidence intervals of the parameters entering the models fitted by these authors. We note that we are not accounting for the correlation between the parameters, which may result in an over-estimation of the computed uncertainties. In B15, the XLF is measured in a non-parametric approach based on the Bayesian spectral analysis of individual sources and a smoothness prior connecting bins of the XLF. The associated interval presented in Fig. \ref{fig:xlf} is obtained from a tabulated version of the non-parametric XLF, allowing intrinsic absorption to vary over $\mathrm{log}\, N_\mathrm{H} = 21-26$. This function was initially derived in the $z=4-7$ range within which it is constant. We have re-scaled it to the cosmological volume in the range $z=5.7-6.4$. The hard luminosity cut-off displayed in Fig. \ref{fig:xlf} is due to the luminosity range on which B15 evaluated the XLF. The model by A15 shown in Fig. \ref{fig:xlf} corresponds to the best fit to the unobscured sample ($20< \mathrm{log}\, N_\mathrm{H}< 22$) from this work. 

The error bars of $\phi_{est}$ appear only marginally consistent with the extrapolations of the functions of V14, U14, M15, and G15, while tending noticeably to higher space densities. The function by B15 shows the best agreement with our data. $\phi_{est}$ is not corrected for redshift-completeness. The 30\% spectroscopic completeness of eFEDS makes this single detection a lower limit on the number of bright high-redshift quasars in the field. The $1\sigma$ Poisson uncertainty on $\phi_{est}$ indicates that expected number counts for $z\sim 6$ X-ray-selected AGNs based on the integration of current XLF models \citep[for eROSITA expected counts, see][]{kolodzig13} may in fact represent quite a conservative estimate of the true number of luminous, high-redshift AGNs that eROSITA will detect.

\subsection{Comparison to eFEDS expected number counts}
\label{sec:efeds_counts}
We can compare the single detection in eFEDS to the number of counts expected from extrapolated parametric XLFs (for the same type of object, in the same field).
The expected number of detected AGNs in a survey for an XLF model $\phi_m$ with parameters $\theta$ can be written as:

\begin{equation}
    N= \int \int A(log \, L_\mathrm{X}, z)\frac{\mathrm{d}V}{\mathrm{d}z} \phi_m(\theta) \, \mathrm{d}\,z \, \mathrm{d}\, \mathrm{log} L_\mathrm{X}
    \label{eq:integrate}
.\end{equation}

\begin{figure}
\centering
\includegraphics[width= 9 cm]{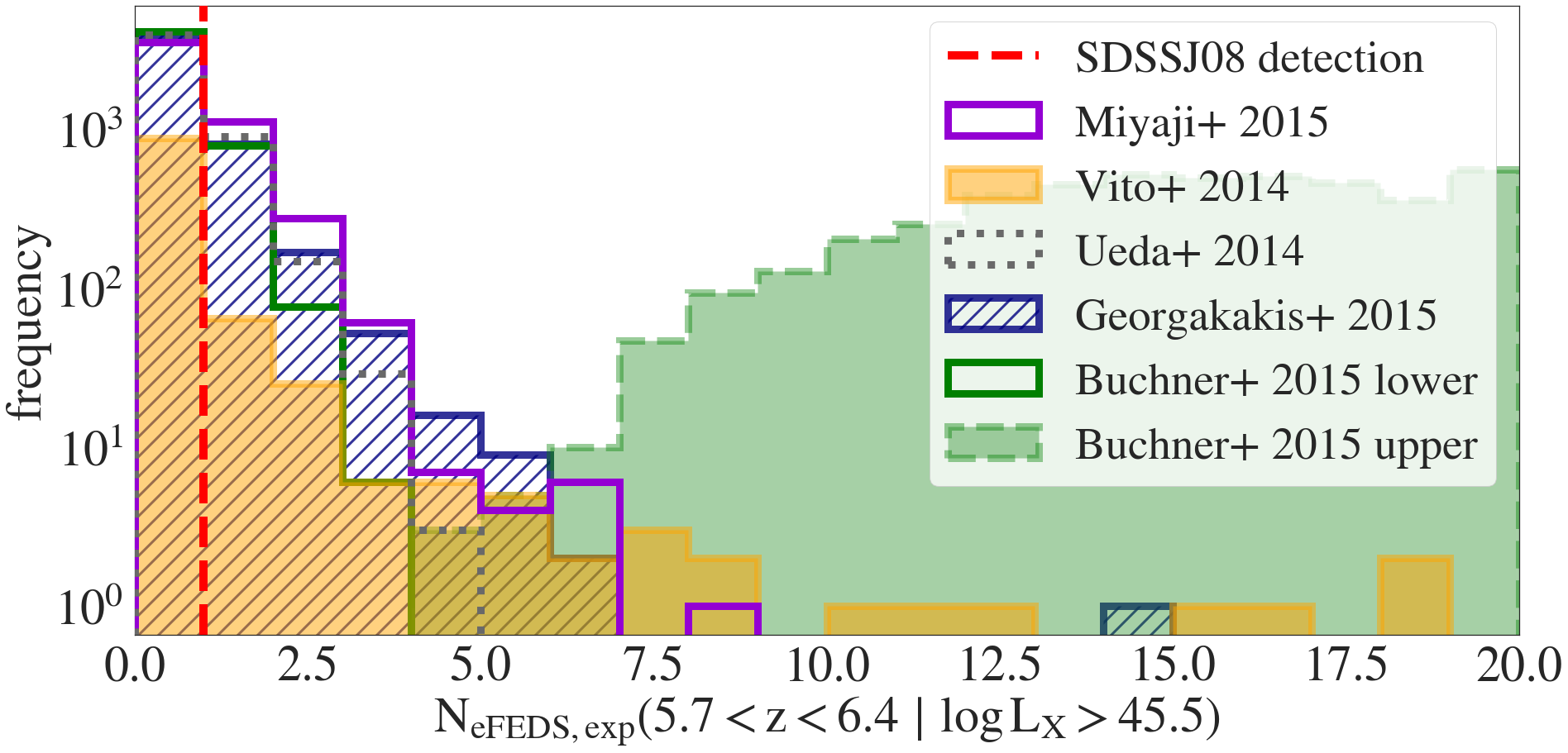}
\caption{ Frequency of expected eFEDS AGN source counts.
 The black bar is the lower limit imposed by the detection of eFEDSU J083644.0+005459.}
\label{fig:counts}
\centering
\end{figure}

To account for the area sensitive to a luminosity $L_\mathrm{X}$ at a redshift $z$, we  used the model presented in Appendix \ref{section:sensitivity}.
We compute $N$ in the bin $z=5.7-6.4$ and allow  luminosities $\mathrm{log} \, (L_{2-10 \, \mathrm{keV}}/\mathrm{(erg/s)}) =45.5-50$. The lower bound $\mathrm{log} \, (L_{2-10 \, \mathrm{keV}}/\mathrm{(erg/s)})=45.5$ is roughly the $1 \sigma$ lower edge of the confidence interval calculated for SDSSJ08 in Section \ref{section:fit}. With these bounds we effectively probe whether or not the extrapolated XLFs are consistent with at least one detection of a quasar, which is at least as X-ray luminous as SDSSJ08, in the redshift range covered by the SDSS selection in eFEDS (given its area and sensitivity). We first perform the integration on the nominal models of U14, V14, M15, B15, A15, and G15. For B15, we use the upper and lower bounds of the non-parametric XLF, because these delimit the $90\%$ credible interval between their constant-slope model (upper bound) and constant-value model (lower  bound).  The uncertainties of the expected number of AGN counts from U14, V14, M15, and G15 are derived from the 1 $\sigma$ errors of the parameters of the XLFs. Sampling from these, we generate 5000 counts with Eq. \ref{eq:integrate}. For each integration result $\bar{N,}$ we draw a random integer from a Poisson distribution with rate $\lambda = \bar{N}$. Similarly, we perform 5000 Poisson draws using the expected counts from the B15 lower and upper bounds and the nominal A15 model. The generated count distributions are well approximated by the log-normal distribution as can be verified using quantile-quantile plots \citep{waller}. The resulting count distributions are shown in Fig. \ref{fig:counts}. They are compared to the lower limit imposed by the detection of SDSSJ08 in eFEDS. The expected counts from the nominal as well as the 50th and 84.1th percentiles of the Poisson count distributions are presented in Table \ref{tab:counts_efeds}.

\begin{table}[]
\begin{tabular}{@{}lcccc@{}}
\toprule
XLF & \multicolumn{1}{l}{$\bar{N}^{nominal}_\mathrm{eFEDS}$} &
\multicolumn{1}{l}{method} &
\multicolumn{1}{l}{$N^{50-th}_\mathrm{eFEDS}$} &
\multicolumn{1}{l}{$N^{84.1-th}_\mathrm{eFEDS}$} \\ \midrule
V14        & 0.0 & sampled & 0 & 0 \\
U14        & 0.2 & sampled & 0 & 1 \\
M15      & 0.4 & sampled & 0 & 1 \\
G15 & 0.2 & sampled & 0 & 1 \\ \midrule
B15 lower     & 0.2  & nominal &  0 & 1 \\
B15 upper    & 15.9  & nominal & 15 & 19 \\ 
A15     & 0.0  & nominal & 0 & 0 \\ \bottomrule

\end{tabular}
\caption{Source count predictions. $\bar{N}^{nominal}_\mathrm{eFEDS}$ are expected source counts in eFEDS obtained from the integration of various nominal best-fitting XLF models. We also list the 50th and 84.1th percentiles of the distributions of expected Poisson counts in eFEDS (accounting for the XLF fit uncertainties). The investigated intervals are $z=5.7-6.4$ and $\mathrm{log} \, (L_{2-10 \, \mathrm{keV}}/\mathrm{(erg/s)}) =45.5-50$. The \textit{method} indicates whether the $1\sigma$ uncertainties on the XLF parameters were accounted for (sampled) or if the nominal model was used (nominal) for the derivation of the percentiles of the count distributions.}
\label{tab:counts_efeds}
\end{table}

\begin{figure}
\includegraphics[width=9 cm]{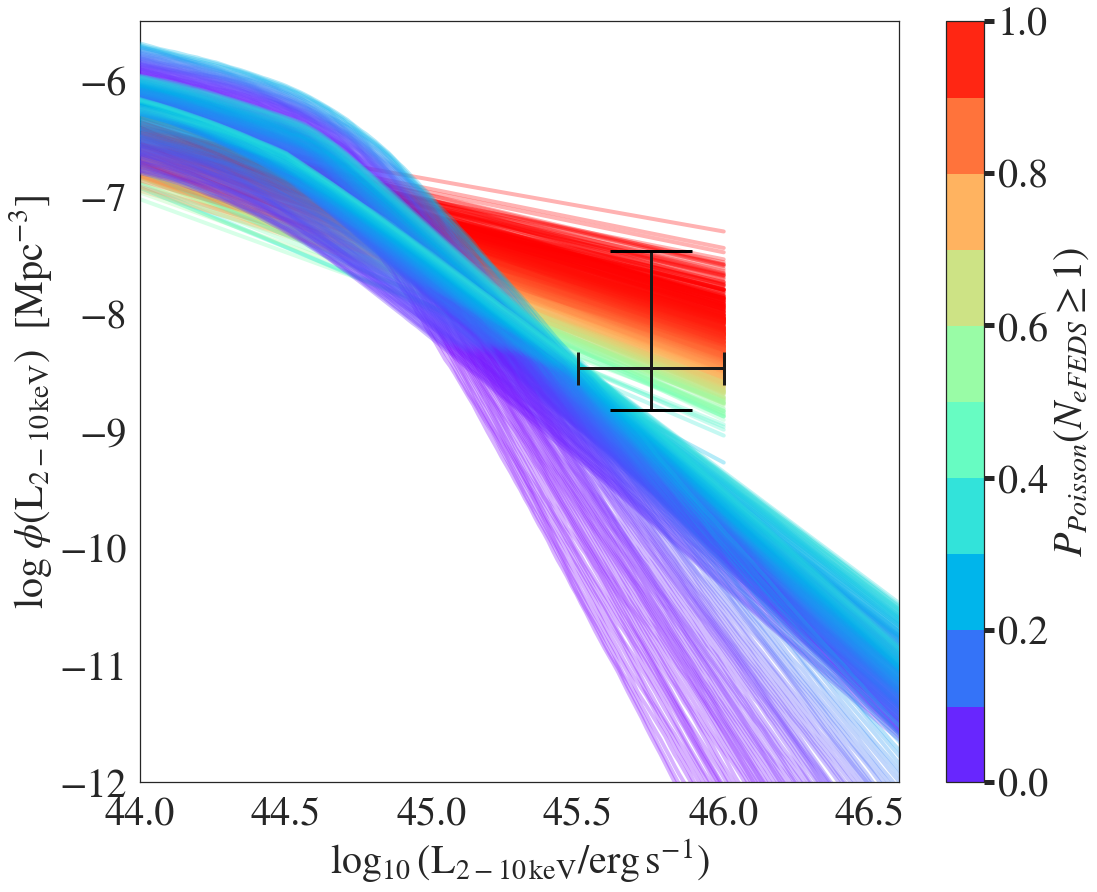}
\caption{Samples drawn from the $1\sigma$ uncertainties on the XLFs from V14, U14, M15, G15, and B15, colour-coded according to their Poisson probability of supporting at least one detection in eFEDS. Flatter slopes of the XLF are preferred.}
\label{fig:xlf_weighted}
\centering
\end{figure}

 Except for the higher bound of the XLF of B15, the nominal parametric XLFs investigated in this paper all predict less than one count in the probed redshift--luminosity bin. Accounting for the parameter uncertainties, the count expectations computed from the extrapolated XLFs from M15, G15, and U14 are consistent with the detection of SDSSJ08 at the 84.1th percentile. The PDE model by V14 is rejected at the 88th percentile. This indicates that a shallower slope on the XLF beyond the break luminosity is favoured at $z\sim 6$. This result is best illustrated in Fig. \ref{fig:xlf_weighted}, in which we show samples drawn from the $1\sigma$ uncertainties on the XLF models from B15, M15, G15, V14, and U14. For each of these functions, expected counts were computed with Eq. \ref{eq:integrate} over the ranges  $z=5.7-6.4$ and $L_{2-10 \, \mathrm{keV}}/\mathrm{(erg/s)}) =45.5-46$. We colour-code the sampled XLFs according to the Poisson probability of detecting at least one source in this bin in eFEDS: $1-P_{Poisson}(0,k_{pred})$, where $k_{pred}$ is the expected value obtained from Eq. \ref{eq:integrate}. It clearly appears that a milder decrease in AGN space density beyond $L_*$ (e.g. B15) is more consistent with our data.

\section{Discussion}

Here, we report the X-ray detection of SDSSJ08 at z=5.81 in eFEDS. \citet{medvedev20a} reported the detection of CFHQJ14 at z=6.18 in the eROSITA all-sky survey. While both quasars were previously known from optical surveys, they were both detected as X-ray sources in `blind' X-ray surveys, distinguishing them from distant quasars detected in dedicated follow-up observations \citep[e.g.][]{brandt02,vito19,pons20}.
As such these two objects can be considered the highest redshift X-ray-selected AGNs discovered thus far, and demonstrate the power of eROSITA to push the boundaries of our knowledge of accretion power at high redshift. 

Even with the detection of this single object, we are able to set constraints on the evolution of the high-redshift XLF, given the very well-characterised selection function in the eFEDS field. We furthermore argue in this paper that SDSSJ08 has X-ray emission dominated by the X-ray corona. As a result, the X-ray luminosity should be representative of the bolometric luminosity. Extrapolating from the results in Section \ref{section:AGNspacedens}, we predict the number of $z>5.7$ quasars that will be detected by eROSITA in the full-sky survey. The resulting constraints on the XLF can provide important information on the accretion power in the early Universe.

Here, we first discuss the physical nature of SDSSJ08 as inferred from our results, and thereafter the implications for AGN demographics and in particular the eROSITA all-sky survey. 

\subsection{The radio core of SDSSJ08}

 \citet{banados15} classified SDSSJ08 as radio loud ($\rm R=11.9\pm 0.3$), although close to the threshold of the definition by \citet[][]{kellerman89}: $ \rm R=f_{5\, GHz}/f_{4400 \si{\angstrom}}>10$. To date, nine radio-loud $z>5.5$ quasars have been discovered \citep[e.g.][]{banados15,coppejans16}, of which only three have been detected in X-rays: SDSSJ08, CFHQJ14 \citep{medvedev20a,medvedev20b} and PSO J030947.49+271757.3 \citep{belladitta20}. CFHQJ14 is also classified as a steep-spectrum radio source \citep{coppejans17}, while PSO J030947.49+271757.3 has a flat radio spectrum, typically observed in blazars. Besides its spectral shape, CFHQJ14 is also similar to SDSSJ08 in terms of morphology. Indeed, \citet{frey11} showed with VLBI images that its radio core is confined to scales <100 pc.
  
  The steep radio spectral slopes of SDSSJ08 and CFHQJ14, as well as their compact morphologies are characteristic of compact steep-spectrum sources and peaked spectrum sources (i.e. GHz-peaked sources and MHz-peaked sources). The steep radio spectra exclude the possibility of a jetted AGN seen at a small inclination, i.e. relativistic beaming \citep{padovani92}.   
  While variability cannot be excluded, the spectral flattening at lower frequencies reported in this work for SDSSJ08 hints at the peaked nature of its radio spectrum. 
  MHz-peaked sources and GHz-peaked sources are thought to be at the very beginning of their evolution into large-scale radio sources (Fanaroff-Riley I or II, \citealt{fanaroff74}), a stage at which their jets are still contained within the $\rm \sim \, 1 \, kpc$ of their narrow line region \citep[e.g.][ and references therein]{orienti16}. 
  Measurements of the hot-spot separation velocities of compact sources with steep spectra and the associated kinematic age support the youth hypothesis \citep{giroletti09}. We note here that radio spectral variability in SDSSJ08 would not necessarily contradict this scenario, as the adiabatic expansion of young jets would result in the shift of the peak towards lower frequencies (see Section \ref{section:flattening}). An alternative to the young radio source scenario is confinement through the surrounding dense interstellar medium  \citep[e.g.][ and references therein]{odea98,odea20}.
  The turnover observed in the spectra of these objects is thought to be due to synchrotron self-absorption or free-free absorption through shocks in the dense environment surrounding the quasar \citep[for a review see][]{odea20}, to name just a few. However, these mechanisms have difficulty in describing MHz-to-GHz spectra of both compact radio sources and extended structures of radio galaxies \citep[e.g.][]{2003AJ....126..723T}. More recently, jet energy dissipation and a change in acceleration mechanism have been put forward as alternative explanations \citep{2009ApJ...695..707G}. \cite{Harris_2019} discovered spectral curvature in a blazar using LOFAR long-baseline observations and showed how radio observations in the MHz energy range can improve estimates of source parameters such as the equipartition magnetic field. 

\subsection{Origin of the X-ray emission}
\label{section:nature_discussion}

  For SDSSJ08, our tentative estimation of the photon index yielded $2.20^{+0.49}_{-0.60}$. \citet{medvedev20b} performed a 20ks \textit{XMM-Newton} DDT follow-up observation of CFHQJ14.
  Their absorbed power-law fit to CFHQJ14 yielded $\rm \Gamma=2.5 \pm 0.2$. The source PSO J030947.49+271757.3 was observed with \textit{SWIFT}/XRT and its spectral analysis returned $\rm \Gamma =1.6\pm 0.6$, consistent with typical blazar photon indices. While the uncertainty on $\Gamma$ is too large to unambiguously demonstrate the absence (or presence) of an additional X-ray component in the spectrum of SDSSJ08 due to a potential jet contribution, the posterior distribution of the photon index clearly tends towards higher values, typical for radio loud quasars at these redshifts \citep{vito19}. 
  
 In summary, the eROSITA detected quasars SDSSJ08 and CFHQJ14 are both X-ray luminous and have steep radio spectra. They differ in two aspects:

 (a) {Their photon indices}: CFHQJ14 has a well-constrained, steep $\Gamma$, while SDSSJ08 tends to a slightly flatter value, consistent with the population of X-ray-detected, radio-quiet z>6 quasars presented in \citet{vito19}.
 
 (b) {The relative strengths of their optical and X-ray emissions}: Unlike CFHQJ14,  SDSSJ08 does not show any significant X-ray excess luminosity with respect to the bulk of the AGN population  (Fig. \ref{fig:aox}).

These observations lead us to believe that the X-ray output of SDSSJ08 is dominated by classical accretion processes and is not boosted by the relativistic bulk motion of a jet \citep[e.g.][]{siemiginowska08}. Its radio core bears the typical spectral signature of confined jets, which nevertheless do not appear to contribute strongly to the overall X-ray output of the quasar.  With its mass of $\rm \sim 3\times 10^{9} \, M_\odot$, SDSSJ08 lies at the high end of the $z>5.8$ quasar mass distribution \citep[e.g. Figure 7 in ][]{onoue19}. Estimating the bolometric luminosity of SDSSJ08 from $\rm M_{1450 \, \si{\angstrom}}$ \citep{runnoe12}, we find that it accretes at $\rm \sim 0.1\,L_{Edd}$. In summary, the emergent picture for SDSSJ08 is that of a relatively massive and moderately accreting black hole powering a young and expanding radio core.  

Investigating the properties of a large sample of radio-loud quasars, \citet{zhu20} showed that steep-spectrum radio quasars follow a similar $\mathrm{\alpha_{OX}}-L_\mathrm{ 2500 \, \si{\angstrom}}$ relation to that of radio quiet quasars, indicating that the X-ray emission of these sources originates from the corona. Parametrising the corona--jet relation and performing model fitting, these latter authors find no evidence for a significant jet contribution to the X-ray output of steep spectrum quasars. Our findings for SDSSJ08 fit well within this picture. The coronal origin of the X-ray emission confirms that by deriving constraints on the XLF from the detection of SDSSJ08, we are truly tracing black hole accretion at high redshifts.

\subsection{$z\sim 6$ quasar demographics from optical surveys}

With an absolute UV magnitude of $\rm M_{1450 \, \si{\angstrom}}=-27.86$, SDSSJ08 is the brightest SDSS $z>5.7$ quasar found to date \citep{jiang16}. As such, it belongs to a class of extremely rare objects, given the steep decline of the quasar luminosity function (QLF) beyond the break luminosity \citep[e.g.][]{shen20}. The complete sample of $5.7<z<6.4$ quasars  found in $11240 \, \mathrm{deg^2}$ of the SDSS main survey \citet{jiang16} contains 29 extremely bright sources ($M_{1450 \, \si{\angstrom}}<-26.22$). We investigate here if the detection of SDSSJ08 is consistent with the space density of bright quasars inferred from optical surveys. We first note that the choice of the location of the eFEDS field was not driven by the presence of spectroscopically confirmed high-redshift quasars; the main motivation was the availability of a large array of complementary multi-wavelength surveys. A second observation is that SDSSJ08 was initially discovered in the main single-epoch imaging survey and not in deeper fields such as overlap regions and SDSS Stripe 82. Therefore, eFEDS is not biased towards a higher density of bright SDSS $z\sim 6$ quasars.

The colour and magnitude incompleteness of high-redshift quasar surveys are encoded in well-defined selection functions \citep[e.g.][]{fan01,jiang16}. These selection biases are accounted for in the QLF fit.
In Fig. \ref{fig:aox}, we show contours of the eFEDS normalised sensitive area. The sensitive area is a function of the net count rate \citep[e.g.][]{georgakakis08}. Its dependency on $\alpha_{OX}$, i.e. $L_{2\,\mathrm{keV}}$, was computed by simulating X-ray spectra with a redshifted power law as baseline model: \textit{clumin*tbabs*zpowerlw}. The convolutional model \textit{clumin} enabled us to generate spectra for configurations of $L_{2-10\,\mathrm{keV}}$ and $z$. We fixed $\Gamma = 2$, $z=6$ and assumed a Galactic absorption of $N_\mathrm{H}=3\times 10^{20} \mathrm{cm^{-2}}$. By converting the broad-band restframe luminosities $L_{2-10\,\mathrm{keV}}$ to the monochromatic $L_{2\,\mathrm{keV}}$ we were able to derive the eFEDS sensitivity for a grid of $\alpha_{OX}$ and $M_{1450 \, \si{\angstrom}}$ using Eq. \ref{eq:aox} (a similar procedure was applied in Appendix \ref{section:sensitivity}).   
 eROSITA is sensitive to $z\sim 6$ quasars which have an $\alpha_{OX}(L_{2500\, \si{\angstrom}})$ within $1\sigma$ of the $\alpha_{ox}-L_{2500\,\si{\angstrom}}$ relation of \citet{lusso10} beyond $M_{1450\,\si{\angstrom}}<-24$.
 
\begin{figure}
\includegraphics[width=8 cm]{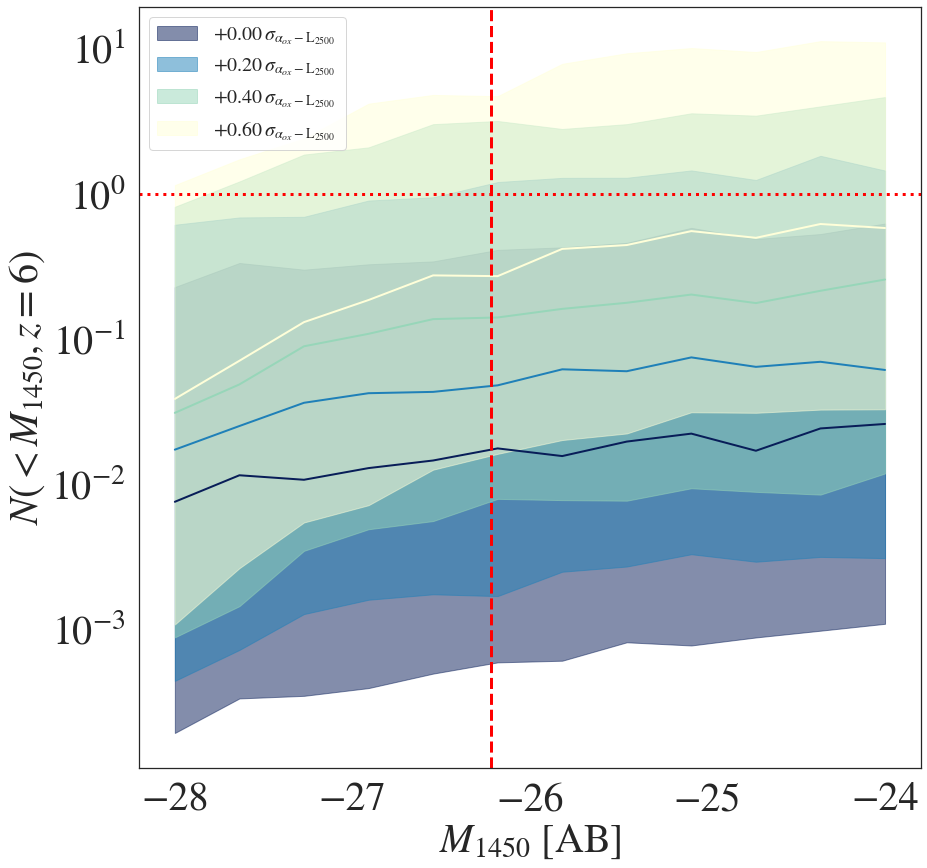}
\caption{Confidence intervals for the expected number of $z\sim 6$ optically selected quasars detectable by  eROSITA  in eFEDS. These predictions are derived from the UV QLF presented in \citet{kulkarni19}, for a fixed $\alpha_{ox}-L_{2500}$ relation \citep{lusso10} and deviations therefrom. The eFEDS sensitivity is accounted for. The vertical line shows the magnitude limit of the SDSS main survey sample of \citet{jiang16}. The horizontal line marks the single detection limit. Within $1\sigma$ of the typical $\alpha_{OX}$ the detection of SDSSJ08 in eFEDS is consistent with predictions from optical surveys.} 
\label{fig:uvqlf}
\centering
\end{figure}

We can obtain the expected number of sources beyond a certain UV luminosity threshold at a given redshift by computing:

\begin{equation}
    N(<M_{1450},z+ \Delta z) =\int^{\Delta z}_{z} \int^{M_{1450}}_{-\infty} \phi_{UV}(M,z) \, \Omega \frac{\mathrm{d}V(z)}{\mathrm{d}z}\,\mathrm{d}M \, \mathrm{d}z,
    \label{eq:manti}
\end{equation}

where $\phi_{UV}$ is the quasar UV luminosity function, $\Omega$ is the solid angle subtended by the survey,  and $\frac{\mathrm{d}V(z)}{\mathrm{d}z}$ is the differential comoving volume \citep[e.g.][]{manti17}. From the best-fitting double power-law model for the UV luminosity function derived by \citet{kulkarni19}, which was fitted on a sample including the high-redshift SDSS quasars of \citet{jiang16}, we find that we expect $0.79^{+13.19}_{-0.76}$ sources at $M_{1450\si{\angstrom}}<-26.22$ (i.e. the absolute magnitude of the faintest high-redshift quasar from the main survey) and $z \in [5.7,6.4]$  in an eFEDS-sized field. These calculations account for the $1\sigma$ uncertainties on the best-fitting parameters of the broken power law. We note that, at $z\sim 6$, the scatter on the fit parameters is large. From the same calculation using the best-fitting nominal double power-law model of \citet{jiang16} we obtain approximately one expected source count in eFEDS.     
In order to compute the number of bright quasars that we expect to detect with eROSITA in eFEDS, we assume a fixed $\alpha_{ox}-L_{2500\si{\angstrom}}$ \citep{lusso10} and positive offsets therefrom (in fractions of its $1\sigma$ uncertainty). Fixing $\alpha_{ox}(L_{2500\si{\angstrom}})$, we can derive the eFEDS sensitivity as a function of $M_{1450\si{\angstrom}}$ and account for it in the integration of Eq. \ref{eq:manti}. 
A spectral slope of $\alpha \sim - 0.3$ was assumed to convert $L_{2500\si{\angstrom}}$ to $M_{1450\si{\angstrom}}$. The cumulative expected integrated counts at $z=6$ in eFEDS for the UV QLF fitted by \citet{kulkarni19} are shown in Fig. \ref{fig:uvqlf}. The $\alpha_{ox}-L_{2500\si{\angstrom}}$ relation from \citet{lusso10} has been assumed.
At $M_{1450\si{\angstrom}}<-26.22$, we can expect to detect one source assuming at least a $+0.2\sigma$ deviation from the \citet{lusso10} $\alpha_{ox}-L_{2500\si{\angstrom}}$ scaling relation.

The detection of SDSSJ08 is therefore consistent with the findings of optical surveys and does not require a significant deviation from the $\alpha_{ox}-L_{2500}$ relation. 
We note that \citet{vito19} found no significant evolution of $\alpha_{OX}$ with redshift. A larger sample of X-ray-selected quasars will be needed to further characterise the corona-to-disc relation at high redshifts. The detection of sources such as CFHQJ14 with additional jet-driven X-ray-emission components may point to a greater diversity in terms of optical to X-ray properties.

At high redshifts, cosmic variance is another important source of uncertainty in the space density measurement of quasars  \citep{trenti08,robertson10,moster11,bhomwhick20}. This is particularly true at the bright end of the QLF, because the most luminous sources are expected to populate the most massive haloes. However, the relatively large sky area covered by the eFEDS significantly reduces the effect of clustering in the large-scale structure on the expected space density of high-z quasars. Using the method of \citet{trenti08}, which is based on mock observations of dark-matter-only simulations, we compute the cosmic variance in the redshift selection window detailed in \citet{jiang16} over the $140\,\mathrm{deg^2}$ of eFEDS. A duty cycle of 0.5 was assumed. We obtain a vanishing relative cosmic variance, that is, the effect of large-scale structure is negligible compared to the Poisson noise.

\subsection{eRASS:8 count prediction}

We investigate how the detection of an X-ray source associated with SDSSJ08 can be used to predict number counts of high-redshift AGNs in eRASS:8. Given that eFEDS was initially designed to reach the average final depth of the all-sky survey, a first-order approach would consist in re-scaling the single detection in eFEDS to the full-extragalactic sky (i.e. eROSITA-DE+eROITA-RU; $\rm 34100 \, deg^2$, $\mid b \mid >10\degree$). The exact sky-area of eFEDS is: $\rm 142 deg^2$. The area-scale factor is therefore: $s_{AREA}=244$. Considering one detection in eFEDS at $z>5.7$ (the lowest redshift of all spectroscopically confirmed high-redshift quasars in eFEDS), error bars on the number of expected detections in a field of eFEDS-like depth and area can be obtained by inverting the Poisson probability distribution. The lower (upper) bounds are 0.17 (1.8). Multiplying these by the scale factor, we obtain $N_{scaled}=244^{+195}_{-202}$. We note that the prediction assumes that the SDSSJ08 is the only source in eFEDS in this redshift regime. However, the average exposure of eFEDS is near-uniformly $\rm \sim 2.3 \, ks$ while the eROSITA scan pattern makes the exposure of eRASS:8 non-uniform, with an average of $\rm \sim 1.6 ks$ in the equatorial region \citep{clerc18}. $N_{scaled}$ therefore possibly over-predicts the actual number of detectable $z>5.7$ sources in eRASS:8.

Alternatively, using Eq. \ref{eq:integrate}, we can also make predictions for eRASS:8 using XLF extrapolations which are consistent with the eFEDS detection. \citet{kolodzig13} followed a similar approach to compute pre-mission estimates using the LDDE XLF parametrisation of \citet{hasinger05}. We use the methodology detailed in Appendix \ref{section:sensitivity}, accounting this time for the predicted eRASS:8 sensitivity to point sources. \citet{clerc18} generated sensitivity curves for eRASS:8, by computing the selection function of point sources from a simulated eROSITA sky in three exposure modes: equatorial ($\rm \sim 1.6 \, ks$), intermediate ($\rm \sim 4 \, ks$), and deep ($\rm \sim 9.7 \, ks$). 
For a more conservative estimate, we select the sensitivity curve computed from the shallower equatorial simulation. We integrate the XLFs over a redshift range $z=5.7-6.4$ and luminosity range $\mathrm{log} \, (L_{2-10 \, \mathrm{keV}}/\mathrm{(erg/s)}) = 45.5-50$. Sampling from the uncertainties on the fit parameters, we integrate XLFs by M15, U14, and G15 and obtain distributions of expected source counts. We compute weighted percentiles of these distributions. The weight we ascribe to each count is the Poisson probability of detecting at least one source in the redshift-luminosity bin and in eFEDS given the sampled XLF model (see Section \ref{sec:efeds_counts}). The resulting weighted percentiles are reported in Table \ref{tab:erass}. From the extrapolated XLFs we obtain an average of 88 detections in the probed redshift--luminosity bin. The confidence intervals spanned by the 15.9th and 84.1th percentiles are large and right-skewed because of the poor constraints on the parameters governing the shape of XLF models. This prediction is higher than the one presented in \citet{kolodzig13} by a factor of about three. The expected value from simple area scaling is more optimistic with 244 counts, but the error-bars obtained from the inversion of the Poisson distribution are larger than the confidence intervals from the XLF predictions. We underline the conservative nature of the XLF estimates, which is due to the following: (1) eFEDS is currently $30 \%$ redshift-complete, (2) the integration of extrapolated XLFs favours a non-detection, with the eFEDS detection only being supported at the $+1\sigma$ limit, and (3) we have not accounted for regions of deeper exposure in the all-sky survey.

\begin{table}[]
\centering
\begin{tabular}{@{}llll@{}}
\toprule
Mod. & $N_{eRASS:8}^{(5.7<z<6.4)}$ & $N_{eRASS:8}^{15.9-th}$ & $N_{eRASS:8}^{84.1-th}$ \\ \midrule
G15 & \textbf{99}          & 38                 & 245                 \\
U14 &       \textbf{73}        &       38               &  136                    \\ 
M15  &       \textbf{92}        &       41               &  204                    \\ \midrule

$s_{AREA}$ &       \textbf{244}        &       42               &  439                    \\\bottomrule
\end{tabular}
\caption{The 50th, 15.9th, and 84.1th percentiles of the count predictions of 5.7<z<6.4 and $\mathrm{log} \, (L_{2-10 \, \mathrm{keV}}/\mathrm{(erg/s)}) > 45.5$ AGNs that will be detected with eROSITA in eRASS:8 (restricted to the extragalactic sky, $\rm 34100 \, deg^2$). The counts are obtained by sampling from the XLF models of G15, U14, and M15. The distributions are weighted by the probability of the sampled XLFs of supporting at least one detection in eFEDS. The results for $s_{AREA}$ are obtained from naive area-scaling. The edges of the $1\sigma$ confidence interval are estimated from the percentile method.}
\label{tab:erass}

\end{table}
\section{Conclusions}

We report the blind detection of eFEDSU J083644.0+005459, an eROSITA X-ray source matched to the well-known quasar SDSS J083643.85+005453.3 (z=5.81). The detection is robust in terms of X-ray photon counts, astrometry, and multi-wavelength counterpart association. The eROSITA flux of the source is consistent with previous X-ray observations carried out with Chandra \citep{brandt02}.

From GHz radio surveys,  SDSS J083643.85+005453.3 is known to host a steep spectrum radio core within its central 40 pc. With the LOFAR 145 MHz and ASKAP 888 MHz observations, we confirm a spectral flattening at frequencies below 1 GHz. The shape of its radio spectrum indicates that this quasar has (possibly young) jets confined in its central region. Alternatively, the observed flattening of the spectral slope could be the spectral signature of adiabatically expanding jets.

The multi-wavelength properties of the quasar are evidence against relativistic beaming or iC-CMB boosting of the X-ray emission, suggesting that it originates in the X-ray-emitting corona. We examined the constraints on the XLF implied by this detection which favour a relatively shallow slope of the XLF beyond the break luminosity at $z\sim 6$. The population of X-ray-luminous high-redshift quasars may therefore be larger than previously thought. From the parametric XLFs presented by G15, M15, and U14 we predict the detection of $\sim 90 $ AGNs at the bright end of the XLF ($z=5.7-6.4$ and $\mathrm{log} \, (L_{2-10 \, \mathrm{keV}}/\mathrm{(erg/s)}) > 45.5$) in the eROSITA full-sky survey by the end of 2023.


\begin{acknowledgements}

    We thank the referee for carefully reading the present work and their efforts towards improving it.

    \newline

    We thank Paolo Padovani for his valuable and constructive comments in the preparatory phase of this paper.
    
    \newline 
    
    JW acknowledges support by the Deutsche Forschungsgemeinschaft (DFG, German Research Foundation) under Germany's Excellence Strategy - EXC-2094 - 390783311. 
    
    \newline
    
    DNH acknowledges support from the ERC StG DRANOEL 714245. MB acknowledges support from the Deutsche Forschungsgemeinschaft under Germany's Excellence Strategy - EXC 2121 "Quantum Universe" - 390833306.

    This work is based on data from eROSITA, the primary instrument aboard SRG, a joint Russian-German science mission supported by the Russian Space Agency (Roskosmos), in the interests of the Russian Academy of Sciences represented by its Space Research Institute (IKI), and the Deutsches Zentrum für Luft- und Raumfahrt (DLR). The SRG spacecraft was built by Lavochkin Association (NPOL) and its subcontractors, and is operated by NPOL with support from the Max Planck Institute for Extraterrestrial Physics (MPE).
    
    The development and construction of the eROSITA X-ray instrument was led by MPE, with contributions from the Dr. Karl Remeis Observatory Bamberg \& ECAP (FAU Erlangen-Nuernberg), the University of Hamburg Observatory, the Leibniz Institute for Astrophysics Potsdam (AIP), and the Institute for Astronomy and Astrophysics of the University of Tübingen, with the support of DLR and the Max Planck Society. The Argelander Institute for Astronomy of the University of Bonn and the Ludwig Maximilians Universität Munich also participated in the science preparation for eROSITA.
    
    The eROSITA data shown here were processed using the eSASS software system developed by the German eROSITA consortium.
    
    \newline
    
    LOFAR is the Low Frequency Array designed and constructed by ASTRON. It has observing, data processing, and data storage facilities in several countries, which are owned by various parties (each with their own funding sources), and which are collectively operated by the ILT foundation under a joint scientific policy. The ILT resources have benefited from the following recent major funding sources: CNRS-INSU, Observatoire de Paris and Université d'Orléans, France; BMBF, MIWF-NRW, MPG, Germany; Science Foundation Ireland (SFI), Department of Business, Enterprise and Innovation (DBEI), Ireland; NWO, The Netherlands; The Science and Technology Facilities Council, UK; Ministry of Science and Higher Education, Poland; The Istituto Nazionale di Astrofisica (INAF), Italy. \\
    
    LOFAR data products were provided by the LOFAR Surveys Key Science project (LSKSP; https://lofar-surveys.org/)
    The efforts of the LSKSP have benefited from funding from the European Research Council, NOVA, NWO, CNRS-INSU, the SURF Co-operative, the UK Science and Technology Funding Council and the J\"ulich Supercomputing Centre. 
    
\newline

The Australian SKA Pathfinder is part of the Australia Telescope National Facility which is managed by CSIRO. Operation of ASKAP is funded by the Australian Government with support from the National Collaborative Research Infrastructure Strategy. ASKAP uses the resources of the Pawsey Supercomputing Centre. Establishment of ASKAP, the Murchison Radio-astronomy Observatory and the Pawsey Supercomputing Centre are initiatives of the Australian Government, with support from the Government of Western Australia and the Science and Industry Endowment Fund. We acknowledge the Wajarri Yamatji people as the traditional owners of the Observatory site.


    \newline
    
    The Legacy Surveys consist of three individual and complementary projects: the Dark Energy Camera Legacy Survey (DECaLS; NOAO Proposal ID $\#$ 2014B-0404; PIs: David Schlegel and Arjun Dey), the Beijing-Arizona Sky Survey (BASS; NOAO Proposal ID $\#$ 2015A-0801; PIs: Zhou Xu and Xiaohui Fan), and the Mayall z-band Legacy Survey (MzLS; NOAO Proposal ID $\#$ 2016A-0453; PI: Arjun Dey). DECaLS, BASS and MzLS together include data obtained, respectively, at the Blanco telescope, Cerro Tololo Inter-American Observatory, National Optical Astronomy Observatory (NOAO); the Bok telescope, Steward Observatory, University of Arizona; and the Mayall telescope, Kitt Peak National Observatory, NOAO. The Legacy Surveys project is honored to be permitted to conduct astronomical research on Iolkam Du'ag (Kitt Peak), a mountain with particular significance to the Tohono O'odham Nation.

NOAO is operated by the Association of Universities for Research in Astronomy (AURA) under a cooperative agreement with the National Science Foundation.

This project used data obtained with the Dark Energy Camera (DECam), which was constructed by the Dark Energy Survey (DES) collaboration. Funding for the DES Projects has been provided by the U.S. Department of Energy, the U.S. National Science Foundation, the Ministry of Science and Education of Spain, the Science and Technology Facilities Council of the United Kingdom, the Higher Education Funding Council for England, the National Center for Supercomputing Applications at the University of Illinois at Urbana-Champaign, the Kavli Institute of Cosmological Physics at the University of Chicago, Center for Cosmology and Astro-Particle Physics at the Ohio State University, the Mitchell Institute for Fundamental Physics and Astronomy at Texas A\&M University, Financiadora de Estudos e Projetos, Fundacao Carlos Chagas Filho de Amparo, Financiadora de Estudos e Projetos, Fundacao Carlos Chagas Filho de Amparo a Pesquisa do Estado do Rio de Janeiro, Conselho Nacional de Desenvolvimento Cientifico e Tecnologico and the Ministerio da Ciencia, Tecnologia e Inovacao, the Deutsche Forschungsgemeinschaft and the Collaborating Institutions in the Dark Energy Survey. The Collaborating Institutions are Argonne National Laboratory, the University of California at Santa Cruz, the University of Cambridge, Centro de Investigaciones Energeticas, Medioambientales y Tecnologicas-Madrid, the University of Chicago, University College London, the DES-Brazil Consortium, the University of Edinburgh, the Eidgenossische Technische Hochschule (ETH) Zurich, Fermi National Accelerator Laboratory, the University of Illinois at Urbana-Champaign, the Institut de Ciencies de l'Espai (IEEC/CSIC), the Institut de Fisica d'Altes Energies, Lawrence Berkeley National Laboratory, the Ludwig-Maximilians Universitat Munchen and the associated Excellence Cluster Universe, the University of Michigan, the National Optical Astronomy Observatory, the University of Nottingham, the Ohio State University, the University of Pennsylvania, the University of Portsmouth, SLAC National Accelerator Laboratory, Stanford University, the University of Sussex, and Texas A\&M University.

BASS is a key project of the Telescope Access Program (TAP), which has been funded by the National Astronomical Observatories of China, the Chinese Academy of Sciences (the Strategic Priority Research Program "The Emergence of Cosmological Structures" Grant \# XDB09000000), and the Special Fund for Astronomy from the Ministry of Finance. The BASS is also supported by the External Cooperation Program of Chinese Academy of Sciences (Grant \# 114A11KYSB20160057), and Chinese National Natural Science Foundation (Grant \# 11433005).

The Legacy Survey team makes use of data products from the Near-Earth Object Wide-field Infrared Survey Explorer (NEOWISE), which is a project of the Jet Propulsion Laboratory/California Institute of Technology. NEOWISE is funded by the National Aeronautics and Space Administration.

The Legacy Surveys imaging of the DESI footprint is supported by the Director, Office of Science, Office of High Energy Physics of the U.S. Department of Energy under Contract No. DE-AC02-05CH1123, by the National Energy Research Scientific Computing Center, a DOE Office of Science User Facility under the same contract; and by the U.S. National Science Foundation, Division of Astronomical Sciences under Contract No. AST-0950945 to NOAO.

\newline

The Hyper Suprime-Cam (HSC) collaboration includes the astronomical communities of Japan and Taiwan, and Princeton University. The HSC instrumentation and software were developed by the National Astronomical Observatory of Japan (NAOJ), the Kavli Institute for the Physics and Mathematics of the Universe (Kavli IPMU), the University of Tokyo, the High Energy Accelerator Research Organization (KEK), the Academia Sinica Institute for Astronomy and Astrophysics in Taiwan (ASIAA), and Princeton University. Funding was contributed by the FIRST program from Japanese Cabinet Office, the Ministry of Education, Culture, Sports, Science and Technology (MEXT), the Japan Society for the Promotion of Science (JSPS), Japan Science and Technology Agency (JST), the Toray Science Foundation, NAOJ, Kavli IPMU, KEK, ASIAA, and Princeton University. 

This paper makes use of software developed for the Large Synoptic Survey Telescope. We thank the LSST Project for making their code available as free software at  http://dm.lsst.org

The Pan-STARRS1 Surveys (PS1) have been made possible through contributions of the Institute for Astronomy, the University of Hawaii, the Pan-STARRS Project Office, the Max-Planck Society and its participating institutes, the Max Planck Institute for Astronomy, Heidelberg and the Max Planck Institute for Extraterrestrial Physics, Garching, The Johns Hopkins University, Durham University, the University of Edinburgh, Queen’s University Belfast, the Harvard-Smithsonian Center for Astrophysics, the Las Cumbres Observatory Global Telescope Network Incorporated, the National Central University of Taiwan, the Space Telescope Science Institute, the National Aeronautics and Space Administration under Grant No. NNX08AR22G issued through the Planetary Science Division of the NASA Science Mission Directorate, the National Science Foundation under Grant No. AST-1238877, the University of Maryland, and Eotvos Lorand University (ELTE) and the Los Alamos National Laboratory.
    
\end{acknowledgements}

%
%

\bibliographystyle{aa} 
\bibliography{bibliography.bib}

\begin{thebibliography}{136}
\expandafter\ifx\csname natexlab\endcsname\relax\def\natexlab#1{#1}\fi

\bibitem[{Ai {et~al.}(2016)Ai, Dou, Fan, Wang, Wu, \& Bian}]{ai16}
Ai, Y., Dou, L., Fan, X., {et~al.} 2016, The Astrophysical Journal Letters,
  823, L37

\bibitem[{{Aihara} {et~al.}(2018){Aihara}, {Arimoto}, {Armstrong}, {Arnouts},
  {Bahcall}, {Bickerton}, {Bosch}, {Bundy}, {Capak}, {Chan}, {Chiba}, {Coupon},
  {Egami}, {Enoki}, {Finet}, {Fujimori}, {Fujimoto}, {Furusawa}, {Furusawa},
  {Goto}, {Goulding}, {Greco}, {Greene}, {Gunn}, {Hamana}, {Harikane},
  {Hashimoto}, {Hattori}, {Hayashi}, {Hayashi}, {He{\l}miniak}, {Higuchi},
  {Hikage}, {Ho}, {Hsieh}, {Huang}, {Huang}, {Ikeda}, {Imanishi}, {Inoue},
  {Iwasawa}, {Iwata}, {Jaelani}, {Jian}, {Kamata}, {Karoji}, {Kashikawa},
  {Katayama}, {Kawanomoto}, {Kayo}, {Koda}, {Koike}, {Kojima}, {Komiyama},
  {Konno}, {Koshida}, {Koyama}, {Kusakabe}, {Leauthaud}, {Lee}, {Lin}, {Lin},
  {Lupton}, {Mand elbaum}, {Matsuoka}, {Medezinski}, {Mineo}, {Miyama},
  {Miyatake}, {Miyazaki}, {Momose}, {More}, {More}, {Moritani}, {Moriya},
  {Morokuma}, {Mukae}, {Murata}, {Murayama}, {Nagao}, {Nakata}, {Niida},
  {Niikura}, {Nishizawa}, {Obuchi}, {Oguri}, {Oishi}, {Okabe}, {Okamoto},
  {Okura}, {Ono}, {Onodera}, {Onoue}, {Osato}, {Ouchi}, {Price}, {Pyo}, {Sako},
  {Sawicki}, {Shibuya}, {Shimasaku}, {Shimono}, {Shirasaki}, {Silverman},
  {Simet}, {Speagle}, {Spergel}, {Strauss}, {Sugahara}, {Sugiyama}, {Suto},
  {Suyu}, {Suzuki}, {Tait}, {Takada}, {Takata}, {Tamura}, {Tanaka}, {Tanaka},
  {Tanaka}, {Tanaka}, {Terai}, {Terashima}, {Toba}, {Tominaga}, {Toshikawa},
  {Turner}, {Uchida}, {Uchiyama}, {Umetsu}, {Uraguchi}, {Urata}, {Usuda},
  {Utsumi}, {Wang}, {Wang}, {Wong}, {Yabe}, {Yamada}, {Yamanoi}, {Yasuda},
  {Yeh}, {Yonehara}, \& {Yuma}}]{aihara18}
{Aihara}, H., {Arimoto}, N., {Armstrong}, R., {et~al.} 2018, \pasj, 70, S4

\bibitem[{Aird {et~al.}(2015)Aird, Coil, Georgakakis, Nandra, Barro, \&
  Pérez-González}]{aird15}
Aird, J., Coil, A.~L., Georgakakis, A., {et~al.} 2015, Monthly Notices of the
  Royal Astronomical Society, 451, 1892

\bibitem[{Ananna {et~al.}(2019)Ananna, Treister, Urry, Ricci, Kirkpatrick,
  LaMassa, Buchner, Civano, Tremmel, \& Marchesi}]{ananna19}
Ananna, T.~T., Treister, E., Urry, C.~M., {et~al.} 2019, The Astrophysical
  Journal, 871, 240

\bibitem[{{Arnaud}(1996)}]{arnaud96}
{Arnaud}, K.~A. 1996, in Astronomical Society of the Pacific Conference Series,
  Vol. 101, Astronomical Data Analysis Software and Systems V, ed. G.~H.
  {Jacoby} \& J.~{Barnes}, 17

\bibitem[{{Arnouts} {et~al.}(1999){Arnouts}, {Cristiani}, {Moscardini},
  {Matarrese}, {Lucchin}, {Fontana}, \& {Giallongo}}]{arnouts99}
{Arnouts}, S., {Cristiani}, S., {Moscardini}, L., {et~al.} 1999, \mnras, 310,
  540

\bibitem[{{Avni} \& {Tananbaum}(1986)}]{avni86}
{Avni}, Y. \& {Tananbaum}, H. 1986, \apj, 305, 83

\bibitem[{{Ba{\~n}ados} {et~al.}(2018){Ba{\~n}ados}, {Venemans},
  {Mazzucchelli}, {Farina}, {Walter}, {Wang}, {Decarli}, {Stern}, {Fan},
  {Davies}, {Hennawi}, {Simcoe}, {Turner}, {Rix}, {Yang}, {Kelson}, {Rudie}, \&
  {Winters}}]{banados18}
{Ba{\~n}ados}, E., {Venemans}, B.~P., {Mazzucchelli}, C., {et~al.} 2018, \nat,
  553, 473

\bibitem[{Barger {et~al.}(2003)Barger, Cowie, Capak, Alexander, Bauer, Brandt,
  Garmire, \& Hornschemeier}]{barger03}
Barger, A.~J., Cowie, L.~L., Capak, P., {et~al.} 2003, The Astrophysical
  Journal Letters, 584, L61

\bibitem[{Bañados {et~al.}(2016)Bañados, Venemans, Decarli, Farina,
  Mazzucchelli, Walter, Fan, Stern, Schlafly, Chambers, Rix, Jiang, McGreer,
  Simcoe, Wang, Yang, Morganson, De~Rosa, Greiner, Baloković, Burgett, Cooper,
  Draper, Flewelling, Hodapp, Jun, Kaiser, Kudritzki, Magnier, Metcalfe,
  Miller, Schindler, Tonry, Wainscoat, Waters, \& Yang}]{banados16}
Bañados, E., Venemans, B.~P., Decarli, R., {et~al.} 2016, The Astrophysical
  Journal Supplement Series, 227, 11

\bibitem[{Bañados {et~al.}(2015)Bañados, Venemans, Morganson, Hodge, Decarli,
  Walter, Stern, Schlafly, Farina, Greiner, Chambers, Fan, Rix, Burgett,
  Draper, Flewelling, Kaiser, Metcalfe, Morgan, Tonry, \&
  Wainscoat}]{banados15}
Bañados, E., Venemans, B.~P., Morganson, E., {et~al.} 2015, The Astrophysical
  Journal, 804, 118

\bibitem[{Becker {et~al.}(1995)Becker, White, \& Helfand}]{becker95}
Becker, R.~H., White, R.~L., \& Helfand, D.~J. 1995, The Astrophysical Journal,
  450, 559

\bibitem[{{Belladitta} {et~al.}(2020){Belladitta}, {Moretti}, {Caccianiga},
  {Spingola}, {Severgnini}, {Della Ceca}, {Ghisellini}, {Dallacasa},
  {Sbarrato}, {Cicone}, {Cassar{\`a}}, \& {Pedani}}]{belladitta20}
{Belladitta}, S., {Moretti}, A., {Caccianiga}, A., {et~al.} 2020, \aap, 635, L7

\bibitem[{{Bhowmick} {et~al.}(2020){Bhowmick}, {Somerville}, {Di Matteo},
  {Wilkins}, {Feng}, \& {Tenneti}}]{bhomwhick20}
{Bhowmick}, A.~K., {Somerville}, R.~S., {Di Matteo}, T., {et~al.} 2020, \mnras,
  496, 754

\bibitem[{{Boller} {et~al.}(2016){Boller}, {Freyberg}, {Tr{\"u}mper}, {Haberl},
  {Voges}, \& {Nandra}}]{boller16}
{Boller}, T., {Freyberg}, M.~J., {Tr{\"u}mper}, J., {et~al.} 2016, \aap, 588,
  A103

\bibitem[{Brandt {et~al.}(2002)Brandt, Schneider, Fan, Strauss, Gunn, Richards,
  Anderson, Vanden~Berk, Bahcall, Brinkmann, Brunner, Chen, Hennessy, Lamb,
  Voges, \& York}]{brandt02}
Brandt, W.~N., Schneider, D.~P., Fan, X., {et~al.} 2002, The Astrophysical
  Journal Letters, 569, L5

\bibitem[{{Brusa} {et~al.}(2009){Brusa}, {Comastri}, {Gilli}, {Hasinger},
  {Iwasawa}, {Mainieri}, {Mignoli}, {Salvato}, {Zamorani}, {Bongiorno},
  {Cappelluti}, {Civano}, {Fiore}, {Merloni}, {Silverman}, {Trump}, {Vignali},
  {Capak}, {Elvis}, {Ilbert}, {Impey}, \& {Lilly}}]{brusa09}
{Brusa}, M., {Comastri}, A., {Gilli}, R., {et~al.} 2009, \apj, 693, 8

\bibitem[{Buchner {et~al.}(2015)Buchner, Georgakakis, Nandra, Brightman,
  Menzel, Liu, Hsu, Salvato, Rangel, Aird, Merloni, \& Ross}]{buchner15}
Buchner, J., Georgakakis, A., Nandra, K., {et~al.} 2015, The Astrophysical
  Journal, 802, 89

\bibitem[{{Buchner} {et~al.}(2014){Buchner}, {Georgakakis}, {Nandra}, {Hsu},
  {Rangel}, {Brightman}, {Merloni}, {Salvato}, {Donley}, \&
  {Kocevski}}]{buchner14}
{Buchner}, J., {Georgakakis}, A., {Nandra}, K., {et~al.} 2014, \aap, 564, A125

\bibitem[{Capak {et~al.}(2011)Capak, Riechers, Scoville, Carilli, Cox, Neri,
  Robertson, Salvato, Schinnerer, Yan, Wilson, Yun, Civano, Elvis, Karim,
  Mobasher, \& Staguhn}]{capak11}
Capak, P.~L., Riechers, D., Scoville, N.~Z., {et~al.} 2011, Nature, 470, 233

\bibitem[{{Cash}(1979)}]{cash79}
{Cash}, W. 1979, \apj, 228, 939

\bibitem[{{Celotti} {et~al.}(2001){Celotti}, {Ghisellini}, \&
  {Chiaberge}}]{cellotti00}
{Celotti}, A., {Ghisellini}, G., \& {Chiaberge}, M. 2001, \mnras, 321, L1

\bibitem[{{Chambers} {et~al.}(2016){Chambers}, {Magnier}, {Metcalfe},
  {Flewelling}, {Huber}, {Waters}, {Denneau}, {Draper}, {Farrow}, {Finkbeiner},
  {Holmberg}, {Koppenhoefer}, {Price}, {Rest}, {Saglia}, {Schlafly}, {Smartt},
  {Sweeney}, {Wainscoat}, {Burgett}, {Chastel}, {Grav}, {Heasley}, {Hodapp},
  {Jedicke}, {Kaiser}, {Kudritzki}, {Luppino}, {Lupton}, {Monet}, {Morgan},
  {Onaka}, {Shiao}, {Stubbs}, {Tonry}, {White}, {Ba{\~n}ados}, {Bell},
  {Bender}, {Bernard}, {Boegner}, {Boffi}, {Botticella}, {Calamida},
  {Casertano}, {Chen}, {Chen}, {Cole}, {Deacon}, {Frenk}, {Fitzsimmons},
  {Gezari}, {Gibbs}, {Goessl}, {Goggia}, {Gourgue}, {Goldman}, {Grant},
  {Grebel}, {Hambly}, {Hasinger}, {Heavens}, {Heckman}, {Henderson}, {Henning},
  {Holman}, {Hopp}, {Ip}, {Isani}, {Jackson}, {Keyes}, {Koekemoer}, {Kotak},
  {Le}, {Liska}, {Long}, {Lucey}, {Liu}, {Martin}, {Masci}, {McLean}, {Mindel},
  {Misra}, {Morganson}, {Murphy}, {Obaika}, {Narayan}, {Nieto-Santisteban},
  {Norberg}, {Peacock}, {Pier}, {Postman}, {Primak}, {Rae}, {Rai}, {Riess},
  {Riffeser}, {Rix}, {R{\"o}ser}, {Russel}, {Rutz}, {Schilbach}, {Schultz},
  {Scolnic}, {Strolger}, {Szalay}, {Seitz}, {Small}, {Smith}, {Soderblom},
  {Taylor}, {Thomson}, {Taylor}, {Thakar}, {Thiel}, {Thilker}, {Unger},
  {Urata}, {Valenti}, {Wagner}, {Walder}, {Walter}, {Watters}, {Werner},
  {Wood-Vasey}, \& {Wyse}}]{chambers16}
{Chambers}, K.~C., {Magnier}, E.~A., {Metcalfe}, N., {et~al.} 2016, arXiv
  e-prints, arXiv:1612.05560

\bibitem[{{Civano} {et~al.}(2011){Civano}, {Brusa}, {Comastri}, {Elvis},
  {Salvato}, {Zamorani}, {Capak}, {Fiore}, {Gilli}, {Hao}, {Ikeda}, {Kakazu},
  {Kartaltepe}, {Masters}, {Miyaji}, {Mignoli}, {Puccetti}, {Shankar},
  {Silverman}, {Vignali}, {Zezas}, \& {Koekemoer}}]{civano11}
{Civano}, F., {Brusa}, M., {Comastri}, A., {et~al.} 2011, \apj, 741, 91

\bibitem[{{Clerc} {et~al.}(2018){Clerc}, {Ramos-Ceja}, {Ridl}, {Lamer},
  {Brunner}, {Hofmann}, {Comparat}, {Pacaud}, {K{\"a}fer}, {Reiprich},
  {Merloni}, {Schmid}, {Brand}, {Wilms}, {Friedrich}, {Finoguenov}, {Dauser},
  \& {Kreykenbohm}}]{clerc18}
{Clerc}, N., {Ramos-Ceja}, M.~E., {Ridl}, J., {et~al.} 2018, \aap, 617, A92

\bibitem[{{Condon} {et~al.}(1998){Condon}, {Cotton}, {Greisen}, {Yin},
  {Perley}, {Taylor}, \& {Broderick}}]{condon98}
{Condon}, J.~J., {Cotton}, W.~D., {Greisen}, E.~W., {et~al.} 1998, \aj, 115,
  1693

\bibitem[{Coppejans {et~al.}(2016)Coppejans, Frey, Cseh, Müller, Paragi,
  Falcke, Gabányi, Gurvits, An, \& Titov}]{coppejans16}
Coppejans, R., Frey, S., Cseh, D., {et~al.} 2016, Monthly Notices of the Royal
  Astronomical Society, 463, 3260

\bibitem[{Coppejans {et~al.}(2017)Coppejans, van Velzen, Intema, Müller, Frey,
  Coppejans, Cseh, Williams, Falcke, Körding, Orrú, Paragi, \&
  Gabányi}]{coppejans17}
Coppejans, R., van Velzen, S., Intema, H.~T., {et~al.} 2017, Monthly Notices of
  the Royal Astronomical Society, stx215

\bibitem[{{Cutri} {et~al.}(2013){Cutri}, {Wright}, {Conrow}, {Fowler},
  {Eisenhardt}, {Grillmair}, {Kirkpatrick}, {Masci}, {McCallon}, {Wheelock},
  {Fajardo-Acosta}, {Yan}, {Benford}, {Harbut}, {Jarrett}, {Lake}, {Leisawitz},
  {Ressler}, {Stanford}, {Tsai}, {Liu}, {Helou}, {Mainzer}, {Gettings},
  {Gonzalez}, {Hoffman}, {Marsh}, {Padgett}, {Skrutskie}, {Beck}, {Papin}, \&
  {Wittman}}]{cutri13}
{Cutri}, R.~M., {Wright}, E.~L., {Conrow}, T., {et~al.} 2013, {Explanatory
  Supplement to the AllWISE Data Release Products}, Explanatory Supplement to
  the AllWISE Data Release Products

\bibitem[{de~Gasperin {et~al.}(2019)de~Gasperin, Dijkema, Drabent, Mevius,
  Rafferty, van Weeren, Br{\"{u}}ggen, Callingham, Emig, Heald, Intema,
  Morabito, Offringa, Oonk, Orr{\`{u}}, R{\"{o}}ttgering, Sabater, Shimwell,
  Shulevski, \& Williams}]{deGasperin2019}
de~Gasperin, F., Dijkema, T.~J., Drabent, A., {et~al.} 2019, A{\&}A, 622, A5

\bibitem[{{Evans} {et~al.}(2020){Evans}, {Primini}, {Miller}, {Evans}, {Allen},
  {Anderson}, {Becker}, {Budynkiewicz}, {Burke}, {Chen}, {Civano}, {D'Abrusco},
  {Doe}, {Fabbiano}, {Martinez Galarza}, {Gibbs}, {Glotfelty}, {Graessle},
  {Grier}, {Hain}, {Hall}, {Harbo}, {Houck}, {Lauer}, {Laurino}, {Lee},
  {McCollough}, {McDowell}, {McLaughlin}, {Morgan}, {Mossman}, {Nguyen},
  {Nichols}, {Nowak}, {Paxson}, {Perdikeas}, {Plummer}, {Rots},
  {Siemiginowska}, {Sundheim}, {Thong}, {Tibbetts}, {Van Stone}, {Winkelman},
  \& {Zografou}}]{evans20}
{Evans}, I.~N., {Primini}, F.~A., {Miller}, J.~B., {et~al.} 2020, in American
  Astronomical Society Meeting Abstracts, American Astronomical Society Meeting
  Abstracts, 154.05

\bibitem[{Fan {et~al.}(2001)Fan, Narayanan, Lupton, Strauss, Knapp, Becker,
  White, Pentericci, Leggett, Haiman, Gunn, Ivezić, Schneider, Anderson,
  Brinkmann, Bahcall, Connolly, Csabai, Doi, Fukugita, Geballe, Grebel,
  Harbeck, Hennessy, Lamb, Miknaitis, Munn, Nichol, Okamura, Pier, Prada,
  Richards, Szalay, \& York}]{fan01}
Fan, X., Narayanan, V.~K., Lupton, R.~H., {et~al.} 2001, The Astronomical
  Journal, 122, 2833

\bibitem[{{Fanaroff} \& {Riley}(1974)}]{fanaroff74}
{Fanaroff}, B.~L. \& {Riley}, J.~M. 1974, \mnras, 167, 31P

\bibitem[{{Feroz} {et~al.}(2009){Feroz}, {Hobson}, \& {Bridges}}]{feroz09}
{Feroz}, F., {Hobson}, M.~P., \& {Bridges}, M. 2009, \mnras, 398, 1601

\bibitem[{Frey {et~al.}(2003)Frey, Mosoni, Paragi, \& Gurvits}]{frey03}
Frey, S., Mosoni, L., Paragi, Z., \& Gurvits, L.~I. 2003, MNRAS, 343, 20

\bibitem[{{Frey} {et~al.}(2011){Frey}, {Paragi}, {Gurvits}, {Gab{\'a}nyi}, \&
  {Cseh}}]{frey11}
{Frey}, S., {Paragi}, Z., {Gurvits}, L.~I., {Gab{\'a}nyi}, K.~{\'E}., \&
  {Cseh}, D. 2011, \aap, 531, L5

\bibitem[{{Frey} {et~al.}(2005){Frey}, {Paragi}, {Mosoni}, \&
  {Gurvits}}]{frey05}
{Frey}, S., {Paragi}, Z., {Mosoni}, L., \& {Gurvits}, L.~I. 2005, \aap, 436,
  L13

\bibitem[{{Gaia Collaboration} {et~al.}(2018){Gaia Collaboration}, {Brown},
  {Vallenari}, {Prusti}, {de Bruijne}, {Babusiaux}, {Bailer-Jones}, {Biermann},
  {Evans}, {Eyer}, {Jansen}, {Jordi}, {Klioner}, {Lammers}, {Lindegren},
  {Luri}, {Mignard}, {Panem}, {Pourbaix}, {Randich}, {Sartoretti}, {Siddiqui},
  {Soubiran}, {van Leeuwen}, {Walton}, {Arenou}, {Bastian}, {Cropper},
  {Drimmel}, {Katz}, {Lattanzi}, {Bakker}, {Cacciari}, {Casta{\~n}eda},
  {Chaoul}, {Cheek}, {De Angeli}, {Fabricius}, {Guerra}, {Holl}, {Masana},
  {Messineo}, {Mowlavi}, {Nienartowicz}, {Panuzzo}, {Portell}, {Riello},
  {Seabroke}, {Tanga}, {Th{\'e}venin}, {Gracia-Abril}, {Comoretto},
  {Garcia-Reinaldos}, {Teyssier}, {Altmann}, {Andrae}, {Audard},
  {Bellas-Velidis}, {Benson}, {Berthier}, {Blomme}, {Burgess}, {Busso},
  {Carry}, {Cellino}, {Clementini}, {Clotet}, {Creevey}, {Davidson}, {De
  Ridder}, {Delchambre}, {Dell'Oro}, {Ducourant},
  {Fern{\'a}ndez-Hern{\'a}ndez}, {Fouesneau}, {Fr{\'e}mat}, {Galluccio},
  {Garc{\'\i}a-Torres}, {Gonz{\'a}lez-N{\'u}{\~n}ez}, {Gonz{\'a}lez-Vidal},
  {Gosset}, {Guy}, {Halbwachs}, {Hambly}, {Harrison}, {Hern{\'a}ndez},
  {Hestroffer}, {Hodgkin}, {Hutton}, {Jasniewicz}, {Jean-Antoine-Piccolo},
  {Jordan}, {Korn}, {Krone-Martins}, {Lanzafame}, {Lebzelter}, {L{\"o}ffler},
  {Manteiga}, {Marrese}, {Mart{\'\i}n-Fleitas}, {Moitinho}, {Mora}, {Muinonen},
  {Osinde}, {Pancino}, {Pauwels}, {Petit}, {Recio-Blanco}, {Richards},
  {Rimoldini}, {Robin}, {Sarro}, {Siopis}, {Smith}, {Sozzetti}, {S{\"u}veges},
  {Torra}, {van Reeven}, {Abbas}, {Abreu Aramburu}, {Accart}, {Aerts},
  {Altavilla}, {{\'A}lvarez}, {Alvarez}, {Alves}, {Anderson}, {Andrei},
  {Anglada Varela}, {Antiche}, {Antoja}, {Arcay}, {Astraatmadja}, {Bach},
  {Baker}, {Balaguer-N{\'u}{\~n}ez}, {Balm}, {Barache}, {Barata}, {Barbato},
  {Barblan}, {Barklem}, {Barrado}, {Barros}, {Barstow}, {Bartholom{\'e}
  Mu{\~n}oz}, {Bassilana}, {Becciani}, {Bellazzini}, {Berihuete}, {Bertone},
  {Bianchi}, {Bienaym{\'e}}, {Blanco-Cuaresma}, {Boch}, {Boeche}, {Bombrun},
  {Borrachero}, {Bossini}, {Bouquillon}, {Bourda}, {Bragaglia}, {Bramante},
  {Breddels}, {Bressan}, {Brouillet}, {Br{\"u}semeister}, {Brugaletta},
  {Bucciarelli}, {Burlacu}, {Busonero}, {Butkevich}, {Buzzi}, {Caffau},
  {Cancelliere}, {Cannizzaro}, {Cantat-Gaudin}, {Carballo}, {Carlucci},
  {Carrasco}, {Casamiquela}, {Castellani}, {Castro-Ginard}, {Charlot},
  {Chemin}, {Chiavassa}, {Cocozza}, {Costigan}, {Cowell}, {Crifo}, {Crosta},
  {Crowley}, {Cuypers}, {Dafonte}, {Damerdji}, {Dapergolas}, {David}, {David},
  {de Laverny}, {De Luise}, {De March}, {de Martino}, {de Souza}, {de Torres},
  {Debosscher}, {del Pozo}, {Delbo}, {Delgado}, {Delgado}, {Di Matteo},
  {Diakite}, {Diener}, {Distefano}, {Dolding}, {Drazinos}, {Dur{\'a}n},
  {Edvardsson}, {Enke}, {Eriksson}, {Esquej}, {Eynard Bontemps}, {Fabre},
  {Fabrizio}, {Faigler}, {Falc{\~a}o}, {Farr{\`a}s Casas}, {Federici},
  {Fedorets}, {Fernique}, {Figueras}, {Filippi}, {Findeisen}, {Fonti},
  {Fraile}, {Fraser}, {Fr{\'e}zouls}, {Gai}, {Galleti}, {Garabato},
  {Garc{\'\i}a-Sedano}, {Garofalo}, {Garralda}, {Gavel}, {Gavras}, {Gerssen},
  {Geyer}, {Giacobbe}, {Gilmore}, {Girona}, {Giuffrida}, {Glass}, {Gomes},
  {Granvik}, {Gueguen}, {Guerrier}, {Guiraud}, {Guti{\'e}rrez-S{\'a}nchez},
  {Haigron}, {Hatzidimitriou}, {Hauser}, {Haywood}, {Heiter}, {Helmi}, {Heu},
  {Hilger}, {Hobbs}, {Hofmann}, {Holland}, {Huckle}, {Hypki}, {Icardi},
  {Jan{\ss}en}, {Jevardat de Fombelle}, {Jonker}, {Juh{\'a}sz}, {Julbe},
  {Karampelas}, {Kewley}, {Klar}, {Kochoska}, {Kohley}, {Kolenberg},
  {Kontizas}, {Kontizas}, {Koposov}, {Kordopatis}, {Kostrzewa-Rutkowska},
  {Koubsky}, {Lambert}, {Lanza}, {Lasne}, {Lavigne}, {Le Fustec}, {Le
  Poncin-Lafitte}, {Lebreton}, {Leccia}, {Leclerc}, {Lecoeur-Taibi},
  {Lenhardt}, {Leroux}, {Liao}, {Licata}, {Lindstr{\o}m}, {Lister}, {Livanou},
  {Lobel}, {L{\'o}pez}, {Managau}, {Mann}, {Mantelet}, {Marchal}, {Marchant},
  {Marconi}, {Marinoni}, {Marschalk{\'o}}, {Marshall}, {Martino}, {Marton},
  {Mary}, {Massari}, {Matijevi{\v{c}}}, {Mazeh}, {McMillan}, {Messina},
  {Michalik}, {Millar}, {Molina}, {Molinaro}, {Moln{\'a}r}, {Montegriffo},
  {Mor}, {Morbidelli}, {Morel}, {Morris}, {Mulone}, {Muraveva}, {Musella},
  {Nelemans}, {Nicastro}, {Noval}, {O'Mullane}, {Ord{\'e}novic},
  {Ord{\'o}{\~n}ez-Blanco}, {Osborne}, {Pagani}, {Pagano}, {Pailler},
  {Palacin}, {Palaversa}, {Panahi}, {Pawlak}, {Piersimoni}, {Pineau}, {Plachy},
  {Plum}, {Poggio}, {Poujoulet}, {Pr{\v{s}}a}, {Pulone}, {Racero}, {Ragaini},
  {Rambaux}, {Ramos-Lerate}, {Regibo}, {Reyl{\'e}}, {Riclet}, {Ripepi}, {Riva},
  {Rivard}, {Rixon}, {Roegiers}, {Roelens}, {Romero-G{\'o}mez}, {Rowell},
  {Royer}, {Ruiz-Dern}, {Sadowski}, {Sagrist{\`a} Sell{\'e}s}, {Sahlmann},
  {Salgado}, {Salguero}, {Sanna}, {Santana-Ros}, {Sarasso}, {Savietto},
  {Schultheis}, {Sciacca}, {Segol}, {Segovia}, {S{\'e}gransan}, {Shih},
  {Siltala}, {Silva}, {Smart}, {Smith}, {Solano}, {Solitro}, {Sordo}, {Soria
  Nieto}, {Souchay}, {Spagna}, {Spoto}, {Stampa}, {Steele},
  {Steidelm{\"u}ller}, {Stephenson}, {Stoev}, {Suess}, {Surdej}, {Szabados},
  {Szegedi-Elek}, {Tapiador}, {Taris}, {Tauran}, {Taylor}, {Teixeira},
  {Terrett}, {Teyssand ier}, {Thuillot}, {Titarenko}, {Torra Clotet}, {Turon},
  {Ulla}, {Utrilla}, {Uzzi}, {Vaillant}, {Valentini}, {Valette}, {van Elteren},
  {Van Hemelryck}, {van Leeuwen}, {Vaschetto}, {Vecchiato}, {Veljanoski},
  {Viala}, {Vicente}, {Vogt}, {von Essen}, {Voss}, {Votruba}, {Voutsinas},
  {Walmsley}, {Weiler}, {Wertz}, {Wevers}, {Wyrzykowski}, {Yoldas},
  {{\v{Z}}erjal}, {Ziaeepour}, {Zorec}, {Zschocke}, {Zucker}, {Zurbach}, \&
  {Zwitter}}]{gaia18}
{Gaia Collaboration}, {Brown}, A.~G.~A., {Vallenari}, A., {et~al.} 2018, \aap,
  616, A1

\bibitem[{{Gehrels}(1986)}]{gehrels86}
{Gehrels}, N. 1986, \apj, 303, 336

\bibitem[{Georgakakis {et~al.}(2015)Georgakakis, Aird, Buchner, Salvato,
  Menzel, Brandt, McGreer, Dwelly, Mountrichas, Koki, Georgantopoulos, Hsu,
  Merloni, Liu, Nandra, \& Ross}]{georgakakis15}
Georgakakis, A., Aird, J., Buchner, J., {et~al.} 2015, Monthly Notices of the
  Royal Astronomical Society, 453, 1946

\bibitem[{{Georgakakis} {et~al.}(2008){Georgakakis}, {Nandra}, {Laird}, {Aird},
  \& {Trichas}}]{georgakakis08}
{Georgakakis}, A., {Nandra}, K., {Laird}, E.~S., {Aird}, J., \& {Trichas}, M.
  2008, \mnras, 388, 1205

\bibitem[{{Giroletti} \& {Polatidis}(2009)}]{giroletti09}
{Giroletti}, M. \& {Polatidis}, A. 2009, Astronomische Nachrichten, 330, 193

\bibitem[{{Godfrey} {et~al.}(2009){Godfrey}, {Bicknell}, {Lovell}, {Jauncey},
  {Gelbord}, {Schwartz}, {Marshall}, {Birkinshaw}, {Georganopoulos}, {Murphy},
  {Perlman}, \& {Worrall}}]{2009ApJ...695..707G}
{Godfrey}, L.~E.~H., {Bicknell}, G.~V., {Lovell}, J.~E.~J., {et~al.} 2009,
  \apj, 695, 707

\bibitem[{{Greiner} {et~al.}(2008){Greiner}, {Bornemann}, {Clemens}, {Deuter},
  {Hasinger}, {Honsberg}, {Huber}, {Huber}, {Krauss}, {Kr{\"u}hler},
  {K{\"u}pc{\"u} Yolda{\textcommabelow s}}, {Mayer-Hasselwand er}, {Mican},
  {Primak}, {Schrey}, {Steiner}, {Szokoly}, {Th{\"o}ne}, {Yolda{\textcommabelow
  s}}, {Klose}, {Laux}, \& {Winkler}}]{grainer08}
{Greiner}, J., {Bornemann}, W., {Clemens}, C., {et~al.} 2008, \pasp, 120, 405

\bibitem[{Harris {et~al.}(2019)Harris, Mold{\'{o}}n, Oonk, Massaro, Paggi,
  Deller, Godfrey, Morganti, \& Jorstad}]{Harris_2019}
Harris, D.~E., Mold{\'{o}}n, J., Oonk, J. R.~R., {et~al.} 2019, The
  Astrophysical Journal, 873, 21

\bibitem[{Hasinger {et~al.}(2005)Hasinger, Miyaji, \& Schmidt}]{hasinger05}
Hasinger, G., Miyaji, T., \& Schmidt, M. 2005, Astronomy and Astrophysics, 441,
  417

\bibitem[{{HI4PI Collaboration} {et~al.}(2016){HI4PI Collaboration}, {Ben
  Bekhti}, {Fl{\"o}er}, {Keller}, {Kerp}, {Lenz}, {Winkel}, {Bailin},
  {Calabretta}, {Dedes}, {Ford}, {Gibson}, {Haud}, {Janowiecki}, {Kalberla},
  {Lockman}, {McClure-Griffiths}, {Murphy}, {Nakanishi}, {Pisano}, \&
  {Staveley-Smith}}]{hi4pi16}
{HI4PI Collaboration}, {Ben Bekhti}, N., {Fl{\"o}er}, L., {et~al.} 2016, \aap,
  594, A116

\bibitem[{{Hiroi} {et~al.}(2012){Hiroi}, {Ueda}, {Akiyama}, \&
  {Watson}}]{hiroi12}
{Hiroi}, K., {Ueda}, Y., {Akiyama}, M., \& {Watson}, M.~G. 2012, \apj, 758, 49

\bibitem[{{Ilbert} {et~al.}(2006){Ilbert}, {Arnouts}, {McCracken},
  {Bolzonella}, {Bertin}, {Le F{\`e}vre}, {Mellier}, {Zamorani}, {Pell{\`o}},
  {Iovino}, {Tresse}, {Le Brun}, {Bottini}, {Garilli}, {Maccagni}, {Picat},
  {Scaramella}, {Scodeggio}, {Vettolani}, {Zanichelli}, {Adami}, {Bardelli},
  {Cappi}, {Charlot}, {Ciliegi}, {Contini}, {Cucciati}, {Foucaud}, {Franzetti},
  {Gavignaud}, {Guzzo}, {Marano}, {Marinoni}, {Mazure}, {Meneux}, {Merighi},
  {Paltani}, {Pollo}, {Pozzetti}, {Radovich}, {Zucca}, {Bondi}, {Bongiorno},
  {Busarello}, {de La Torre}, {Gregorini}, {Lamareille}, {Mathez}, {Merluzzi},
  {Ripepi}, {Rizzo}, \& {Vergani}}]{ilbert06}
{Ilbert}, O., {Arnouts}, S., {McCracken}, H.~J., {et~al.} 2006, \aap, 457, 841

\bibitem[{{Intema} {et~al.}(2017){Intema}, {Jagannathan}, {Mooley}, \&
  {Frail}}]{Intema2017}
{Intema}, H.~T., {Jagannathan}, P., {Mooley}, K.~P., \& {Frail}, D.~A. 2017,
  \aap, 598, A78

\bibitem[{Jiang {et~al.}(2016)Jiang, McGreer, Fan, Strauss, Bañados, Becker,
  Bian, Farnsworth, Shen, Wang, Wang, Wang, White, Wu, Wu, Yang, \&
  Yang}]{jiang16}
Jiang, L., McGreer, I.~D., Fan, X., {et~al.} 2016, The Astrophysical Journal,
  833, 222

\bibitem[{{Johnston} {et~al.}(2008){Johnston}, {Taylor}, {Bailes}, {Bartel},
  {Baugh}, {Bietenholz}, {Blake}, {Braun}, {Brown}, {Chatterjee}, {Darling},
  {Deller}, {Dodson}, {Edwards}, {Ekers}, {Ellingsen}, {Feain}, {Gaensler},
  {Haverkorn}, {Hobbs}, {Hopkins}, {Jackson}, {James}, {Joncas}, {Kaspi},
  {Kilborn}, {Koribalski}, {Kothes}, {Landecker}, {Lenc}, {Lovell}, {Macquart},
  {Manchester}, {Matthews}, {McClure-Griffiths}, {Norris}, {Pen}, {Phillips},
  {Power}, {Protheroe}, {Sadler}, {Schmidt}, {Stairs}, {Staveley-Smith},
  {Stil}, {Tingay}, {Tzioumis}, {Walker}, {Wall}, \& {Wolleben}}]{johnston08}
{Johnston}, S., {Taylor}, R., {Bailes}, M., {et~al.} 2008, Experimental
  Astronomy, 22, 151

\bibitem[{{Just} {et~al.}(2007){Just}, {Brandt}, {Shemmer}, {Steffen},
  {Schneider}, {Chartas}, \& {Garmire}}]{just07}
{Just}, D.~W., {Brandt}, W.~N., {Shemmer}, O., {et~al.} 2007, \apj, 665, 1004

\bibitem[{{Kellermann} {et~al.}(1989){Kellermann}, {Sramek}, {Schmidt},
  {Shaffer}, \& {Green}}]{kellerman89}
{Kellermann}, K.~I., {Sramek}, R., {Schmidt}, M., {Shaffer}, D.~B., \& {Green},
  R. 1989, \aj, 98, 1195

\bibitem[{{Khorunzhev} {et~al.}(2018){Khorunzhev}, {Sazonov}, \&
  {Burenin}}]{Khorunzhev18}
{Khorunzhev}, G.~A., {Sazonov}, S.~Y., \& {Burenin}, R.~A. 2018, Astronomy
  Letters, 44, 500

\bibitem[{{Kolodzig} {et~al.}(2013){Kolodzig}, {Gilfanov}, {Sunyaev},
  {Sazonov}, \& {Brusa}}]{kolodzig13}
{Kolodzig}, A., {Gilfanov}, M., {Sunyaev}, R., {Sazonov}, S., \& {Brusa}, M.
  2013, \aap, 558, A89

\bibitem[{{Kuijken} {et~al.}(2019){Kuijken}, {Heymans}, {Dvornik},
  {Hildebrandt}, {de Jong}, {Wright}, {Erben}, {Bilicki}, {Giblin}, {Shan},
  {Getman}, {Grado}, {Hoekstra}, {Miller}, {Napolitano}, {Paolilo}, {Radovich},
  {Schneider}, {Sutherland }, {Tewes}, {Tortora}, {Valentijn}, \& {Verdoes
  Kleijn}}]{kuijken19}
{Kuijken}, K., {Heymans}, C., {Dvornik}, A., {et~al.} 2019, \aap, 625, A2

\bibitem[{{Kulkarni} {et~al.}(2019){Kulkarni}, {Worseck}, \&
  {Hennawi}}]{kulkarni19}
{Kulkarni}, G., {Worseck}, G., \& {Hennawi}, J.~F. 2019, \mnras, 488, 1035

\bibitem[{Kurk {et~al.}(2009)Kurk, Walter, Fan, Jiang, Jester, Rix, \&
  Riechers}]{kurk09}
Kurk, J.~D., Walter, F., Fan, X., {et~al.} 2009, The Astrophysical Journal,
  702, 833

\bibitem[{{Lawrence} {et~al.}(2007){Lawrence}, {Warren}, {Almaini}, {Edge},
  {Hambly}, {Jameson}, {Lucas}, {Casali}, {Adamson}, {Dye}, {Emerson},
  {Foucaud}, {Hewett}, {Hirst}, {Hodgkin}, {Irwin}, {Lodieu}, {McMahon},
  {Simpson}, {Smail}, {Mortlock}, \& {Folger}}]{lawrence07}
{Lawrence}, A., {Warren}, S.~J., {Almaini}, O., {et~al.} 2007, \mnras, 379,
  1599

\bibitem[{{Lusso} {et~al.}(2010){Lusso}, {Comastri}, {Vignali}, {Zamorani},
  {Brusa}, {Gilli}, {Iwasawa}, {Salvato}, {Civano}, {Elvis}, {Merloni},
  {Bongiorno}, {Trump}, {Koekemoer}, {Schinnerer}, {Le Floc'h}, {Cappelluti},
  {Jahnke}, {Sargent}, {Silverman}, {Mainieri}, {Fiore}, {Bolzonella}, {Le
  F{\`e}vre}, {Garilli}, {Iovino}, {Kneib}, {Lamareille}, {Lilly}, {Mignoli},
  {Scodeggio}, \& {Vergani}}]{lusso10}
{Lusso}, E., {Comastri}, A., {Vignali}, C., {et~al.} 2010, \aap, 512, A34

\bibitem[{{Lusso} \& {Risaliti}(2016)}]{lusso16}
{Lusso}, E. \& {Risaliti}, G. 2016, \apj, 819, 154

\bibitem[{{Madau} \& {Rees}(2000)}]{madau00}
{Madau}, P. \& {Rees}, M.~J. 2000, \apjl, 542, L69

\bibitem[{Manti {et~al.}(2017)Manti, Gallerani, Ferrara, Greig, \&
  Feruglio}]{manti17}
Manti, S., Gallerani, S., Ferrara, A., Greig, B., \& Feruglio, C. 2017, Monthly
  Notices of the Royal Astronomical Society, 466, 1160

\bibitem[{Marchesi {et~al.}(2016)Marchesi, Lanzuisi, Civano, Iwasawa, Suh,
  Comastri, Zamorani, Allevato, Griffiths, Miyaji, Ranalli, Salvato,
  Schawinski, Silverman, Treister, Urry, \& Vignali}]{marchesi16}
Marchesi, S., Lanzuisi, G., Civano, F., {et~al.} 2016, The Astrophysical
  Journal, 830, 100

\bibitem[{Matsuoka {et~al.}(2019)Matsuoka, Iwasawa, Onoue, Kashikawa, Strauss,
  Lee, Imanishi, Nagao, Akiyama, Asami, Bosch, Furusawa, Goto, Gunn, Harikane,
  Ikeda, Izumi, Kawaguchi, Kato, Kikuta, Kohno, Komiyama, Koyama, Lupton,
  Minezaki, Miyazaki, Murayama, Niida, Nishizawa, Noboriguchi, Oguri, Ono,
  Ouchi, Price, Sameshima, Schulze, Silverman, Sugiyama, Tait, Takada, Takata,
  Tanaka, Tang, Toba, Utsumi, Wang, \& Yamashita}]{matsuoka19}
Matsuoka, Y., Iwasawa, K., Onoue, M., {et~al.} 2019, The Astrophysical Journal,
  883, 183

\bibitem[{Matsuoka {et~al.}(2018{\natexlab{a}})Matsuoka, Iwasawa, Onoue,
  Kashikawa, Strauss, Lee, Imanishi, Nagao, Akiyama, Asami, Bosch, Furusawa,
  Goto, Gunn, Harikane, Ikeda, Izumi, Kawaguchi, Kato, Kikuta, Kohno, Komiyama,
  Lupton, Minezaki, Miyazaki, Morokuma, Murayama, Niida, Nishizawa, Oguri, Ono,
  Ouchi, Price, Sameshima, Schulze, Shirakata, Silverman, Sugiyama, Tait,
  Takada, Takata, Tanaka, Tang, Toba, Utsumi, Wang, \& Yamashita}]{matsuoka18b}
Matsuoka, Y., Iwasawa, K., Onoue, M., {et~al.} 2018{\natexlab{a}}, The
  Astrophysical Journal Supplement Series, 237, 5

\bibitem[{Matsuoka {et~al.}(2018{\natexlab{b}})Matsuoka, Onoue, Kashikawa,
  Iwasawa, Strauss, Nagao, Imanishi, Lee, Akiyama, Asami, Bosch, Foucaud,
  Furusawa, Goto, Gunn, Harikane, Ikeda, Izumi, Kawaguchi, Kikuta, Kohno,
  Komiyama, Lupton, Minezaki, Miyazaki, Morokuma, Murayama, Niida, Nishizawa,
  Oguri, Ono, Ouchi, Price, Sameshima, Schulze, Shirakata, Silverman, Sugiyama,
  Tait, Takada, Takata, Tanaka, Tang, Toba, Utsumi, \& Wang}]{matsuoka18a}
Matsuoka, Y., Onoue, M., Kashikawa, N., {et~al.} 2018{\natexlab{b}},
  Publications of the Astronomical Society of Japan, 70, S35

\bibitem[{{Matsuoka} {et~al.}(2016){Matsuoka}, {Onoue}, {Kashikawa}, {Iwasawa},
  {Strauss}, {Nagao}, {Imanishi}, {Niida}, {Toba}, {Akiyama}, {Asami}, {Bosch},
  {Foucaud}, {Furusawa}, {Goto}, {Gunn}, {Harikane}, {Ikeda}, {Kawaguchi},
  {Kikuta}, {Komiyama}, {Lupton}, {Minezaki}, {Miyazaki}, {Morokuma},
  {Murayama}, {Nishizawa}, {Ono}, {Ouchi}, {Price}, {Sameshima}, {Silverman},
  {Sugiyama}, {Tait}, {Takada}, {Takata}, {Tanaka}, {Tang}, \&
  {Utsumi}}]{matsuoka16}
{Matsuoka}, Y., {Onoue}, M., {Kashikawa}, N., {et~al.} 2016, \apj, 828, 26

\bibitem[{{Medvedev} {et~al.}(2020{\natexlab{a}}){Medvedev},
  {{\mockalph{bbbbb}}Sazonov}, {Gilfanov}, {Burenin}, {Khorunzhev},
  {Meshcheryakov}, {Sunyaev}, {Bikmaev}, \& {Irtuganov}}]{medvedev20a}
{Medvedev}, P., {{\mockalph{bbbbb}}Sazonov}, S., {Gilfanov}, M., {et~al.}
  2020{\natexlab{a}}, \mnras, 497, 1842

\bibitem[{{Medvedev} {et~al.}(2020{\natexlab{b}}){Medvedev},
  {{\mockalph{cccccc}}Gilfanov}, {Sazonov}, {Schartel}, \&
  {Sunyaev}}]{medvedev20b}
{Medvedev}, P., {{\mockalph{cccccc}}Gilfanov}, M., {Sazonov}, S., {Schartel},
  N., \& {Sunyaev}, R. 2020{\natexlab{b}}, arXiv e-prints, arXiv:2011.13724

\bibitem[{Menzel {et~al.}(2016)Menzel, Merloni, Georgakakis, Salvato, Aubourg,
  Brandt, Brusa, Buchner, Dwelly, Nandra, Pâris, Petitjean, \&
  Schwope}]{menzel16}
Menzel, M.-L., Merloni, A., Georgakakis, A., {et~al.} 2016, Monthly Notices of
  the Royal Astronomical Society, 457, 110

\bibitem[{{Mingaliev} {et~al.}(2012){Mingaliev}, {Sotnikova}, {Torniainen},
  {Tornikoski}, \& {Udovitskiy}}]{mingaliev12}
{Mingaliev}, M.~G., {Sotnikova}, Y.~V., {Torniainen}, I., {Tornikoski}, M., \&
  {Udovitskiy}, R.~Y. 2012, \aap, 544, A25

\bibitem[{Miyaji {et~al.}(2015)Miyaji, Hasinger, Salvato, Brusa, Cappelluti,
  Civano, Puccetti, Elvis, Brunner, Fotopoulou, Ueda, Griffiths, Koekemoer,
  Akiyama, Comastri, Gilli, Lanzuisi, Merloni, \& Vignali}]{miyaji15}
Miyaji, T., Hasinger, G., Salvato, M., {et~al.} 2015, The Astrophysical
  Journal, 804, 104

\bibitem[{{Moster} {et~al.}(2011){Moster}, {Somerville}, {Newman}, \&
  {Rix}}]{moster11}
{Moster}, B.~P., {Somerville}, R.~S., {Newman}, J.~A., \& {Rix}, H.-W. 2011,
  \apj, 731, 113

\bibitem[{Nanni {et~al.}(2017)Nanni, Vignali, Gilli, Moretti, \&
  Brandt}]{nanni17}
Nanni, R., Vignali, C., Gilli, R., Moretti, A., \& Brandt, W.~N. 2017,
  Astronomy and Astrophysics, 603, A128

\bibitem[{Nelder \& Wedderburn(1972)}]{nelder72}
Nelder, J.~A. \& Wedderburn, R. W.~M. 1972, Journal of the Royal Statistical
  Society: Series A (General), 135, 370

\bibitem[{{O'Dea}(1998)}]{odea98}
{O'Dea}, C.~P. 1998, \pasp, 110, 493

\bibitem[{{O'Dea} \& {Saikia}(2020)}]{odea20}
{O'Dea}, C.~P. \& {Saikia}, D.~J. 2020, arXiv e-prints, arXiv:2009.02750

\bibitem[{{Onoue} {et~al.}(2019){Onoue}, {Kashikawa}, {Matsuoka}, {Kato},
  {Izumi}, {Nagao}, {Strauss}, {Harikane}, {Imanishi}, {Ito}, {Iwasawa},
  {Kawaguchi}, {Lee}, {Noboriguchi}, {Suh}, {Tanaka}, \& {Toba}}]{onoue19}
{Onoue}, M., {Kashikawa}, N., {Matsuoka}, Y., {et~al.} 2019, \apj, 880, 77

\bibitem[{{Orienti}(2016)}]{orienti16}
{Orienti}, M. 2016, Astronomische Nachrichten, 337, 9

\bibitem[{{Orienti} \& {Dallacasa}(2008)}]{orienti08}
{Orienti}, M. \& {Dallacasa}, D. 2008, \aap, 477, 807

\bibitem[{{Orienti} {et~al.}(2007){Orienti}, {Dallacasa}, \&
  {Stanghellini}}]{orienti07}
{Orienti}, M., {Dallacasa}, D., \& {Stanghellini}, C. 2007, \aap, 475, 813

\bibitem[{{Padovani} \& {Urry}(1992)}]{padovani92}
{Padovani}, P. \& {Urry}, C.~M. 1992, \apj, 387, 449

\bibitem[{Page \& Carrera(2000)}]{page00}
Page, M.~J. \& Carrera, F.~J. 2000, Monthly Notices of the Royal Astronomical
  Society, 311, 433, publisher: Oxford Academic

\bibitem[{Petric {et~al.}(2003)Petric, Carilli, Bertoldi, Fan, Cox, Strauss,
  Omont, \& Schneider}]{petric03}
Petric, A.~O., Carilli, C.~L., Bertoldi, F., {et~al.} 2003, The Astronomical
  Journal, 126, 15

\bibitem[{{Planck Collaboration} {et~al.}(2020){Planck Collaboration},
  {Aghanim}, {Akrami}, {Ashdown}, {Aumont}, {Baccigalupi}, {Ballardini},
  {Banday}, {Barreiro}, {Bartolo}, {Basak}, {Battye}, {Benabed}, {Bernard},
  {Bersanelli}, {Bielewicz}, {Bock}, {Bond}, {Borrill}, {Bouchet}, {Boulanger},
  {Bucher}, {Burigana}, {Butler}, {Calabrese}, {Cardoso}, {Carron},
  {Challinor}, {Chiang}, {Chluba}, {Colombo}, {Combet}, {Contreras}, {Crill},
  {Cuttaia}, {de Bernardis}, {de Zotti}, {Delabrouille}, {Delouis}, {Di
  Valentino}, {Diego}, {Dor{\'e}}, {Douspis}, {Ducout}, {Dupac}, {Dusini},
  {Efstathiou}, {Elsner}, {En{\ss}lin}, {Eriksen}, {Fantaye}, {Farhang},
  {Fergusson}, {Fernandez-Cobos}, {Finelli}, {Forastieri}, {Frailis},
  {Fraisse}, {Franceschi}, {Frolov}, {Galeotta}, {Galli}, {Ganga},
  {G{\'e}nova-Santos}, {Gerbino}, {Ghosh}, {Gonz{\'a}lez-Nuevo}, {G{\'o}rski},
  {Gratton}, {Gruppuso}, {Gudmundsson}, {Hamann}, {Handley}, {Hansen},
  {Herranz}, {Hildebrandt}, {Hivon}, {Huang}, {Jaffe}, {Jones}, {Karakci},
  {Keih{\"a}nen}, {Keskitalo}, {Kiiveri}, {Kim}, {Kisner}, {Knox},
  {Krachmalnicoff}, {Kunz}, {Kurki-Suonio}, {Lagache}, {Lamarre}, {Lasenby},
  {Lattanzi}, {Lawrence}, {Le Jeune}, {Lemos}, {Lesgourgues}, {Levrier},
  {Lewis}, {Liguori}, {Lilje}, {Lilley}, {Lindholm}, {L{\'o}pez-Caniego},
  {Lubin}, {Ma}, {Mac{\'\i}as-P{\'e}rez}, {Maggio}, {Maino}, {Mandolesi},
  {Mangilli}, {Marcos-Caballero}, {Maris}, {Martin}, {Martinelli},
  {Mart{\'\i}nez-Gonz{\'a}lez}, {Matarrese}, {Mauri}, {McEwen}, {Meinhold},
  {Melchiorri}, {Mennella}, {Migliaccio}, {Millea}, {Mitra},
  {Miville-Desch{\^e}nes}, {Molinari}, {Montier}, {Morgante}, {Moss}, {Natoli},
  {N{\o}rgaard-Nielsen}, {Pagano}, {Paoletti}, {Partridge}, {Patanchon},
  {Peiris}, {Perrotta}, {Pettorino}, {Piacentini}, {Polastri}, {Polenta},
  {Puget}, {Rachen}, {Reinecke}, {Remazeilles}, {Renzi}, {Rocha}, {Rosset},
  {Roudier}, {Rubi{\~n}o-Mart{\'\i}n}, {Ruiz-Granados}, {Salvati}, {Sandri},
  {Savelainen}, {Scott}, {Shellard}, {Sirignano}, {Sirri}, {Spencer},
  {Sunyaev}, {Suur-Uski}, {Tauber}, {Tavagnacco}, {Tenti}, {Toffolatti},
  {Tomasi}, {Trombetti}, {Valenziano}, {Valiviita}, {Van Tent}, {Vibert},
  {Vielva}, {Villa}, {Vittorio}, {Wandelt}, {Wehus}, {White}, {White},
  {Zacchei}, \& {Zonca}}]{planck18}
{Planck Collaboration}, {Aghanim}, N., {Akrami}, Y., {et~al.} 2020, \aap, 641,
  A6

\bibitem[{Pons {et~al.}(2020)Pons, McMahon, Banerji, \& Reed}]{pons20}
Pons, E., McMahon, R.~G., Banerji, M., \& Reed, S.~L. 2020, Monthly Notices of
  the Royal Astronomical Society, 491, 3884

\bibitem[{{Predehl} {et~al.}(2020){Predehl}, {Andritschke}, {Arefiev},
  {Babyshkin}, {Batanov}, {Becker}, {B{\"o}hringer}, {Bogomolov}, {Boller},
  {Borm}, {Bornemann}, {Br{\"a}uninger}, {Br{\"u}ggen}, {Brunner}, {Brusa},
  {Bulbul}, {Buntov}, {Burwitz}, {Burkert}, {Clerc}, {Churazov}, {Coutinho},
  {Dauser}, {Dennerl}, {Doroshenko}, {Eder}, {Emberger}, {Eraerds},
  {Finoguenov}, {Freyberg}, {Friedrich}, {Friedrich}, {F{\"u}rmetz},
  {Georgakakis}, {Gilfanov}, {Granato}, {Grossberger}, {Gueguen}, {Gureev},
  {Haberl}, {H{\"a}lker}, {Hartner}, {Hasinger}, {Huber}, {Ji}, {Kienlin},
  {Kink}, {Korotkov}, {Kreykenbohm}, {Lamer}, {Lomakin}, {Lapshov}, {Liu},
  {Maitra}, {Meidinger}, {Menz}, {Merloni}, {Mernik}, {Mican}, {Mohr},
  {M{\"u}ller}, {Nandra}, {Nazarov}, {Pacaud}, {Pavlinsky}, {Perinati},
  {Pfeffermann}, {Pietschner}, {Ramos-Ceja}, {Rau}, {Reiffers}, {Reiprich},
  {Robrade}, {Salvato}, {Sanders}, {Santangelo}, {Sasaki}, {Scheuerle},
  {Schmid}, {Schmitt}, {Schwope}, {Shirshakov}, {Steinmetz}, {Stewart},
  {Str{\"u}der}, {Sunyaev}, {Tenzer}, {Tiedemann}, {Tr{\"u}mper}, {Voron},
  {Weber}, {Wilms}, \& {Yaroshenko}}]{predehl20}
{Predehl}, P., {Andritschke}, R., {Arefiev}, V., {et~al.} 2020, arXiv e-prints,
  arXiv:2010.03477

\bibitem[{Reed {et~al.}(2015)Reed, McMahon, Banerji, Becker, Gonzalez-Solares,
  Martini, Ostrovski, Rauch, Abbott, Abdalla, Allam, Benoit-Levy, Bertin,
  Buckley-Geer, Burke, Carnero~Rosell, da~Costa, D'Andrea, DePoy, Desai, Diehl,
  Doel, Cunha, Estrada, Evrard, Fausti~Neto, Finley, Fosalba, Frieman, Gruen,
  Honscheid, James, Kent, Kuehn, Kuropatkin, Lahav, Maia, Makler, Marshall,
  Merritt, Miquel, Mohr, Nord, Ogando, Plazas, Romer, Roodman, Rykoff, Sako,
  Sanchez, Santiago, Schubnell, Sevilla, Smith, Soares-Santos, Suchyta,
  Swanson, Tarle, Thomas, Tucker, Walker, \& Wechsler}]{reed15}
Reed, S.~L., McMahon, R.~G., Banerji, M., {et~al.} 2015, Monthly Notices of the
  Royal Astronomical Society, 454, 3952

\bibitem[{{Reeves} {et~al.}(1997){Reeves}, {Turner}, {Ohashi}, \&
  {Kii}}]{reeves97}
{Reeves}, J.~N., {Turner}, M.~J.~L., {Ohashi}, T., \& {Kii}, T. 1997, \mnras,
  292, 468

\bibitem[{{Robertson}(2010)}]{robertson10}
{Robertson}, B.~E. 2010, \apjl, 716, L229

\bibitem[{{Rosen} {et~al.}(2016){Rosen}, {Webb}, {Watson}, {Ballet}, {Barret},
  {Braito}, {Carrera}, {Ceballos}, {Coriat}, {Della Ceca}, {Denkinson},
  {Esquej}, {Farrell}, {Freyberg}, {Gris{\'e}}, {Guillout}, {Heil},
  {Koliopanos}, {Law-Green}, {Lamer}, {Lin}, {Martino}, {Michel}, {Motch},
  {Nebot Gomez-Moran}, {Page}, {Page}, {Page}, {Pakull}, {Pye}, {Read},
  {Rodriguez}, {Sakano}, {Saxton}, {Schwope}, {Scott}, {Sturm}, {Traulsen},
  {Yershov}, \& {Zolotukhin}}]{rosen16}
{Rosen}, S.~R., {Webb}, N.~A., {Watson}, M.~G., {et~al.} 2016, \aap, 590, A1

\bibitem[{{Runnoe} {et~al.}(2012){Runnoe}, {Brotherton}, \& {Shang}}]{runnoe12}
{Runnoe}, J.~C., {Brotherton}, M.~S., \& {Shang}, Z. 2012, \mnras, 422, 478

\bibitem[{Salvato {et~al.}(2018)Salvato, Buchner, Budavári, Dwelly, Merloni,
  Brusa, Rau, Fotopoulou, \& Nandra}]{salvato18}
Salvato, M., Buchner, J., Budavári, T., {et~al.} 2018, Monthly Notices of the
  Royal Astronomical Society, 473, 4937

\bibitem[{{Schlafly} {et~al.}(2019){Schlafly}, {Meisner}, \&
  {Green}}]{schlafly19}
{Schlafly}, E.~F., {Meisner}, A.~M., \& {Green}, G.~M. 2019, \apjs, 240, 30

\bibitem[{Seabold \& Perktold(2010)}]{seabold10}
Seabold, S. \& Perktold, J. 2010, in 9th Python in Science Conference

\bibitem[{{Shang} {et~al.}(2011){Shang}, {Brotherton}, {Wills}, {Wills},
  {Cales}, {Dale}, {Green}, {Runnoe}, {Nemmen}, {Gallagher}, {Ganguly},
  {Hines}, {Kelly}, {Kriss}, {Li}, {Tang}, \& {Xie}}]{shang2011}
{Shang}, Z., {Brotherton}, M.~S., {Wills}, B.~J., {et~al.} 2011, \apjs, 196, 2

\bibitem[{{Shastri} {et~al.}(1993){Shastri}, {Wilkes}, {Elvis}, \&
  {McDowell}}]{shastri93}
{Shastri}, P., {Wilkes}, B.~J., {Elvis}, M., \& {McDowell}, J. 1993, \apj, 410,
  29

\bibitem[{{Shen} {et~al.}(2020){Shen}, {Hopkins}, {Faucher-Gigu{\`e}re},
  {Alexander}, {Richards}, {Ross}, \& {Hickox}}]{shen20}
{Shen}, X., {Hopkins}, P.~F., {Faucher-Gigu{\`e}re}, C.-A., {et~al.} 2020,
  \mnras, 495, 3252

\bibitem[{Shimwell {et~al.}(2017)Shimwell, R{\"{o}}ttgering, Best, Williams,
  Dijkema, de~Gasperin, Hardcastle, Heald, Hoang, Horneffer, Intema, Mahony,
  Mandal, Mechev, Morabito, Oonk, Rafferty, Retana-Montenegro, Sabater, Tasse,
  van Weeren, Br{\"{u}}ggen, Brunetti, Chy{\.{z}}y, Conway, Haverkorn, Jackson,
  Jarvis, McKean, Miley, Morganti, White, Wise, van Bemmel, Beck, Brienza,
  Bonafede, {Calistro Rivera}, Cassano, Clarke, Cseh, Deller, Drabent, van
  Driel, Engels, Falcke, Ferrari, Fr{\"{o}}hlich, Garrett, Harwood, Heesen,
  Hoeft, Horellou, Israel, Kapi{\'{n}}ska, Kunert-Bajraszewska, McKay, Mohan,
  Orr{\'{u}}, Pizzo, Prandoni, Schwarz, Shulevski, Sipior, Smith, Sridhar,
  Steinmetz, Stroe, Varenius, van~der Werf, Zensus, \& Zwart}]{Shimwell2017}
Shimwell, T.~W., R{\"{o}}ttgering, H. J.~A., Best, P.~N., {et~al.} 2017,
  A{\&}A, 598, A104

\bibitem[{Shimwell {et~al.}(2019)Shimwell, Tasse, Hardcastle, Mechev, Williams,
  Best, R{\"{o}}ttgering, Callingham, Dijkema, {De Gasperin}, \&
  Others}]{Shimwell2019}
Shimwell, T.~W., Tasse, C., Hardcastle, M.~J., {et~al.} 2019, A{\&}A, 622, A1

\bibitem[{{Siemiginowska} {et~al.}(2008){Siemiginowska}, {LaMassa}, {Aldcroft},
  {Bechtold}, \& {Elvis}}]{siemiginowska08}
{Siemiginowska}, A., {LaMassa}, S., {Aldcroft}, T.~L., {Bechtold}, J., \&
  {Elvis}, M. 2008, \apj, 684, 811

\bibitem[{{Simmonds} {et~al.}(2018){Simmonds}, {Buchner}, {Salvato}, {Hsu}, \&
  {Bauer}}]{simmonds18}
{Simmonds}, C., {Buchner}, J., {Salvato}, M., {Hsu}, L.~T., \& {Bauer}, F.~E.
  2018, \aap, 618, A66

\bibitem[{Smirnov \& Tasse(2015)}]{Smirnov2015}
Smirnov, O.~M. \& Tasse, C. 2015, MNRAS, 449, 2668

\bibitem[{{Stark} {et~al.}(1992){Stark}, {Gammie}, {Wilson}, {Bally}, {Linke},
  {Heiles}, \& {Hurwitz}}]{stark92}
{Stark}, A.~A., {Gammie}, C.~F., {Wilson}, R.~W., {et~al.} 1992, \apjs, 79, 77

\bibitem[{Stern {et~al.}(2003)Stern, Hall, Barrientos, Bunker, Elston, Ledlow,
  Raines, \& Willis}]{stern03}
Stern, D., Hall, P.~B., Barrientos, L.~F., {et~al.} 2003, The Astrophysical
  Journal Letters, 596, L39

\bibitem[{{Strateva} {et~al.}(2005){Strateva}, {Brandt}, {Schneider}, {Vanden
  Berk}, \& {Vignali}}]{strateva05}
{Strateva}, I.~V., {Brandt}, W.~N., {Schneider}, D.~P., {Vanden Berk}, D.~G.,
  \& {Vignali}, C. 2005, \aj, 130, 387

\bibitem[{Sutherland \& Saunders(1992)}]{sutherland92}
Sutherland, W. \& Saunders, W. 1992, Monthly Notices of the Royal Astronomical
  Society, 259, 413

\bibitem[{{Tananbaum} {et~al.}(1979){Tananbaum}, {Avni}, {Branduardi}, {Elvis},
  {Fabbiano}, {Feigelson}, {Giacconi}, {Henry}, {Pye}, {Soltan}, \&
  {Zamorani}}]{tananbaum79}
{Tananbaum}, H., {Avni}, Y., {Branduardi}, G., {et~al.} 1979, \apjl, 234, L9

\bibitem[{{Tasse}(2014{\natexlab{a}})}]{Tasse2014a}
{Tasse}, C. 2014{\natexlab{a}}, arXiv e-prints, arXiv:1410.8706

\bibitem[{{Tasse}(2014{\natexlab{b}})}]{Tasse2014b}
{Tasse}, C. 2014{\natexlab{b}}, \aap, 566, A127

\bibitem[{Tasse {et~al.}(2018)Tasse, Hugo, Mirmont, Smirnov, Atemkeng, Bester,
  Hardcastle, Lakhoo, Perkins, \& Shimwell}]{Tasse2018}
Tasse, C., Hugo, B., Mirmont, M., {et~al.} 2018, A{\&}A, 611, 1

\bibitem[{{Tavecchio} {et~al.}(2000){Tavecchio}, {Maraschi}, {Sambruna}, \&
  {Urry}}]{tavecchio00}
{Tavecchio}, F., {Maraschi}, L., {Sambruna}, R.~M., \& {Urry}, C.~M. 2000,
  \apjl, 544, L23

\bibitem[{{Tingay} \& {de Kool}(2003)}]{2003AJ....126..723T}
{Tingay}, S.~J. \& {de Kool}, M. 2003, \aj, 126, 723

\bibitem[{{Trenti} \& {Stiavelli}(2008)}]{trenti08}
{Trenti}, M. \& {Stiavelli}, M. 2008, \apj, 676, 767

\bibitem[{Ueda {et~al.}(2014)Ueda, Akiyama, Hasinger, Miyaji, \&
  Watson}]{ueda14}
Ueda, Y., Akiyama, M., Hasinger, G., Miyaji, T., \& Watson, M.~G. 2014, The
  Astrophysical Journal, 786, 104

\bibitem[{{van Haarlem} {et~al.}(2013){van Haarlem}, {Wise}, {Gunst}, {Heald},
  {McKean}, {Hessels}, {de Bruyn}, {Nijboer}, {Swinbank}, {Fallows},
  {Brentjens}, {Nelles}, {Beck}, {Falcke}, {Fender}, {H{\"o}randel},
  {Koopmans}, {Mann}, {Miley}, {R{\"o}ttgering}, {Stappers}, {Wijers},
  {Zaroubi}, {van den Akker}, {Alexov}, {Anderson}, {Anderson}, {van Ardenne},
  {Arts}, {Asgekar}, {Avruch}, {Batejat}, {B{\"a}hren}, {Bell}, {Bell}, {van
  Bemmel}, {Bennema}, {Bentum}, {Bernardi}, {Best}, {B{\^\i}rzan}, {Bonafede},
  {Boonstra}, {Braun}, {Bregman}, {Breitling}, {van de Brink}, {Broderick},
  {Broekema}, {Brouw}, {Br{\"u}ggen}, {Butcher}, {van Cappellen}, {Ciardi},
  {Coenen}, {Conway}, {Coolen}, {Corstanje}, {Damstra}, {Davies}, {Deller},
  {Dettmar}, {van Diepen}, {Dijkstra}, {Donker}, {Doorduin}, {Dromer}, {Drost},
  {van Duin}, {Eisl{\"o}ffel}, {van Enst}, {Ferrari}, {Frieswijk}, {Gankema},
  {Garrett}, {de Gasperin}, {Gerbers}, {de Geus}, {Grie{\ss}meier}, {Grit},
  {Gruppen}, {Hamaker}, {Hassall}, {Hoeft}, {Holties}, {Horneffer}, {van der
  Horst}, {van Houwelingen}, {Huijgen}, {Iacobelli}, {Intema}, {Jackson},
  {Jelic}, {de Jong}, {Juette}, {Kant}, {Karastergiou}, {Koers}, {Kollen},
  {Kondratiev}, {Kooistra}, {Koopman}, {Koster}, {Kuniyoshi}, {Kramer},
  {Kuper}, {Lambropoulos}, {Law}, {van Leeuwen}, {Lemaitre}, {Loose}, {Maat},
  {Macario}, {Markoff}, {Masters}, {McFadden}, {McKay-Bukowski}, {Meijering},
  {Meulman}, {Mevius}, {Middelberg}, {Millenaar}, {Miller-Jones}, {Mohan},
  {Mol}, {Morawietz}, {Morganti}, {Mulcahy}, {Mulder}, {Munk}, {Nieuwenhuis},
  {van Nieuwpoort}, {Noordam}, {Norden}, {Noutsos}, {Offringa}, {Olofsson},
  {Omar}, {Orr{\'u}}, {Overeem}, {Paas}, {Pand ey-Pommier}, {Pandey}, {Pizzo},
  {Polatidis}, {Rafferty}, {Rawlings}, {Reich}, {de Reijer}, {Reitsma},
  {Renting}, {Riemers}, {Rol}, {Romein}, {Roosjen}, {Ruiter}, {Scaife}, {van
  der Schaaf}, {Scheers}, {Schellart}, {Schoenmakers}, {Schoonderbeek},
  {Serylak}, {Shulevski}, {Sluman}, {Smirnov}, {Sobey}, {Spreeuw}, {Steinmetz},
  {Sterks}, {Stiepel}, {Stuurwold}, {Tagger}, {Tang}, {Tasse}, {Thomas},
  {Thoudam}, {Toribio}, {van der Tol}, {Usov}, {van Veelen}, {van der Veen},
  {ter Veen}, {Verbiest}, {Vermeulen}, {Vermaas}, {Vocks}, {Vogt}, {de Vos},
  {van der Wal}, {van Weeren}, {Weggemans}, {Weltevrede}, {White}, {Wijnholds},
  {Wilhelmsson}, {Wucknitz}, {Yatawatta}, {Zarka}, {Zensus}, \& {van
  Zwieten}}]{vanhaarlem13}
{van Haarlem}, M.~P., {Wise}, M.~W., {Gunst}, A.~W., {et~al.} 2013, \aap, 556,
  A2

\bibitem[{van Weeren {et~al.}(2016)van Weeren, Williams, Hardcastle, Shimwell,
  Rafferty, Sabater, Heald, Sridhar, Dijkema, Brunetti, Br{\"{u}}ggen,
  Andrade-Santos, Ogrean, R{\"{o}}ttgering, Dawson, Forman, de~Gasperin, Jones,
  Miley, Rudnick, Sarazin, Bonafede, Best, BiÌrzan, Cassano, Chy{\.{z}}y,
  Croston, Ensslin, Ferrari, Hoeft, Horellou, Jarvis, Kraft, Mevius, Intema,
  Murray, Orr{\'{u}}, Pizzo, Simionescu, Stroe, van~der Tol, \&
  White}]{VanWeeren2016a}
van Weeren, R.~J., Williams, W.~L., Hardcastle, M.~J., {et~al.} 2016,
  Astrophys. J. Suppl. Ser., 223, 2

\bibitem[{Venemans {et~al.}(2013)Venemans, Findlay, Sutherland, Rosa, McMahon,
  Simcoe, González-Solares, Kuijken, \& Lewis}]{venemans13}
Venemans, B.~P., Findlay, J.~R., Sutherland, W.~J., {et~al.} 2013, The
  Astrophysical Journal, 779, 24, publisher: IOP Publishing

\bibitem[{{Venemans} {et~al.}(2015){Venemans}, {Verdoes Kleijn}, {Mwebaze},
  {Valentijn}, {Ba{\~n}ados}, {Decarli}, {de Jong}, {Findlay}, {Kuijken}, {La
  Barbera}, {McFarland}, {McMahon}, {Napolitano}, {Sikkema}, \&
  {Sutherland}}]{venemans15}
{Venemans}, B.~P., {Verdoes Kleijn}, G.~A., {Mwebaze}, J., {et~al.} 2015,
  \mnras, 453, 2259

\bibitem[{{Vignali} {et~al.}(2002){Vignali}, {Bauer}, {Alexander}, {Brand t},
  {Hornschemeier}, {Schneider}, \& {Garmire}}]{vignali02}
{Vignali}, C., {Bauer}, F.~E., {Alexander}, D.~M., {et~al.} 2002, \apjl, 580,
  L105

\bibitem[{Vito {et~al.}(2019)Vito, Brandt, Bauer, Calura, Gilli, Luo, Shemmer,
  Vignali, Zamorani, Brusa, Civano, Comastri, \& Nanni}]{vito19}
Vito, F., Brandt, W.~N., Bauer, F.~E., {et~al.} 2019, Astronomy and
  Astrophysics, 630, A118

\bibitem[{Vito {et~al.}(2018)Vito, Brandt, Yang, Gilli, Luo, Vignali, Xue,
  Comastri, Koekemoer, Lehmer, Liu, Paolillo, Ranalli, Schneider, Shemmer,
  Volonteri, \& Wang}]{vito18}
Vito, F., Brandt, W.~N., Yang, G., {et~al.} 2018, Monthly Notices of the Royal
  Astronomical Society, 473, 2378

\bibitem[{Vito {et~al.}(2014)Vito, Gilli, Vignali, Comastri, Brusa, Cappelluti,
  \& Iwasawa}]{vito14}
Vito, F., Gilli, R., Vignali, C., {et~al.} 2014, Monthly Notices of the Royal
  Astronomical Society, 445, 3557

\bibitem[{Waller \& Turnbull(1992)}]{waller}
Waller, L. \& Turnbull, B. 1992, American Statistician - AMER STATIST, 46, 5

\bibitem[{Wang {et~al.}(2017)Wang, Fan, Yang, Wu, Yang, Bian, McGreer, Li, Li,
  Ding, Dey, Dye, Findlay, Green, James, Jiang, Lang, Lawrence, Myers, Ross,
  Schlegel, \& Shanks}]{wang17}
Wang, F., Fan, X., Yang, J., {et~al.} 2017, The Astrophysical Journal, 839, 27

\bibitem[{{White} {et~al.}(1997){White}, {Becker}, {Helfand}, \&
  {Gregg}}]{white97}
{White}, R.~L., {Becker}, R.~H., {Helfand}, D.~J., \& {Gregg}, M.~D. 1997,
  \apj, 475, 479

\bibitem[{Whiting \& Humphreys(2012)}]{whiting12}
Whiting, M. \& Humphreys, B. 2012, Publications of the Astronomical Society of
  Australia, 29, 371–381

\bibitem[{{Wilkes} \& {Elvis}(1987)}]{wilkes87}
{Wilkes}, B.~J. \& {Elvis}, M. 1987, \apj, 323, 243

\bibitem[{Williams {et~al.}(2016)Williams, van Weeren, R{\"{o}}ttgering, Best,
  Dijkema, de~Gasperin, Hardcastle, Heald, Prandoni, Sabater, Shimwell, Tasse,
  van Bemmel, Br{\"{u}}ggen, Brunetti, Conway, En{\ss}lin, Engels, Falcke,
  Ferrari, Haverkorn, Jackson, Jarvis, Kapi{\'{n}}ska, Mahony, Miley, Morabito,
  Morganti, Orr{\'{u}}, Retana-Montenegro, Sridhar, Toribio, White, Wise, \&
  Zwart}]{Williams2016a}
Williams, W.~L., van Weeren, R.~J., R{\"{o}}ttgering, H. J.~A., {et~al.} 2016,
  MNRAS, 460, 2385

\bibitem[{{Willott} {et~al.}(2009){Willott}, {Delorme}, {Reyl{\'e}}, {Albert},
  {Bergeron}, {Crampton}, {Delfosse}, {Forveille}, {Hutchings}, {McLure},
  {Omont}, \& {Schade}}]{willott09}
{Willott}, C.~J., {Delorme}, P., {Reyl{\'e}}, C., {et~al.} 2009, \aj, 137, 3541

\bibitem[{Willott {et~al.}(2010)Willott, Delorme, Reylé, Albert, Bergeron,
  Crampton, Delfosse, Forveille, Hutchings, McLure, Omont, \&
  Schade}]{willott10}
Willott, C.~J., Delorme, P., Reylé, C., {et~al.} 2010, The Astronomical
  Journal, 139, 906

\bibitem[{{Wright} {et~al.}(2010){Wright}, {Eisenhardt}, {Mainzer}, {Ressler},
  {Cutri}, {Jarrett}, {Kirkpatrick}, {Padgett}, {McMillan}, {Skrutskie},
  {Stanford}, {Cohen}, {Walker}, {Mather}, {Leisawitz}, {Gautier}, {McLean},
  {Benford}, {Lonsdale}, {Blain}, {Mendez}, {Irace}, {Duval}, {Liu}, {Royer},
  {Heinrichsen}, {Howard}, {Shannon}, {Kendall}, {Walsh}, {Larsen}, {Cardon},
  {Schick}, {Schwalm}, {Abid}, {Fabinsky}, {Naes}, \& {Tsai}}]{wright10}
{Wright}, E.~L., {Eisenhardt}, P. R.~M., {Mainzer}, A.~K., {et~al.} 2010, \aj,
  140, 1868

\bibitem[{Wu {et~al.}(2015)Wu, Wang, Fan, Yi, Zuo, Bian, Jiang, McGreer, Wang,
  Yang, Yang, Thompson, \& Beletsky}]{wu15}
Wu, X.-B., Wang, F., Fan, X., {et~al.} 2015, Nature, 518, 512

\bibitem[{{Zhu} {et~al.}(2020){Zhu}, {Brandt}, {Luo}, {Wu}, {Xue}, \&
  {Yang}}]{zhu20}
{Zhu}, S.~F., {Brandt}, W.~N., {Luo}, B., {et~al.} 2020, \mnras, 496, 245

\end{thebibliography}

\begin{appendix} 
\section{Accounting for eFEDS sensitivity}
\label{section:sensitivity}

\begin{figure}
\includegraphics[width=8 cm]{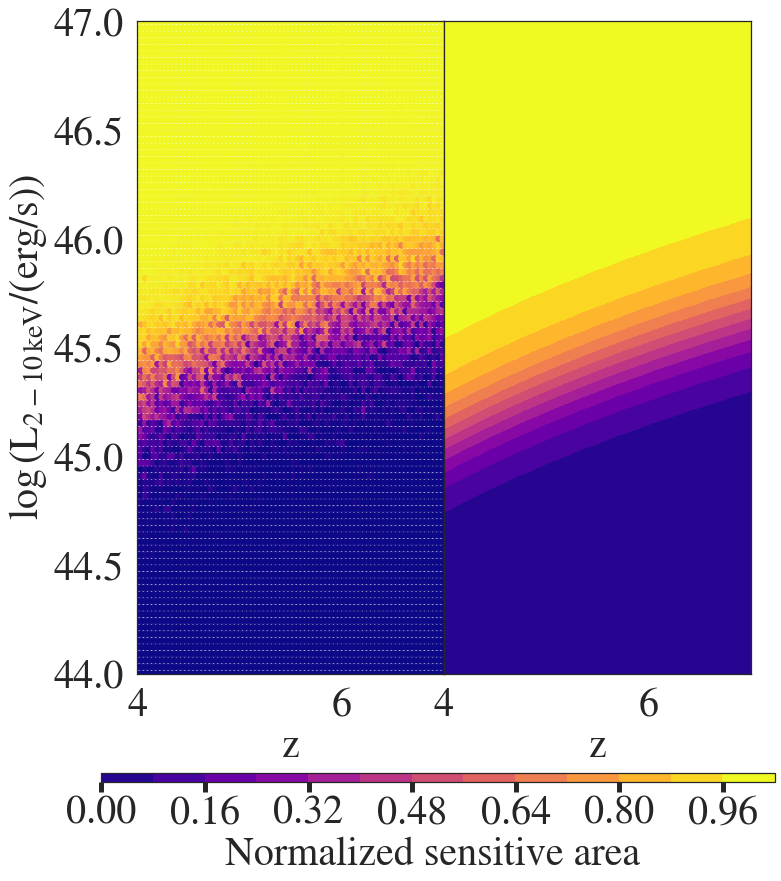}
\caption{\textit{Left}: Sensitive area as a function of luminosity and redshift from synthetic spectra. \textit{Right}: Fit to the $(L_{2-10 \, \mathrm{keV}},z,A(L_{2-10 \, \mathrm{keV}},\mathrm{log}(1+z)))$ surface (used for efficient integration).}
\label{fig:sim_sens}
\centering
\end{figure}

The sensitive area $ A(\mathrm{log} \, L_\mathrm{X}, z)$ is derived from the eFEDS sensitivity curve which expresses the sensitive area as a function of counts in the range $\rm 0.2 - 2.3 \, keV$. It was generated with the eSASS task \texttt{APETOOL} \citep[for more details on the sensitivity determination please see][]{georgakakis08}. In Eq. \ref{eq:estimate}, we integrate over redshifts and luminosity. In order to convert a given redshift--luminosity in soft band counts, we simulated X-ray spectra with \texttt{XSPEC} using a model \textit{clumin*tbabs*zpowerlw}. This model corresponds to a redshifted power law with Galactic absorption. We have frozen the photon-index to $\Gamma = 2$, a value which is consistent with results from spectral analysis carried out at $z>5.7$ \citep{nanni17,vito19}. The Galactic column density was fixed to the values: $3\times 10^{20} \mathrm{cm^{-2}}$. The convolution model \textit{clumin} was used to fix the values of redshift and luminosity for a given $z-L_\mathrm{X}$ configuration. ARF and RMF files for a standard eFEDS source were used.
We have generated a grid of spectra over the ranges $z=4-8$ and $\mathrm{log} \, (L_{2-10 \, \mathrm{keV}}/\mathrm{(erg/s)})=44-47$. For each spectrum, the count-rate in the range $\rm 0.2 - 2.3 \, keV$, was computed and the normalised area sensitivity was evaluated with the \texttt{APETOOL} sensitivity curve. This sensitivity grid is shown in Fig. 1. The synthetic $ (L_{2-10 \, \mathrm{keV}},z,A(L_{2-10 \, \mathrm{keV}},z))$ data display a sharp break and are distributed as a multivariate sigmoid. We fitted $ A(L_{2-10 \mathrm{keV}},\mathrm{log}(1+z))$ using logistic regression. We note that we fit in $\mathrm{log}(1+z)$ and not $z$ in order to capture the evolution of the flux limit with $\sim L/(1+z)^{-4}$. The regression was implemented with a generalised linear model \citep[GLM, e.g.][]{nelder72} using the \texttt{statsmodels} Python library \citep{seabold10}. A GLM is a regression model for which the probability density function of the outcome variable can be specified. In the case of $A(L_{2-10 \, \mathrm{keV}},\mathrm{log}(1+z))$, we select the binomial distribution with \textit{logit} as link function. The resulting fitted surface is shown in Fig. \ref{fig:sim_sens} This function can be used in the integral of Eq. \ref{eq:estimate}.

\end{appendix}

\end{document}